\documentclass[11pt,a4paper]{article}
\usepackage{jheppub}
\usepackage{epsf}
\usepackage{amsfonts}
\usepackage{bbold}
\usepackage{graphicx}
\usepackage{graphics}
\usepackage{epstopdf}
\usepackage{url}
\usepackage{axodraw2}
\usepackage{pstricks}
\usepackage{color}
\usepackage{relsize,exscale,scalefnt,anyfontsize,cases}
\usepackage[utf8]{inputenc}

\newcommand{\unit}{1\!\!1}
\newcommand{\nn}{\nonumber} 
\newcommand{\beq}{\begin{equation}}
\newcommand{\eeq}{\end{equation}} 
\newcommand{\beqa}{\begin{eqnarray}} 
\newcommand{\eeqa}{\end{eqnarray}}

\def\bc{\begin{center}}
\def\ec{\end{center}}
\def\bi{\begin{itemize}}
\def\ei{\end{itemize}}
\def\be{\begin{equation}}
 \def\ee{\end{equation}}
\def\ben{\begin{equation*}}
 \def\een{\end{equation*}}
 \def\bea{\begin{eqnarray}}
 \def\eea{\end{eqnarray}}
 \def\bean{\begin{eqnarray*}}
 \def\eean{\end{eqnarray*}}
\newcommand{\ie}{{\em i.e.}}  \newcommand{\eg}{{\em e.g.}}

\newcommand{\lqcd}{\Lambda_{_{\rm QCD}}} 
  
\newcommand{\morder}[1]{{\cal O}\left(#1 \right)}
\newcommand{\eq}[1]{(\ref{#1})}
\newcommand{\ed}{\end{document}}

\newcommand{\ket}[1]{\vert{#1}\rangle}
\newcommand{\bra}[1]{\langle{#1}\vert}
\newcommand{\ave}[1]{\langle{#1}\rangle}

\newcommand{\tr}{\mathrm{Tr}\,}  

\newcommand{\B}{{\cal B}}

\newcommand{\zerovec}{{\boldsymbol 0}}
\newcommand{\avec}{{\boldsymbol a}}

\newcommand{\pvec}{{\boldsymbol p}}
\newcommand{\rvec}{{\boldsymbol r}}
\newcommand{\ellvec}{{\boldsymbol \ell}}

\newcommand{\qvec}{{\boldsymbol q}}
\newcommand{\xvec}{{\boldsymbol x}}

\newcommand{\uvec}{{\boldsymbol u}}
\newcommand{\vvec}{{\boldsymbol v}}

\newcommand{\Pvec}{{\boldsymbol P}}
\newcommand{\Rvec}{{\boldsymbol R}}
\newcommand{\nablavec}{{\boldsymbol \nabla}}

\newcommand{\jpsi}{{\mathrm J}/\psi}

\newcommand{\pp}{p--p}
\newcommand{\pA}{p--A}

\newcommand{\dd}{{\rm d}}
\newcommand{\half}{\frac{1}{2}}  \newcommand{\halft}{{\textstyle \frac{1}{2}}}

\newcommand{\lsim}{\lesssim} \newcommand{\gsim}{\gtrsim}

 \def\esim{\,\mathrel{\rlap{\lower0.2em\hbox{$-$}}\raise0.15em\hbox{\footnotesize $\hskip0.04em\sim$}}\,}
 \def\gsim{\mathrel{\rlap{\lower0.2em\hbox{$\sim$}}\raise0.2em\hbox{$>$}}}
 \def\ksim{\mathrel{\rlap{\lower0.2em\hbox{$\sim$}}\raise0.2em\hbox{$<$}}}

\def \OBKet (#1,#2,#3) {
\resizebox{#2 mm}{!}{\raisebox{#3 pt}{
\begin{axopicture}(102,136) (223,-164)
    \SetWidth{2.0}
    \SetColor{Black}
    \Line[arrow,arrowpos=0.5,arrowlength=17.333,arrowwidth=6.933,arrowinset=0.2](288,-120)(224,-120)
    \Line[arrow,arrowpos=0.5,arrowlength=17.333,arrowwidth=6.933,arrowinset=0.2](224,-72)(288,-72)
    \Line(288,-72)(288,-120)
    \SetWidth{1.0}
     \Bezier(224,-40)(223.45,-36.291)(228.496,-31.83)(234.093,-31.079) \Bezier(234.093,-31.079)(245.286,-29.577)(246.812,-44.499)(240.285,-45.317) \Bezier(240.285,-45.317)(233.758,-46.134)(231.556,-31.297)(246.017,-29.906) \Bezier(246.017,-29.906)(260.478,-28.515)(260.96,-43.508)(254.49,-43.942) \Bezier(254.49,-43.942)(248.02,-44.376)(246.495,-29.454)(261.167,-29.158) \Bezier(261.167,-29.158)(275.838,-28.863)(274.564,-43.808)(268.267,-43.642) \Bezier(268.267,-43.642)(261.97,-43.475)(261.488,-28.483)(275.913,-30.138) \Bezier(275.913,-30.138)(290.339,-31.793)(285.775,-46.082)(280.02,-44.932) \Bezier(280.02,-44.932)(274.264,-43.783)(275.539,-28.837)(289.498,-34.3) \Bezier(289.498,-34.3)(303.458,-39.762)(293.231,-50.735)(288.963,-48.236) \Bezier(288.963,-48.236)(284.695,-45.737)(289.259,-31.448)(299.9,-42.096) \Bezier(299.9,-42.096)(310.541,-52.743)(296.073,-56.704)(293.774,-52.9) \Bezier(293.774,-52.9)(291.474,-49.097)(301.701,-38.124)(306.963,-52.699) \Bezier(306.963,-52.699)(312.225,-67.273)(297.228,-66.986)(296.534,-61.418) \Bezier(296.534,-61.418)(295.84,-55.851)(310.308,-51.89)(311.041,-66.912) \Bezier(311.041,-66.912)(311.775,-81.934)(296.789,-81.273)(296.999,-74.621) \Bezier(296.999,-74.621)(297.209,-67.968)(312.207,-68.255)(311.85,-78.292) \Bezier(311.85,-78.292)(311.671,-83.311)(307.746,-88.165)(304,-88)
     \Bezier(304,-104)(307.747,-103.842)(311.693,-108.359)(311.892,-113.033) \Bezier(311.892,-113.033)(312.29,-122.382)(297.305,-123.035)(297.015,-116.286) \Bezier(297.015,-116.286)(296.726,-109.536)(311.713,-108.905)(311.646,-121.957) \Bezier(311.646,-121.957)(311.579,-135.009)(296.701,-133.098)(296.957,-127.02) \Bezier(296.957,-127.02)(297.213,-120.941)(312.199,-120.289)(309.049,-134.77) \Bezier(309.049,-134.77)(305.898,-149.252)(293.454,-140.877)(295.16,-136.345) \Bezier(295.16,-136.345)(296.866,-131.814)(311.744,-133.725)(302.967,-146.171) \Bezier(302.967,-146.171)(294.19,-158.617)(287.748,-145.071)(291.179,-142.114) \Bezier(291.179,-142.114)(294.611,-139.157)(307.055,-147.532)(293.86,-155.064) \Bezier(293.86,-155.064)(280.666,-162.596)(278.316,-147.781)(283.674,-146.121) \Bezier(283.674,-146.121)(289.032,-144.46)(295.475,-158.006)(281.723,-160.791) \Bezier(281.723,-160.791)(267.971,-163.576)(267.723,-148.578)(273.906,-148.039) \Bezier(273.906,-148.039)(280.088,-147.5)(282.437,-162.315)(268.425,-162.765) \Bezier(268.425,-162.765)(254.412,-163.215)(255.407,-148.248)(261.838,-148.408) \Bezier(261.838,-148.408)(268.27,-148.569)(268.518,-163.567)(254.687,-162.763) \Bezier(254.687,-162.763)(240.856,-161.96)(242.612,-147.063)(249.141,-147.664) \Bezier(249.141,-147.664)(255.67,-148.266)(254.675,-163.233)(241.5,-161.74) \Bezier(241.5,-161.74)(228.326,-160.247)(230.588,-145.418)(237.15,-146.305) \Bezier(237.15,-146.305)(243.711,-147.192)(241.955,-162.089)(232.412,-160.752) \Bezier(232.412,-160.752)(227.64,-160.083)(223.434,-155.707)(224,-152)
    \GOval(301,-96)(20,20)(0){0.882}
    \Text(286,-96)[l]{\fontsize{30}{2}\selectfont {$\downarrow \! \! #1$}} 
  \end{axopicture}
}}}

\def \OBKetQbarQ (#1,#2,#3) {
\resizebox{#2 mm}{!}{\raisebox{#3 pt}{
\begin{axopicture}(108,128) (223,-168)
    \SetWidth{2.0}
    \SetColor{Black}
    \Line(312,-48)(312,-160)
    \Line[arrow,arrowpos=0.5,arrowlength=17.333,arrowwidth=6.933,arrowinset=0.2,flip](288,-128)(224,-128)
    \Line[arrow,arrowpos=0.5,arrowlength=17.333,arrowwidth=6.933,arrowinset=0.2,flip](224,-80)(288,-80)
    \Line(288,-80)(288,-128)
    \SetWidth{1.0}
    \GOval(301,-104)(20,20)(0){0.882}
    \SetWidth{2.0}
    \Line[arrow,arrowpos=0.4,arrowlength=17.333,arrowwidth=6.933,arrowinset=0.2,flip](224,-160)(312,-160)
    \Line[arrow,arrowpos=0.35,arrowlength=17.333,arrowwidth=6.933,arrowinset=0.2](224,-48)(312,-48)
    \Text(285,-104)[l]{\fontsize{28}{2}\selectfont {$\downarrow \! \! #1$}} 
  \end{axopicture}
}}}

\def \OBBra (#1,#2,#3) {
\rotatebox[origin=c]{180}{
\resizebox{#2 mm}{!}{\raisebox{#3 pt}{
\begin{axopicture}(102,136) (223,-164)
    \SetWidth{2.0}
    \SetColor{Black}
    \Line[arrow,arrowpos=0.5,arrowlength=17.333,arrowwidth=6.933,arrowinset=0.2](288,-120)(224,-120)
    \Line[arrow,arrowpos=0.5,arrowlength=17.333,arrowwidth=6.933,arrowinset=0.2](224,-72)(288,-72)
    \Line(288,-72)(288,-120)
    \SetWidth{1.0}
     \Bezier(224,-40)(223.45,-36.291)(228.496,-31.83)(234.093,-31.079) \Bezier(234.093,-31.079)(245.286,-29.577)(246.812,-44.499)(240.285,-45.317) \Bezier(240.285,-45.317)(233.758,-46.134)(231.556,-31.297)(246.017,-29.906) \Bezier(246.017,-29.906)(260.478,-28.515)(260.96,-43.508)(254.49,-43.942) \Bezier(254.49,-43.942)(248.02,-44.376)(246.495,-29.454)(261.167,-29.158) \Bezier(261.167,-29.158)(275.838,-28.863)(274.564,-43.808)(268.267,-43.642) \Bezier(268.267,-43.642)(261.97,-43.475)(261.488,-28.483)(275.913,-30.138) \Bezier(275.913,-30.138)(290.339,-31.793)(285.775,-46.082)(280.02,-44.932) \Bezier(280.02,-44.932)(274.264,-43.783)(275.539,-28.837)(289.498,-34.3) \Bezier(289.498,-34.3)(303.458,-39.762)(293.231,-50.735)(288.963,-48.236) \Bezier(288.963,-48.236)(284.695,-45.737)(289.259,-31.448)(299.9,-42.096) \Bezier(299.9,-42.096)(310.541,-52.743)(296.073,-56.704)(293.774,-52.9) \Bezier(293.774,-52.9)(291.474,-49.097)(301.701,-38.124)(306.963,-52.699) \Bezier(306.963,-52.699)(312.225,-67.273)(297.228,-66.986)(296.534,-61.418) \Bezier(296.534,-61.418)(295.84,-55.851)(310.308,-51.89)(311.041,-66.912) \Bezier(311.041,-66.912)(311.775,-81.934)(296.789,-81.273)(296.999,-74.621) \Bezier(296.999,-74.621)(297.209,-67.968)(312.207,-68.255)(311.85,-78.292) \Bezier(311.85,-78.292)(311.671,-83.311)(307.746,-88.165)(304,-88)
     \Bezier(304,-104)(307.747,-103.842)(311.693,-108.359)(311.892,-113.033) \Bezier(311.892,-113.033)(312.29,-122.382)(297.305,-123.035)(297.015,-116.286) \Bezier(297.015,-116.286)(296.726,-109.536)(311.713,-108.905)(311.646,-121.957) \Bezier(311.646,-121.957)(311.579,-135.009)(296.701,-133.098)(296.957,-127.02) \Bezier(296.957,-127.02)(297.213,-120.941)(312.199,-120.289)(309.049,-134.77) \Bezier(309.049,-134.77)(305.898,-149.252)(293.454,-140.877)(295.16,-136.345) \Bezier(295.16,-136.345)(296.866,-131.814)(311.744,-133.725)(302.967,-146.171) \Bezier(302.967,-146.171)(294.19,-158.617)(287.748,-145.071)(291.179,-142.114) \Bezier(291.179,-142.114)(294.611,-139.157)(307.055,-147.532)(293.86,-155.064) \Bezier(293.86,-155.064)(280.666,-162.596)(278.316,-147.781)(283.674,-146.121) \Bezier(283.674,-146.121)(289.032,-144.46)(295.475,-158.006)(281.723,-160.791) \Bezier(281.723,-160.791)(267.971,-163.576)(267.723,-148.578)(273.906,-148.039) \Bezier(273.906,-148.039)(280.088,-147.5)(282.437,-162.315)(268.425,-162.765) \Bezier(268.425,-162.765)(254.412,-163.215)(255.407,-148.248)(261.838,-148.408) \Bezier(261.838,-148.408)(268.27,-148.569)(268.518,-163.567)(254.687,-162.763) \Bezier(254.687,-162.763)(240.856,-161.96)(242.612,-147.063)(249.141,-147.664) \Bezier(249.141,-147.664)(255.67,-148.266)(254.675,-163.233)(241.5,-161.74) \Bezier(241.5,-161.74)(228.326,-160.247)(230.588,-145.418)(237.15,-146.305) \Bezier(237.15,-146.305)(243.711,-147.192)(241.955,-162.089)(232.412,-160.752) \Bezier(232.412,-160.752)(227.64,-160.083)(223.434,-155.707)(224,-152)
    \GOval(301,-96)(20,20)(0){0.882}
    \Text(290,-96)[l]{\rotatebox[origin=c]{180}{\fontsize{30}{2}\selectfont {$\uparrow \! \! #1$}}}
  \end{axopicture}
}}}}

\def \OBBraQQ (#1,#2,#3) {
\rotatebox[origin=c]{180}{
\resizebox{#2 mm}{!}{\raisebox{#3 pt}{
\begin{axopicture}(108,128) (223,-168)
    \SetWidth{2.0}
    \SetColor{Black}
    \Line(312,-48)(312,-160)
    \Line[arrow,arrowpos=0.5,arrowlength=17.333,arrowwidth=6.933,arrowinset=0.2](288,-128)(224,-128)
    \Line[arrow,arrowpos=0.5,arrowlength=17.333,arrowwidth=6.933,arrowinset=0.2](224,-80)(288,-80)
    \Line(288,-80)(288,-128)
    \SetWidth{1.0}
    \GOval(301,-104)(20,20)(0){0.882}
    \SetWidth{2.0}
    \Line[arrow,arrowpos=0.4,arrowlength=17.333,arrowwidth=6.933,arrowinset=0.2,flip](224,-160)(312,-160)
    \Line[arrow,arrowpos=0.35,arrowlength=17.333,arrowwidth=6.933,arrowinset=0.2](224,-48)(312,-48)
    \Text(285,-104)[l]{\rotatebox[origin=c]{180}{\fontsize{30}{2}\selectfont {$\uparrow \! \! #1$}}}
  \end{axopicture}
}}}}

\def \Crossing (#1,#2) {
\resizebox{#1 mm}{!}{\raisebox{#2 pt}{
\begin{axopicture}(108,128) (223,-168)
    \SetWidth{2.0}
    \SetColor{Black}
    \Line(220,-48)(280,-48)
    \Line(220,-160)(280,-160)
    \Line(220,-80)(280,-128)
    \Line(220,-128)(280,-80)   
  \end{axopicture}
}}}

\def \AAidentity (#1,#2) {
\resizebox{#1 mm}{!}{\raisebox{#2 pt}{
\fcolorbox{white}{white}{
  \begin{axopicture}(82,50) (223,-207)
    \SetWidth{2.0}
    \SetColor{Black}
    \Gluon(224,-158)(304,-158){5}{5}
    \Line[dash,dashsize=7.2](224,-206)(304,-206)
  \end{axopicture}
 }}}}

\def \GQQGidentity (#1,#2) {
\resizebox{#1 mm}{!}{\raisebox{#2 pt}{
\fcolorbox{white}{white}{
  \begin{axopicture}(82,126) (223,-169)
    \SetWidth{2.0}
    \SetColor{Black}
    \Line[arrow,arrowpos=0.5,arrowlength=17.333,arrowwidth=6.933,arrowinset=0.2](304,-130)(224,-130)
    \SetWidth{1.0}
    \Gluon(224,-50)(304,-50){6}{5}
    \SetWidth{2.0}
    \Line[arrow,arrowpos=0.5,arrowlength=17.333,arrowwidth=6.933,arrowinset=0.2](224,-82)(304,-82)
    \SetWidth{1.0}
    \Gluon(304,-162)(224,-162){6}{5}
  \end{axopicture}
  }}}}

\def \proj (#1,#2,#3) {
\resizebox{#2 mm}{!}{\raisebox{#3 pt}{
\fcolorbox{white}{white}{
 \begin{axopicture}(156,42) (223,-211)
    \SetWidth{1.0}
    \SetColor{Black}
    \Gluon(314,-180)(378,-180){6}{4}
    \SetWidth{2.0}
    \Line[arrow,arrowpos=0.5,arrowlength=17.333,arrowwidth=6.933,arrowinset=0.2](312,-202)(376,-202)
    \SetWidth{1.0}
    \Gluon(224,-179)(288,-179){6}{4}
    \SetWidth{2.0}
    \Line[arrow,arrowpos=0.5,arrowlength=17.333,arrowwidth=6.933,arrowinset=0.2](224,-202)(288,-202)
    \SetWidth{1.0}
    \GOval(301,-190)(20,20)(0){0.882}
    \Text(291,-191)[l]{\fontsize{30}{2}\selectfont {$#1$}}
  \end{axopicture}}}}}

\def \QGidentitystraight (#1,#2) {
\resizebox{#1 mm}{!}{\raisebox{#2 pt}{
\fcolorbox{white}{white}{
  \begin{axopicture}(114,48) (223,-200)
    \SetWidth{1.0}
    \SetColor{Black}
    \Gluon(224,-165)(336,-165){6}{8}
    \SetWidth{2.0}
    \Line[arrow,arrowpos=0.5,arrowlength=17.333,arrowwidth=6.933,arrowinset=0.2](224,-192)(336,-192) \end{axopicture}}}}}

\def \symmetrymatchinga (#1,#2,#3,#4,#5,#6) {
\resizebox{#5 mm}{!}{\raisebox{#6 pt}{
\fcolorbox{white}{white}{
  \begin{axopicture}(334,194) (167,-191)
    \SetWidth{1.0}
    \SetColor{Black}
    \Gluon(200,-126)(460,-126){4}{18}
    \Gluon(480,-46)(192,-46){4}{19}
    \Gluon(212,-154)(468,-154){4}{18}
    \Gluon(480,-22)(192,-22){4}{19}
    \GOval(272,-34)(24,24)(0){0.882}
    \SetColor{White}
    \GBox(168,-190)(220,2){0.882}
    \SetWidth{2.0}
    \SetColor{Black}
    \Line(328.001,-93.999)(343.999,-78.001)\Line(328.001,-78.001)(343.999,-93.999)
    \SetWidth{1.0}
    \Gluon(336,-122)(336,-86){4}{2}
    \Gluon(336,-86)(336,-26){4}{4}
    \Text(260,-34)[l]{\fontsize{35}{2}\selectfont {$#1$}}
    \GOval(272,-138)(24,24)(0){0.882}
    \GOval(396,-34)(24,24)(0){0.882}
    \GOval(396,-138)(24,24)(0){0.882}
    \Text(260,-138)[l]{\fontsize{35}{2}\selectfont {$#2$}}
    \Text(384,-34)[l]{\fontsize{35}{2}\selectfont {$#3$}}
    \Text(384,-138)[l]{\fontsize{35}{2}\selectfont {$#4$}}
    \SetColor{White}
    \GBox(448,-190)(500,2){0.882}
  \end{axopicture}
}}}}

\def \symmetrymatchingb (#1,#2,#3,#4,#5,#6) {
\resizebox{#5 mm}{!}{\raisebox{#6 pt}{
\fcolorbox{white}{white}{
  \begin{axopicture}(334,194) (167,-191)
    \SetWidth{1.0}
    \SetColor{Black}
    \Gluon(200,-126)(460,-126){4}{18}
    \Gluon(480,-46)(192,-46){4}{19}
    \Gluon(212,-154)(468,-154){4}{18}
    \Gluon(480,-22)(192,-22){4}{19}
    \GOval(272,-34)(24,24)(0){0.882}
    \SetColor{White}
    \GBox(168,-190)(220,2){0.882}
    \SetWidth{2.0}
    \SetColor{Black}
    \Line(328.001,-93.999)(343.999,-78.001)\Line(328.001,-78.001)(343.999,-93.999)
    \SetWidth{1.0}
    \Gluon(336,-150)(336,-86){4}{5}
    \Gluon(336,-86)(336,-50){4}{2}
    \Text(260,-34)[l]{\fontsize{35}{2}\selectfont {$#1$}}
    \GOval(272,-138)(24,24)(0){0.882}
    \GOval(396,-34)(24,24)(0){0.882}
    \GOval(396,-138)(24,24)(0){0.882}
    \Text(260,-138)[l]{\fontsize{35}{2}\selectfont {$#2$}}
    \Text(384,-34)[l]{\fontsize{35}{2}\selectfont {$#3$}}
    \Text(384,-138)[l]{\fontsize{35}{2}\selectfont {$#4$}}
    \SetColor{White}
    \GBox(448,-190)(500,2){0.882}
  \end{axopicture}
}}}}

\def \matchinga (#1,#2) {
\resizebox{#1 mm}{!}{\raisebox{#2 pt}{
\fcolorbox{white}{white}{
  \begin{axopicture}(210,194) (167,-191)
    \SetWidth{3}
    \SetColor{Black}
    \Line[arrow,arrowpos=0.5,arrowlength=20,arrowwidth=10,arrowinset=0.2](191,-50)(351,-50)
    \Line[arrow,arrowpos=0.5,arrowlength=20,arrowwidth=10,arrowinset=0.2,flip](190,-126)(350,-126)
    \SetWidth{1.0}
    \Gluon(200,-154)(364,-154){6}{12}
    \Gluon(360,-22)(188,-22){6}{12}
    \SetColor{White}
    \GBox(168,-190)(220,2){0.882}
    \GBox(324,-190)(376,2){0.882}
  \end{axopicture}
}}}}

\def \matchingb (#1,#2,#3,#4) {
\resizebox{#3 mm}{!}{\raisebox{#4 pt}{
\begin{axopicture}(210,194) (167,-191)
    \SetWidth{3}
    \SetColor{Black}
    \Line(200,-50)(360,-50)
    \Line(196,-126)(356,-126)
    \SetWidth{1.0}
    \Gluon(200,-154)(364,-154){6}{12}
    \Gluon(360,-22)(188,-22){6}{12}
    \GOval(272,-34)(24,24)(0){0.882}
    \SetColor{White}
    \GBox(168,-190)(220,2){0.882}
    \SetColor{Black}
    \Text(260,-34)[l]{\fontsize{35}{2}\selectfont {$#1$}}
    \GOval(272,-138)(24,24)(0){0.882}
    \Text(260,-138)[l]{\fontsize{35}{2}\selectfont {$#2$}}
    \SetColor{White}
    \GBox(324,-190)(376,2){0.882}
  \end{axopicture}}}}

\def \IJexchange (#1,#2) {
\resizebox{#1 mm}{!}{\raisebox{#2 pt}{
\fcolorbox{white}{white}{
  \begin{axopicture}(82,126) (223,-169)
    \SetWidth{2.0}
    \SetColor{Black}
    \Line[arrow,arrowpos=0.5,arrowlength=17.333,arrowwidth=6.933,arrowinset=0.2](304,-130)(224,-130)
    \SetWidth{1.0}
    \Gluon(224,-50)(304,-50){6}{5}
    \Gluon(264,-56)(264,-130){6}{5}  
    \SetWidth{2.0}
    \Line[arrow,arrowpos=0.5,arrowlength=17.333,arrowwidth=6.933,arrowinset=0.2](224,-82)(304,-82)
    \SetWidth{1.0}
    \Gluon(304,-162)(224,-162){6}{5}
    \Text(280,-30)[l]{\fontsize{30}{2}\selectfont {$i$}}
    \Text(280,-113)[l]{\fontsize{30}{2}\selectfont {$j$}} 
  \end{axopicture}
  }}}}

\def \GGoctet (#1,#2) {
\resizebox{#1 mm}{!}{\raisebox{#2 pt}{
\fcolorbox{white}{white}{
  \begin{axopicture}(56,114) (372,-183)
    \SetWidth{2.0}
    \SetColor{Black}
    \Gluon(424,-70)(400,-110){5}{3}
    \Gluon(400,-110)(400,-142){5}{2}
    \Gluon(400,-110)(376,-70){5}{3}
    \Gluon(376,-182)(400,-142){5}{3}
    \Gluon(400,-142)(424,-182){5}{3}
  \end{axopicture}
  }}}}

\def \GGidentity (#1,#2) {
\resizebox{#1 mm}{!}{\raisebox{#2 pt}{
\fcolorbox{white}{white}{
  \begin{axopicture}(60,66) (202,-143)
    \SetWidth{2.0}
    \SetColor{Black}
    \Gluon(256,-142)(256,-78){5}{5}
    \Gluon(208,-142)(208,-78){5}{5}
  \end{axopicture}
  }}}}

\def \GGcross (#1,#2) {
\resizebox{#1 mm}{!}{\raisebox{#2 pt}{
\fcolorbox{white}{white}{
  \begin{axopicture}(60,66) (202,-143)
    \SetWidth{2.0}
    \SetColor{Black}
    \Gluon(256,-142)(208,-78){5}{5}
    \Gluon(208,-142)(256,-78){5}{5}
  \end{axopicture}
  }}}}

\def \QGidentity (#1,#2) {
\resizebox{#1 mm}{!}{\raisebox{#2 pt}{
\fcolorbox{white}{white}{
  \begin{axopicture}(105,140) (526,-162)
    \SetWidth{1.0}
    \SetColor{Black}
     \Bezier(528,-36)(526.861,-32.427)(530.763,-27.455)(535.804,-26.056) \Bezier(535.804,-26.056)(545.885,-23.257)(548.834,-37.964)(542.711,-39.549) \Bezier(542.711,-39.549)(536.588,-41.134)(532.032,-26.843)(545.436,-24.248) \Bezier(545.436,-24.248)(558.839,-21.653)(559.786,-36.624)(553.609,-37.435) \Bezier(553.609,-37.435)(547.432,-38.246)(544.483,-23.538)(558.62,-22.795) \Bezier(558.62,-22.795)(572.757,-22.052)(571.108,-36.961)(565.034,-36.818) \Bezier(565.034,-36.818)(558.961,-36.676)(558.013,-21.706)(572.449,-23.416) \Bezier(572.449,-23.416)(586.884,-25.127)(582.284,-39.404)(576.455,-38.156) \Bezier(576.455,-38.156)(570.626,-36.908)(572.275,-21.999)(585.695,-26.409) \Bezier(585.695,-26.409)(599.115,-30.819)(591.489,-43.736)(586.127,-41.325) \Bezier(586.127,-41.325)(580.765,-38.915)(585.365,-24.637)(597.873,-32.119) \Bezier(597.873,-32.119)(610.381,-39.6)(599.885,-50.316)(595.285,-46.789) \Bezier(595.285,-46.789)(590.685,-43.261)(598.311,-30.345)(609.101,-40.872) \Bezier(609.101,-40.872)(619.891,-51.399)(607.19,-59.38)(603.534,-54.845) \Bezier(603.534,-54.845)(599.879,-50.31)(610.374,-39.593)(618.409,-52.197) \Bezier(618.409,-52.197)(626.443,-64.801)(612.352,-69.941)(609.747,-64.623) \Bezier(609.747,-64.623)(607.143,-59.305)(619.843,-51.324)(625.036,-65.235) \Bezier(625.036,-65.235)(630.229,-79.146)(615.435,-81.625)(613.897,-75.791) \Bezier(613.897,-75.791)(612.358,-69.957)(626.449,-64.817)(628.848,-78.703) \Bezier(628.848,-78.703)(631.246,-92.59)(616.246,-92.59)(615.74,-86.506) \Bezier(615.74,-86.506)(615.234,-80.422)(630.028,-77.943)(630.028,-92) \Bezier(630.028,-92)(630.028,-106.057)(615.234,-103.578)(615.74,-97.494) \Bezier(615.74,-97.494)(616.246,-91.41)(631.246,-91.41)(628.848,-105.297) \Bezier(628.848,-105.297)(626.449,-119.183)(612.358,-114.043)(613.897,-108.209) \Bezier(613.897,-108.209)(615.435,-102.375)(630.229,-104.854)(625.036,-118.765) \Bezier(625.036,-118.765)(619.843,-132.676)(607.143,-124.695)(609.747,-119.377) \Bezier(609.747,-119.377)(612.352,-114.059)(626.443,-119.199)(618.409,-131.803) \Bezier(618.409,-131.803)(610.374,-144.407)(599.879,-133.69)(603.534,-129.155) \Bezier(603.534,-129.155)(607.19,-124.62)(619.891,-132.601)(609.101,-143.128) \Bezier(609.101,-143.128)(598.311,-153.655)(590.685,-140.739)(595.285,-137.211) \Bezier(595.285,-137.211)(599.885,-133.684)(610.381,-144.4)(597.873,-151.881) \Bezier(597.873,-151.881)(585.365,-159.363)(580.765,-145.085)(586.127,-142.675) \Bezier(586.127,-142.675)(591.489,-140.264)(599.115,-153.181)(585.695,-157.591) \Bezier(585.695,-157.591)(572.275,-162.001)(570.626,-147.092)(576.455,-145.844) \Bezier(576.455,-145.844)(582.284,-144.596)(586.884,-158.873)(572.449,-160.584) \Bezier(572.449,-160.584)(558.013,-162.294)(558.961,-147.324)(565.034,-147.182) \Bezier(565.034,-147.182)(571.108,-147.039)(572.757,-161.948)(558.62,-161.205) \Bezier(558.62,-161.205)(544.483,-160.462)(547.432,-145.754)(553.609,-146.565) \Bezier(553.609,-146.565)(559.786,-147.376)(558.839,-162.347)(545.436,-159.752) \Bezier(545.436,-159.752)(532.032,-157.157)(536.588,-142.866)(542.711,-144.451) \Bezier(542.711,-144.451)(548.834,-146.036)(545.885,-160.743)(535.804,-157.944) \Bezier(535.804,-157.944)(530.763,-156.545)(526.861,-151.573)(528,-148)
    \SetWidth{2.0}
    \Line[arrow,arrowpos=0.5,arrowlength=17.333,arrowwidth=6.933,arrowinset=0.2](592,-116)(528,-116)
    \Line[arrow,arrowpos=0.5,arrowlength=17.333,arrowwidth=6.933,arrowinset=0.2](528,-68)(592,-68)
    \Line(592,-68)(592,-116)
  \end{axopicture}
  }}}}

\def \Gidentity (#1,#2) {
\resizebox{#1 mm}{!}{\raisebox{#2 pt}{
\fcolorbox{white}{white}{
  \begin{axopicture}(105,140) (526,-162)
    \SetWidth{1.0}
    \SetColor{Black}
     \Bezier(528,-36)(526.861,-32.427)(530.763,-27.455)(535.804,-26.056) \Bezier(535.804,-26.056)(545.885,-23.257)(548.834,-37.964)(542.711,-39.549) \Bezier(542.711,-39.549)(536.588,-41.134)(532.032,-26.843)(545.436,-24.248) \Bezier(545.436,-24.248)(558.839,-21.653)(559.786,-36.624)(553.609,-37.435) \Bezier(553.609,-37.435)(547.432,-38.246)(544.483,-23.538)(558.62,-22.795) \Bezier(558.62,-22.795)(572.757,-22.052)(571.108,-36.961)(565.034,-36.818) \Bezier(565.034,-36.818)(558.961,-36.676)(558.013,-21.706)(572.449,-23.416) \Bezier(572.449,-23.416)(586.884,-25.127)(582.284,-39.404)(576.455,-38.156) \Bezier(576.455,-38.156)(570.626,-36.908)(572.275,-21.999)(585.695,-26.409) \Bezier(585.695,-26.409)(599.115,-30.819)(591.489,-43.736)(586.127,-41.325) \Bezier(586.127,-41.325)(580.765,-38.915)(585.365,-24.637)(597.873,-32.119) \Bezier(597.873,-32.119)(610.381,-39.6)(599.885,-50.316)(595.285,-46.789) \Bezier(595.285,-46.789)(590.685,-43.261)(598.311,-30.345)(609.101,-40.872) \Bezier(609.101,-40.872)(619.891,-51.399)(607.19,-59.38)(603.534,-54.845) \Bezier(603.534,-54.845)(599.879,-50.31)(610.374,-39.593)(618.409,-52.197) \Bezier(618.409,-52.197)(626.443,-64.801)(612.352,-69.941)(609.747,-64.623) \Bezier(609.747,-64.623)(607.143,-59.305)(619.843,-51.324)(625.036,-65.235) \Bezier(625.036,-65.235)(630.229,-79.146)(615.435,-81.625)(613.897,-75.791) \Bezier(613.897,-75.791)(612.358,-69.957)(626.449,-64.817)(628.848,-78.703) \Bezier(628.848,-78.703)(631.246,-92.59)(616.246,-92.59)(615.74,-86.506) \Bezier(615.74,-86.506)(615.234,-80.422)(630.028,-77.943)(630.028,-92) \Bezier(630.028,-92)(630.028,-106.057)(615.234,-103.578)(615.74,-97.494) \Bezier(615.74,-97.494)(616.246,-91.41)(631.246,-91.41)(628.848,-105.297) \Bezier(628.848,-105.297)(626.449,-119.183)(612.358,-114.043)(613.897,-108.209) \Bezier(613.897,-108.209)(615.435,-102.375)(630.229,-104.854)(625.036,-118.765) \Bezier(625.036,-118.765)(619.843,-132.676)(607.143,-124.695)(609.747,-119.377) \Bezier(609.747,-119.377)(612.352,-114.059)(626.443,-119.199)(618.409,-131.803) \Bezier(618.409,-131.803)(610.374,-144.407)(599.879,-133.69)(603.534,-129.155) \Bezier(603.534,-129.155)(607.19,-124.62)(619.891,-132.601)(609.101,-143.128) \Bezier(609.101,-143.128)(598.311,-153.655)(590.685,-140.739)(595.285,-137.211) \Bezier(595.285,-137.211)(599.885,-133.684)(610.381,-144.4)(597.873,-151.881) \Bezier(597.873,-151.881)(585.365,-159.363)(580.765,-145.085)(586.127,-142.675) \Bezier(586.127,-142.675)(591.489,-140.264)(599.115,-153.181)(585.695,-157.591) \Bezier(585.695,-157.591)(572.275,-162.001)(570.626,-147.092)(576.455,-145.844) \Bezier(576.455,-145.844)(582.284,-144.596)(586.884,-158.873)(572.449,-160.584) \Bezier(572.449,-160.584)(558.013,-162.294)(558.961,-147.324)(565.034,-147.182) \Bezier(565.034,-147.182)(571.108,-147.039)(572.757,-161.948)(558.62,-161.205) \Bezier(558.62,-161.205)(544.483,-160.462)(547.432,-145.754)(553.609,-146.565) \Bezier(553.609,-146.565)(559.786,-147.376)(558.839,-162.347)(545.436,-159.752) \Bezier(545.436,-159.752)(532.032,-157.157)(536.588,-142.866)(542.711,-144.451) \Bezier(542.711,-144.451)(548.834,-146.036)(545.885,-160.743)(535.804,-157.944) \Bezier(535.804,-157.944)(530.763,-156.545)(526.861,-151.573)(528,-148)
  \end{axopicture}
  }}}}

\def \Qidentity (#1,#2) {
\resizebox{#1 mm}{!}{\raisebox{#2 pt}{
\fcolorbox{white}{white}{
  \begin{axopicture}(105,140) (526,-162)
    \SetWidth{1.0}
    \SetColor{Black}
    \SetWidth{2.0}
    \Line[arrow,arrowpos=0.5,arrowlength=17.333,arrowwidth=6.933,arrowinset=0.2](592,-116)(528,-116)
    \Line[arrow,arrowpos=0.5,arrowlength=17.333,arrowwidth=6.933,arrowinset=0.2](528,-68)(592,-68)
    \Line(592,-68)(592,-116)
  \end{axopicture}
  }}}}

\def \qaTt (#1,#2) {
\resizebox{#1 mm}{!}{\raisebox{#2 pt}{
\fcolorbox{white}{white}{
\begin{axopicture}(183,82) (140,-191)
    \SetWidth{2.0}
    \SetColor{Black}
    \Line(288,-126)(288,-174)
    \Line(168,-126)(168,-174)
    \Line[dash,dashsize=8](152,-190)(152,-110)
    \Line[dash,dashsize=8](304,-190)(304,-110)
    \SetWidth{1.0}
    \GOval(297,-150)(20,20)(0){0.882}
    \GOval(161,-150)(20,20)(0){0.882}
    \SetWidth{2.0}
    \Line[dash,dashsize=10](304,-110)(152,-110)
    \Line[dash,dashsize=10](304,-190)(152,-190)
    \Line[arrow,arrowpos=0.3,arrowlength=12.5,arrowwidth=5,arrowinset=0.2](168,-126)(288,-126)
    \Line[arrow,arrowpos=0.3,arrowlength=12.5,arrowwidth=5,arrowinset=0.2,flip](168,-174)(288,-174)
    \SetWidth{1.0}
    \Gluon(228,-126)(228,-174){4.5}{4}
    \Text(152,-150)[l]{\Huge{\Black{$\alpha$}}}
    \Text(288,-150)[l]{\Huge{\Black{$\beta$}}}
  \end{axopicture}}}}}

\def \qaTtId (#1,#2) {
\resizebox{#1 mm}{!}{\raisebox{#2 pt}{
\fcolorbox{white}{white}{
  \begin{axopicture}(183,82) (140,-191)
    \SetWidth{2.0}
    \SetColor{Black}
    \Line(288,-126)(288,-174)
    \Line(168,-126)(168,-174)
    \Line[dash,dashsize=8](152,-190)(152,-110)
    \Line[dash,dashsize=8](304,-190)(304,-110)
    \SetWidth{1.0}
    \GOval(297,-150)(20,20)(0){0.882}
    \GOval(161,-150)(20,20)(0){0.882}
    \SetWidth{2.0}
    \Line[dash,dashsize=10](304,-110)(152,-110)
    \Line[dash,dashsize=10](304,-190)(152,-190)
    \Line(168,-126)(216,-126)
    \Text(152,-150)[l]{\Huge{\Black{$\alpha$}}}
    \Text(288,-150)[l]{\Huge{\Black{$\beta$}}}
    \Line(169,-174)(216,-174)
    \Line(240,-126)(288,-126)
    \Line(240,-174)(288,-174)
    \Line[arrow,arrowpos=0.5,arrowlength=12.5,arrowwidth=5,arrowinset=0.2](216,-126)(216,-174)
    \Line[arrow,arrowpos=0.5,arrowlength=12.5,arrowwidth=5,arrowinset=0.2,flip](240,-126)(240,-174)
  \end{axopicture}}}}}

\def \qaTtSinglet (#1,#2) {
\resizebox{#1 mm}{!}{\raisebox{#2 pt}{
\fcolorbox{white}{white}{
  \begin{axopicture}(183,82) (140,-191)
    \SetWidth{2.0}
    \SetColor{Black}
    \Line(288,-126)(288,-174)
    \Line(168,-126)(168,-174)
    \Line[dash,dashsize=8](152,-190)(152,-110)
    \Line[dash,dashsize=8](304,-190)(304,-110)
    \SetWidth{1.0}
    \GOval(297,-150)(20,20)(0){0.882}
    \GOval(161,-150)(20,20)(0){0.882}
    \SetWidth{2.0}
    \Line[dash,dashsize=10](304,-110)(152,-110)
    \Line[dash,dashsize=10](304,-190)(152,-190)
    \Line[arrow,arrowpos=0.5,arrowlength=12.5,arrowwidth=5,arrowinset=0.2](168,-126)(288,-126)
    \Line[arrow,arrowpos=0.5,arrowlength=12.5,arrowwidth=5,arrowinset=0.2,flip](168,-174)(288,-174)
    \Text(152,-150)[l]{\Huge{\Black{$\alpha$}}}
    \Text(288,-150)[l]{\Huge{\Black{$\beta$}}}
  \end{axopicture}}}}}

\def \FierzOctet (#1,#2) {
\resizebox{#1 mm}{!}{\raisebox{#2 pt}{
\fcolorbox{white}{white}{
  \begin{axopicture}(66,60) (191,-202)
    \SetWidth{2.0}
    \SetColor{Black}
    \Line[arrow,arrowpos=0.3,arrowlength=12.5,arrowwidth=5,arrowinset=0.2](192,-148)(256,-148)
    \Line[arrow,arrowpos=0.3,arrowlength=12.5,arrowwidth=5,arrowinset=0.2,flip](192,-196)(256,-196)
    \SetWidth{1.0}
    \Gluon(228,-148)(228,-196){4.5}{4}
  \end{axopicture}
}}}}

\def \FierzIdentity (#1,#2) {
\resizebox{#1 mm}{!}{\raisebox{#2 pt}{
\fcolorbox{white}{white}{
  \begin{axopicture}(66,60) (223,-202)
    \SetWidth{2.0}
    \SetColor{Black}
    \Line (224,-148)(240,-148)
    \Line (224,-196)(240,-196)
    \Line (288,-148)(272,-148)
    \Line (288,-196)(272,-196)
    \Line[arrow,arrowpos=0.5,arrowlength=12.5,arrowwidth=5,arrowinset=0.2](240,-148)(240,-196)
    \Line[arrow,arrowpos=0.5,arrowlength=12.5,arrowwidth=5,arrowinset=0.2,flip](272,-148)(272,-196)
  \end{axopicture}
}}}}

\def \FierzSinglet (#1,#2) {
\resizebox{#1 mm}{!}{\raisebox{#2 pt}{
\fcolorbox{white}{white}{
  \begin{axopicture}(66,60) (223,-202)
    \SetWidth{2.0}
    \SetColor{Black}
    \Line[arrow,arrowpos=0.5,arrowlength=12.5,arrowwidth=5,arrowinset=0.2](224,-148)(288,-148)
    \Line[arrow,arrowpos=0.5,arrowlength=12.5,arrowwidth=5,arrowinset=0.2,flip](224,-196)(288,-196)
  \end{axopicture}
}}}}

\def \aabar (#1,#2) {
\resizebox{#1 mm}{!}{\raisebox{#2 pt}{
\fcolorbox{white}{white}{
  \begin{axopicture}(82,50) (223,-207)
    \SetWidth{2.0}
    \SetColor{Black}
    \Line[dash,dashsize=7.2](224,-158)(304,-158)
    \Line[dash,dashsize=7.2](224,-206)(304,-206)
  \end{axopicture}}}}}

\def \aabarSinglet (#1,#2) {
\resizebox{#1 mm}{!}{\raisebox{#2 pt}{
\fcolorbox{white}{white}{
  \begin{axopicture}(82,50) (239,-207)
    \SetWidth{2.0}
    \SetColor{Black}
    \Line[dash,dashsize=7.2](240,-158)(272,-158)
    \Line[dash,dashsize=7.2](240,-206)(272,-206)
    \Line[dash,dashsize=7.2](272,-158)(272,-206)
    \Line[dash,dashsize=7.2](288,-158)(288,-206)
    \Line[dash,dashsize=7.2](288,-158)(320,-158)
    \Line[dash,dashsize=7.2](288,-206)(320,-206)
  \end{axopicture}}}}}

\def \ggTtbc (#1,#2) {
\resizebox{#1 mm}{!}{\raisebox{#2 pt}{
\fcolorbox{white}{white}{
  \begin{axopicture}(183,106) (240,-179)
    \SetWidth{1.5}
    \SetColor{Black}
    \Gluon(400,-78)(400,-174){4}{8}
    \Gluon(384,-94)(384,-158){4}{5}
    \Gluon(256,-174)(256,-78){4}{8}
    \Gluon(272,-158)(272,-94){4}{5}
    \SetWidth{1.0}
    \GOval(397,-126)(20,20)(0){0.882}
    \GOval(261,-126)(20,20)(0){0.882}
    \SetWidth{1.5}
    \Gluon(332,-99)(332,-153){4}{5}
    \Gluon(384,-94)(272,-94){4}{9}
    \Gluon(272,-158)(384,-158){4}{9}
    \Gluon(400,-78)(256,-78){4}{12}
    \Gluon(256,-174)(400,-174){4}{12}
    \Text(252,-126)[l]{\Huge{\Black{$\alpha$}}}
    \Text(388,-126)[l]{\Huge{\Black{$\beta$}}}
  \end{axopicture}
}}}}

\def \ggTtac (#1,#2) {
\resizebox{#1 mm}{!}{\raisebox{#2 pt}{
\fcolorbox{white}{white}{
  \begin{axopicture}(183,106) (240,-179)
    \SetWidth{1.5}
    \SetColor{Black}
    \Gluon(272,-158)(272,-94){4}{5}
    \Gluon(384,-94)(384,-158){4}{5}
    \Gluon(400,-78)(400,-174){4}{8}
    \Gluon(256,-174)(256,-78){4}{8}
    \SetWidth{1.0}
    \GOval(397,-126)(20,20)(0){0.882}
    \GOval(261,-126)(20,20)(0){0.882}
    \SetWidth{1.5}
    \Gluon(330,-82)(331,-154){4}{7}
    \Gluon(384,-94)(272,-94){4}{9}
    \Gluon(272,-158)(384,-158){4}{9}
    \Gluon(400,-78)(256,-78){4}{12}
    \Gluon(256,-174)(400,-174){4}{12}
    \Text(252,-126)[l]{\Huge{\Black{$\alpha$}}}
    \Text(388,-126)[l]{\Huge{\Black{$\beta$}}}
  \end{axopicture}
}}}}

\def \ggIOctet (#1,#2,#3) {
\resizebox{#2 mm}{!}{\raisebox{#3 pt}{
\fcolorbox{white}{white}{
  \begin{axopicture}(297,204) (58,-90)
    \SetWidth{1.0}
    \SetColor{Mahogany}
    \GOval(96,12)(80,48)(0){0.882}
    \Gluon(304,76)(304,-52){5}{10}
    \SetColor{Violet}
    \Gluon(64,76)(64,108){5}{3}
    \Gluon(64,108)(336,108){5}{22}
    \Gluon(336,108)(336,-84){5}{15}
    \Gluon(336,-84)(64,-84){5}{22}
    \Gluon(64,-84)(64,-52){5}{3}
    \SetColor{Mahogany}
    \Gluon(208,-52)(208,76){5}{10}
    \SetColor{Black}
    \GOval(320,12)(32,32)(0){0.882}
    \Text(310,12)[l]{\fontsize{35}{2}\selectfont {$#1$}}
    \SetColor{Mahogany}
    \Gluon(208,76)(304,76){5}{8}
    \Gluon(304,-52)(208,-52){5}{8}
    \SetColor{Violet}
    \Gluon(64,76)(96,44){5}{4}
    \Gluon(96,44)(96,-20){5}{5}
    \Gluon(64,-52)(96,-20){5}{4}
    \SetColor{Black}
     \Bezier(208,-52)(207.999,-54.25)(203.263,-56.492)(198.528,-56.485) \Bezier(198.528,-56.485)(189.059,-56.469)(189.108,-47.469)(193.147,-47.481) \Bezier(193.147,-47.481)(197.186,-47.494)(197.181,-56.494)(185.229,-56.413) \Bezier(185.229,-56.413)(173.277,-56.332)(173.441,-47.333)(177.462,-47.381) \Bezier(177.462,-47.381)(181.483,-47.428)(181.434,-56.428)(170.43,-56.198) \Bezier(170.43,-56.198)(159.425,-55.968)(159.832,-46.977)(163.819,-47.104) \Bezier(163.819,-47.104)(167.806,-47.23)(167.641,-56.229)(156.575,-55.668) \Bezier(156.575,-55.668)(145.51,-55.108)(146.411,-46.154)(150.325,-46.439) \Bezier(150.325,-46.439)(154.239,-46.724)(153.832,-55.715)(142.981,-54.507) \Bezier(142.981,-54.507)(132.129,-53.3)(133.984,-44.493)(137.742,-45.076) \Bezier(137.742,-45.076)(141.5,-45.659)(140.598,-54.614)(130.182,-52.186) \Bezier(130.182,-52.186)(119.766,-49.757)(123.374,-41.512)(126.772,-42.601) \Bezier(126.772,-42.601)(130.171,-43.69)(128.316,-52.497)(118.236,-47.757) \Bezier(118.236,-47.757)(108.156,-43.018)(114.195,-36.345)(116.944,-38.123) \Bezier(116.944,-38.123)(119.693,-39.901)(116.084,-48.146)(107.176,-40.219) \Bezier(107.176,-40.219)(98.268,-32.292)(105.937,-27.583)(108.091,-30.176) \Bezier(108.091,-30.176)(110.245,-32.77)(104.205,-39.443)(98.185,-30.899) \Bezier(98.185,-30.899)(95.175,-26.627)(94.083,-21.177)(96,-20)
     \Bezier(96,44)(94.054,45.129)(94.472,49.852)(96.836,53.446) \Bezier(96.836,53.446)(101.564,60.633)(108.211,54.566)(106.113,51.706) \Bezier(106.113,51.706)(104.016,48.847)(96.232,53.364)(104.083,61.57) \Bezier(104.083,61.57)(111.933,69.776)(116.345,61.932)(113.774,59.888) \Bezier(113.774,59.888)(111.203,57.843)(104.555,63.911)(114.389,69.639) \Bezier(114.389,69.639)(124.223,75.367)(126.54,66.67)(123.342,65.369) \Bezier(123.342,65.369)(120.144,64.068)(115.732,71.913)(126.243,75.013) \Bezier(126.243,75.013)(136.754,78.114)(137.907,69.188)(134.233,68.465) \Bezier(134.233,68.465)(130.559,67.741)(128.242,76.438)(138.592,77.923) \Bezier(138.592,77.923)(148.941,79.409)(149.504,70.427)(145.622,70.055) \Bezier(145.622,70.055)(141.74,69.683)(140.588,78.609)(151.076,79.334) \Bezier(151.076,79.334)(161.565,80.06)(161.824,71.064)(157.854,70.882) \Bezier(157.854,70.882)(153.885,70.701)(153.322,79.684)(163.707,80.018) \Bezier(163.707,80.018)(174.091,80.352)(174.199,71.352)(170.188,71.271) \Bezier(170.188,71.271)(166.177,71.189)(165.918,80.185)(175.903,80.323) \Bezier(175.903,80.323)(185.887,80.46)(185.921,71.46)(181.89,71.428) \Bezier(181.89,71.428)(177.859,71.396)(177.75,80.396)(188.379,80.446) \Bezier(188.379,80.446)(199.007,80.496)(199.011,71.496)(194.969,71.487) \Bezier(194.969,71.487)(190.926,71.479)(190.892,80.479)(199.445,80.489) \Bezier(199.445,80.489)(203.722,80.495)(207.999,78.25)(208,76)
  \end{axopicture}}}}}

\def \ggIdBeta (#1,#2,#3) {
\resizebox{#2 mm}{!}{\raisebox{#3 pt}{
\fcolorbox{white}{white}{
  \begin{axopicture}(308,204) (47,-90)
    \SetWidth{1.0}
    \SetColor{Black}
    \GOval(96,12)(80,48)(0){0.882}
    \Gluon(304,76)(304,-52){5}{10}
    \Gluon(64,76)(64,108){5}{3}
    \Gluon(64,108)(336,108){5}{22}
    \Gluon(336,108)(336,-84){5}{15}
    \Gluon(336,-84)(64,-84){5}{22}
    \Gluon(64,-84)(64,-52){5}{3}
    \Gluon(208,-52)(208,76){5}{10}
    \SetColor{Black}
    \GOval(320,12)(32,32)(0){0.882}
    \Text(310,12)[l]{\fontsize{35}{2}\selectfont {$#1$}} 
    \Gluon(208,76)(304,76){5}{8}
    \Gluon(304,-52)(208,-52){5}{8}
    \SetColor{Black}
    \Gluon(128,76)(208,76){5}{6}
    \Gluon(208,-52)(128,-52){5}{6}
    \Gluon(64,-52)(64,76){5}{10}
    \Gluon(128,-52)(128,76){5}{10}
  \end{axopicture}}}}}

\def \ggXBeta (#1,#2,#3) {
\resizebox{#2 mm}{!}{\raisebox{#3 pt}{
\fcolorbox{white}{white}{
  \begin{axopicture}(308,204) (47,-90)
    \SetWidth{1.0}
    \SetColor{Black}
    \GOval(96,12)(80,48)(0){0.882}
    \Gluon(304,76)(304,-52){5}{10}
    \Gluon(64,76)(64,108){5}{3}
    \Gluon(64,108)(336,108){5}{22}
    \Gluon(336,108)(336,-84){5}{15}
    \Gluon(336,-84)(64,-84){5}{22}
    \Gluon(64,-84)(64,-52){5}{3}
    \Gluon(208,-52)(208,76){5}{10}
    \SetColor{Black}
    \GOval(320,12)(32,32)(0){0.882}
    \Text(310,12)[l]{\fontsize{35}{2}\selectfont {$#1$}} 
    \Gluon(208,76)(304,76){5}{8}
    \Gluon(304,-52)(208,-52){5}{8}
    \SetColor{Black}
    \Gluon(128,76)(208,76){5}{6}
    \Gluon(208,-52)(128,-52){5}{6}
    \Gluon(64,-52)(128,76){5}{10}
    \Gluon(128,-52)(64,76){5}{10}
  \end{axopicture}}}}}

\def \Btwo (#1,#2,#3,#4,#5) {
\resizebox{#1 mm}{!}{\raisebox{#2 pt}{
\fcolorbox{white}{white}{
  \begin{axopicture}(132,196) (46,-46)
    \SetWidth{3.0}
    \SetColor{Black}
    \Line(48,148)(176,148)
    \Line(48,84)(176,84)
    \Line(48,20)(176,20)
    \Line(48,-44)(176,-44)
    \SetWidth{2.0}
    \SetColor{Red}
    \Gluon(112,#3)(112,#4){6}{#5}
  \end{axopicture}
}}}}

\def \Btwobis (#1,#2,#3,#4,#5) {
\resizebox{#1 mm}{!}{\raisebox{#2 pt}{
\fcolorbox{white}{white}{
  \begin{axopicture}(132,196) (46,-46)
    \SetWidth{3.0}
    \SetColor{Black}
    \Line(48,148)(176,148)
    \Line(48,-44)(176,-44)
    \SetWidth{2.0}
    \SetColor{Red}
    \Gluon(112,#3)(112,#4){6}{#5}
  \end{axopicture}
}}}}

\def \Operator_Reduction_2_1 (#1,#2,#3) {
\resizebox{#1 mm}{!}{\raisebox{#2 pt}{
\fcolorbox{white}{white}{
  \begin{axopicture}(132,196) (46,-46)
    \SetWidth{3.0}
    \SetColor{Black}
    \Line(48,148)(176,148)
    \Line(48,84)(128,84)
    \Line(48,20)(128,20)
    \Line(48,-44)(176,-44)
    \Line(128,84)(128,20)
    \Text(140,130)[l]{\fontsize{40}{2}\selectfont {$#3$}}
  \end{axopicture}
}}}}

\def \qgTdTd (#1,#2) {
\resizebox{#1 mm}{!}{\raisebox{#2 pt}{
\begin{axopicture}(114,177) (223,-118)
  \SetWidth{2.0}
  \SetColor{Black}
  \Line[arrow,arrowpos=0.5,arrowlength=17.333,arrowwidth=6.933,arrowinset=0.2](340,-79)(220,-79)
  \SetWidth{1.0}
  \Gluon(220,1)(340,1){6}{7}
  \GluonArc[clock](280,-5)(40,163.74,16.26){6}{7}
  \SetWidth{2.0}
  \Line[arrow,arrowpos=0.5,arrowlength=17.333,arrowwidth=6.933,arrowinset=0.2](220,-31)(340,-31)
  \SetWidth{1.0}
  \Gluon(340,-111)(220,-111){6}{7}
\end{axopicture}
}}}

\def \qgTa (#1,#2) {
\resizebox{#1 mm}{!}{\raisebox{#2 pt}{
\begin{axopicture}(114,177) (223,-118)
  \SetWidth{2.0}
  \SetColor{Black}
  \Line[arrow,arrowpos=0.5,arrowlength=17.333,arrowwidth=6.933,arrowinset=0.2](336,-79)(224,-79)
  \SetWidth{1.0}
  \Gluon(224,1)(336,1){6}{7}
  \SetWidth{2.0}
  \Line[arrow,arrowpos=0.5,arrowlength=17.333,arrowwidth=6.933,arrowinset=0.2](224,-31)(336,-31)
  \SetWidth{1.0}
  \Gluon(336,-111)(224,-111){6}{7}
  \Gluon(300,1)(300,-160){6}{9} 
\end{axopicture}
}}}

\def \qgTb (#1,#2) {
\resizebox{#1 mm}{!}{\raisebox{#2 pt}{
\begin{axopicture}(114,177) (223,-118)
  \SetWidth{2.0}
  \SetColor{Black}
  \Line[arrow,arrowpos=0.5,arrowlength=17.333,arrowwidth=6.933,arrowinset=0.2](336,-79)(224,-79)
  \SetWidth{1.0}
  \Gluon(224,1)(336,1){6}{7}
  \SetWidth{2.0}
  \Line[arrow,arrowpos=0.5,arrowlength=17.333,arrowwidth=6.933,arrowinset=0.2](224,-31)(336,-31)
  \SetWidth{1.0}
  \Gluon(336,-111)(224,-111){6}{7}
  \Gluon(300,-31)(300,-160){6}{7} 
\end{axopicture}
}}}

\def \qgTc (#1,#2) {
\resizebox{#1 mm}{!}{\raisebox{#2 pt}{
\begin{axopicture}(114,177) (223,-118)
  \SetWidth{2.0}
  \SetColor{Black}
  \Line[arrow,arrowpos=0.5,arrowlength=17.333,arrowwidth=6.933,arrowinset=0.2](336,-79)(224,-79)
  \SetWidth{1.0}
  \Gluon(224,1)(336,1){6}{7}
  \SetWidth{2.0}
  \Line[arrow,arrowpos=0.5,arrowlength=17.333,arrowwidth=6.933,arrowinset=0.2](224,-31)(336,-31)
  \SetWidth{1.0}
  \Gluon(336,-111)(224,-111){6}{7}
  \Gluon(300,-79)(300,-160){6}{4} 
\end{axopicture}
}}}

\def \qgTd (#1,#2) {
\resizebox{#1 mm}{!}{\raisebox{#2 pt}{
\begin{axopicture}(114,177) (223,-118)
  \SetWidth{2.0}
  \SetColor{Black}
  \Line[arrow,arrowpos=0.5,arrowlength=17.333,arrowwidth=6.933,arrowinset=0.2](336,-79)(224,-79)
  \SetWidth{1.0}
  \Gluon(224,1)(336,1){6}{7}
  \SetWidth{2.0}
  \Line[arrow,arrowpos=0.5,arrowlength=17.333,arrowwidth=6.933,arrowinset=0.2](224,-31)(336,-31)
  \SetWidth{1.0}
  \Gluon(336,-111)(224,-111){6}{7}
  \Gluon(300,-111)(300,-160){6}{3} 
\end{axopicture}
}}}
  
\def \qgTaTb (#1,#2) {
\resizebox{#1 mm}{!}{\raisebox{#2 pt}{
\begin{axopicture}(114,177) (223,-118)
  \SetWidth{2.0}
  \SetColor{Black}
  \Line[arrow,arrowpos=0.5,arrowlength=17.333,arrowwidth=6.933,arrowinset=0.2](336,-79)(224,-79)
  \SetWidth{1.0}
  \Gluon(224,1)(336,1){6}{7}
  \SetWidth{2.0}
  \Line[arrow,arrowpos=0.5,arrowlength=17.333,arrowwidth=6.933,arrowinset=0.2](224,-31)(336,-31)
  \SetWidth{1.0}
  \Gluon(336,-111)(224,-111){6}{7}
  \Gluon(300,1)(300,-31){6}{2} 
\end{axopicture}
}}}

\def \qgTaTc (#1,#2) {
\resizebox{#1 mm}{!}{\raisebox{#2 pt}{
\begin{axopicture}(114,177) (223,-118)
  \SetWidth{2.0}
  \SetColor{Black}
  \Line[arrow,arrowpos=0.5,arrowlength=17.333,arrowwidth=6.933,arrowinset=0.2](336,-79)(224,-79)
  \SetWidth{1.0}
  \Gluon(224,1)(336,1){6}{7}
  \SetWidth{2.0}
  \Line[arrow,arrowpos=0.5,arrowlength=17.333,arrowwidth=6.933,arrowinset=0.2](224,-31)(336,-31)
  \SetWidth{1.0}
  \Gluon(336,-111)(224,-111){6}{7}
  \Gluon(300,1)(300,-79){6}{4} 
\end{axopicture}
}}}

\def \qgTaTd (#1,#2) {
\resizebox{#1 mm}{!}{\raisebox{#2 pt}{
\begin{axopicture}(114,177) (223,-118)
  \SetWidth{2.0}
  \SetColor{Black}
  \Line[arrow,arrowpos=0.5,arrowlength=17.333,arrowwidth=6.933,arrowinset=0.2](336,-79)(224,-79)
  \SetWidth{1.0}
  \Gluon(224,1)(336,1){6}{7}
  \SetWidth{2.0}
  \Line[arrow,arrowpos=0.5,arrowlength=17.333,arrowwidth=6.933,arrowinset=0.2](224,-31)(336,-31)
  \SetWidth{1.0}
  \Gluon(336,-111)(224,-111){6}{7}
  \Gluon(300,1)(300,-111){6}{6} 
\end{axopicture}
}}}

\def \qgTaTa (#1,#2) {
\resizebox{#1 mm}{!}{\raisebox{#2 pt}{
\begin{axopicture}(114,173) (223,-169)
    \SetWidth{2.0}
    \SetColor{Black}
    \Line[arrow,arrowpos=0.5,arrowlength=17.333,arrowwidth=6.933,arrowinset=0.2](336,-83)(224,-83)
    \SetWidth{1.0}
    \Gluon(224,-3)(336,-3){6}{7}
    \SetWidth{2.0}
    \Line[arrow,arrowpos=0.5,arrowlength=17.333,arrowwidth=6.933,arrowinset=0.2](224,-35)(336,-35)
    \SetWidth{1.0}
    \Gluon(336,-115)(224,-115){6}{7}
     \Bezier(304,-3)(306.495,-2.838)(309.244,-6.845)(309.498,-11.013) \Bezier(309.498,-11.013)(310.006,-19.349)(300.018,-19.853)(299.761,-15.397) \Bezier(299.761,-15.397)(299.504,-10.94)(309.483,-10.293)(310.038,-21.384) \Bezier(310.038,-21.384)(310.593,-32.476)(300.599,-32.815)(300.411,-28.36) \Bezier(300.411,-28.36)(300.223,-23.906)(310.21,-23.401)(310.596,-34.88) \Bezier(310.596,-34.88)(310.981,-46.358)(300.982,-46.515)(300.872,-42.062) \Bezier(300.872,-42.062)(300.762,-37.609)(310.756,-37.27)(310.928,-48.408) \Bezier(310.928,-48.408)(311.099,-59.545)(301.1,-59.509)(301.073,-55.057) \Bezier(301.073,-55.057)(301.046,-50.605)(311.045,-50.449)(311.001,-61.245) \Bezier(311.001,-61.245)(310.957,-72.04)(300.96,-71.79)(301.024,-67.345) \Bezier(301.024,-67.345)(301.087,-62.899)(311.087,-62.935)(310.804,-73.946) \Bezier(310.804,-73.946)(310.52,-84.957)(300.533,-84.457)(300.699,-80.023) \Bezier(300.699,-80.023)(300.865,-75.589)(310.862,-75.839)(310.298,-86.953) \Bezier(310.298,-86.953)(309.734,-98.068)(299.765,-97.284)(300.049,-92.865) \Bezier(300.049,-92.865)(300.333,-88.445)(310.321,-88.945)(309.472,-99.572) \Bezier(309.472,-99.572)(308.624,-110.199)(298.687,-109.077)(299.108,-104.683) \Bezier(299.108,-104.683)(299.528,-100.288)(309.498,-101.072)(308.23,-112.072) \Bezier(308.23,-112.072)(306.962,-123.071)(297.086,-121.506)(297.675,-117.158) \Bezier(297.675,-117.158)(298.265,-112.811)(308.202,-113.933)(306.405,-125.029) \Bezier(306.405,-125.029)(304.608,-136.124)(294.845,-133.963)(295.655,-129.693) \Bezier(295.655,-129.693)(296.465,-125.423)(306.342,-126.988)(303.871,-137.858) \Bezier(303.871,-137.858)(301.399,-148.727)(291.86,-145.726)(292.966,-141.593) \Bezier(292.966,-141.593)(294.071,-137.46)(303.835,-139.621)(300.394,-149.925) \Bezier(300.394,-149.925)(296.953,-160.23)(288.021,-155.734)(289.556,-151.952) \Bezier(289.556,-151.952)(291.092,-148.17)(300.631,-151.171)(294.056,-161.242) \Bezier(294.056,-161.242)(287.481,-171.314)(282.732,-162.514)(284.678,-160.512) \Bezier(284.678,-160.512)(286.624,-158.51)(295.556,-163.006)(283.44,-166.911) \Bezier(283.44,-166.911)(271.323,-170.817)(278.661,-164.023)(279.148,-164.104) \Bezier(279.148,-164.104)(279.635,-164.185)(284.384,-172.985)(273.108,-165.291) \Bezier(273.108,-165.291)(261.831,-157.597)(271.16,-153.995)(273.002,-156.947) \Bezier(273.002,-156.947)(274.843,-159.899)(267.505,-166.693)(262.449,-155.587) \Bezier(262.449,-155.587)(257.394,-144.481)(267.091,-142.04)(268.359,-146.031) \Bezier(268.359,-146.031)(269.626,-150.021)(260.297,-153.623)(257.366,-142.507) \Bezier(257.366,-142.507)(254.435,-131.39)(264.279,-129.631)(265.187,-133.857) \Bezier(265.187,-133.857)(266.096,-138.084)(256.398,-140.525)(254.423,-129.738) \Bezier(254.423,-129.738)(252.448,-118.952)(262.368,-117.691)(263.029,-122.013) \Bezier(263.029,-122.013)(263.689,-126.336)(253.846,-128.095)(252.343,-116.455) \Bezier(252.343,-116.455)(250.84,-104.814)(260.803,-103.957)(261.269,-108.33) \Bezier(261.269,-108.33)(261.735,-112.703)(251.814,-113.963)(250.805,-102.48) \Bezier(250.805,-102.48)(249.795,-90.997)(259.78,-90.452)(260.09,-94.863) \Bezier(260.09,-94.863)(260.4,-99.273)(250.437,-100.13)(249.823,-89.088) \Bezier(249.823,-89.088)(249.21,-78.047)(259.206,-77.766)(259.389,-82.196) \Bezier(259.389,-82.196)(259.572,-86.626)(249.586,-87.17)(249.254,-75.645) \Bezier(249.254,-75.645)(248.921,-64.119)(258.921,-64.063)(258.996,-68.506) \Bezier(258.996,-68.506)(259.07,-72.949)(249.074,-73.23)(249.007,-62.465) \Bezier(249.007,-62.465)(248.94,-51.701)(258.939,-51.839)(258.921,-56.29) \Bezier(258.921,-56.29)(258.903,-60.742)(248.903,-60.797)(249.054,-49.694) \Bezier(249.054,-49.694)(249.205,-38.59)(259.199,-38.911)(259.097,-43.364) \Bezier(259.097,-43.364)(258.995,-47.817)(248.996,-47.679)(249.36,-36.234) \Bezier(249.36,-36.234)(249.724,-24.789)(259.712,-25.277)(259.532,-29.731) \Bezier(259.532,-29.731)(259.351,-34.186)(249.357,-33.865)(249.893,-22.801) \Bezier(249.893,-22.801)(250.43,-11.738)(260.41,-12.377)(260.158,-16.832) \Bezier(260.158,-16.832)(259.907,-21.287)(249.919,-20.799)(250.464,-11.74) \Bezier(250.464,-11.74)(250.737,-7.21)(253.505,-2.84)(256,-3)
\end{axopicture}
}}}

\def \Bfig (#1,#2,#3,#4,#5) {
\resizebox{#1 mm}{!}{\raisebox{#2 pt}{
\fcolorbox{white}{white}{
  \begin{axopicture}(132,196) (46,-46)
    \SetWidth{3.0}
    \SetColor{Black}
    \Line(48,148)(176,148)
    \Line(48,84)(176,84)
    \Line(48,20)(176,20)
    \Line(48,-44)(176,-44)
    \SetWidth{2.0}
    \SetColor{Red}
    \Gluon(112,#3)(112,#4){6}{#5}
  \end{axopicture}
}}}}

\def \OperatorReduction-22-12 (#1,#2,#3) {
\resizebox{#1 mm}{!}{\raisebox{#2 pt}{
\fcolorbox{white}{white}{
  \begin{axopicture}(132,197) (46,-45)
    \SetWidth{3.0}
    \SetColor{Black}
    \Line(48,21)(176,21)
    \Line(48,85)(96,85)
    \Line(48,149)(96,149)
    \Line(48,-43)(176,-43)
    \Line(112,117)(96,149)
    \Line(112,117)(96,85)
    \Line[dash,dashsize=10](112,117)(176,117)
    \Text(135,140)[l]{\fontsize{50}{2}\selectfont {$#3$}}
  \end{axopicture}
}}}}

\def \Bc1_2 (#1,#2,#3,#4,#5) {
\resizebox{#1 mm}{!}{\raisebox{#2 pt}{
\fcolorbox{white}{white}{
  \begin{axopicture}(132,196) (46,-46)
    \SetWidth{3.0}
    \SetColor{Black}
    \Line(48,-12)(176,-12)
    \Line(48,-76)(176,-76)
    \Line[dash,dashsize=10](46,84)(176,84)
    \SetWidth{2.0}
    \SetColor{Red}
    \Gluon(112,#3)(112,#4){6}{#5}
  \end{axopicture}
}}}}

\def \csII (#1,#2) {
\resizebox{#1 mm}{!}{\raisebox{#2 pt}{
\fcolorbox{white}{white}{
  \begin{axopicture}(436,258) (175,-111)
    \SetWidth{1.0}
    \SetColor{Black}
    \GBox(256,-110)(464,146){0.882}
    \SetWidth{2.0}
    \Line[arrow,arrowpos=0.58,arrowlength=10,arrowwidth=4,arrowinset=0.2](176,82)(544,82)
    \Line[arrow,arrowpos=0.42,arrowlength=10,arrowwidth=4,arrowinset=0.2](544,-46)(176,-46)
    \SetWidth{3.0}
    \Line(426,8)(406,28)
    \Line(330,8)(310,28)
    \Line(330,28)(310,8)
    \Line(426,28)(406,8)
    \SetWidth{1.0}
     \Bezier(544,-78)(544.002,-81)(539.389,-84.005)(534.775,-84.01) \Bezier(534.775,-84.01)(525.545,-84.02)(525.521,-72.02)(530.925,-72.013) \Bezier(530.925,-72.013)(536.329,-72.005)(536.337,-84.005)(524.583,-84.029) \Bezier(524.583,-84.029)(512.828,-84.053)(512.788,-72.053)(518.192,-72.038) \Bezier(518.192,-72.038)(523.596,-72.024)(523.62,-84.024)(512.035,-84.063) \Bezier(512.035,-84.063)(500.449,-84.102)(500.394,-72.102)(505.797,-72.08) \Bezier(505.797,-72.08)(511.201,-72.058)(511.242,-84.058)(499.823,-84.112) \Bezier(499.823,-84.112)(488.404,-84.165)(488.332,-72.165)(493.736,-72.136) \Bezier(493.736,-72.136)(499.14,-72.108)(499.196,-84.107)(487.008,-84.181) \Bezier(487.008,-84.181)(474.821,-84.254)(474.732,-72.255)(480.136,-72.218) \Bezier(480.136,-72.218)(485.54,-72.182)(485.612,-84.182)(473.643,-84.27) \Bezier(473.643,-84.27)(461.674,-84.359)(461.569,-72.36)(466.973,-72.316) \Bezier(466.973,-72.316)(472.377,-72.272)(472.466,-84.272)(460.711,-84.375) \Bezier(460.711,-84.375)(448.956,-84.477)(448.836,-72.478)(454.24,-72.427) \Bezier(454.24,-72.427)(459.643,-72.376)(459.748,-84.376)(448.205,-84.492) \Bezier(448.205,-84.492)(436.661,-84.608)(436.526,-72.608)(441.929,-72.551) \Bezier(441.929,-72.551)(447.333,-72.493)(447.453,-84.493)(435.305,-84.63) \Bezier(435.305,-84.63)(423.158,-84.767)(423.008,-72.768)(428.411,-72.703) \Bezier(428.411,-72.703)(433.814,-72.639)(433.95,-84.638)(422.066,-84.787) \Bezier(422.066,-84.787)(410.182,-84.935)(410.019,-72.936)(415.422,-72.866) \Bezier(415.422,-72.866)(420.825,-72.795)(420.975,-84.794)(409.349,-84.952) \Bezier(409.349,-84.952)(397.724,-85.11)(397.55,-73.111)(402.952,-73.035) \Bezier(402.952,-73.035)(408.354,-72.959)(408.517,-84.958)(397.145,-85.123) \Bezier(397.145,-85.123)(385.772,-85.288)(385.589,-73.289)(390.99,-73.209) \Bezier(390.99,-73.209)(396.392,-73.128)(396.567,-85.127)(384.758,-85.307) \Bezier(384.758,-85.307)(372.949,-85.487)(372.759,-73.489)(378.16,-73.405) \Bezier(378.16,-73.405)(383.561,-73.32)(383.745,-85.319)(371.593,-85.511) \Bezier(371.593,-85.511)(359.441,-85.703)(359.247,-73.705)(364.647,-73.619) \Bezier(364.647,-73.619)(370.047,-73.532)(370.238,-85.531)(358.449,-85.719) \Bezier(358.449,-85.719)(346.661,-85.908)(346.47,-73.909)(351.869,-73.823) \Bezier(351.869,-73.823)(357.268,-73.737)(357.461,-85.735)(346.025,-85.915) \Bezier(346.025,-85.915)(334.588,-86.095)(334.409,-74.096)(339.805,-74.013) \Bezier(339.805,-74.013)(345.202,-73.93)(345.392,-85.928)(333.218,-86.107) \Bezier(333.218,-86.107)(321.044,-86.287)(320.888,-74.288)(326.282,-74.212) \Bezier(326.282,-74.212)(331.675,-74.137)(331.855,-86.135)(320.151,-86.285) \Bezier(320.151,-86.285)(308.448,-86.434)(308.329,-74.435)(313.719,-74.373) \Bezier(313.719,-74.373)(319.109,-74.311)(319.266,-86.31)(308.013,-86.417) \Bezier(308.013,-86.417)(296.76,-86.523)(296.701,-74.523)(302.085,-74.483) \Bezier(302.085,-74.483)(307.47,-74.443)(307.59,-86.443)(295.92,-86.491) \Bezier(295.92,-86.491)(284.25,-86.54)(284.291,-74.54)(289.666,-74.536) \Bezier(289.666,-74.536)(295.041,-74.532)(295.1,-86.532)(283.225,-86.478) \Bezier(283.225,-86.478)(271.35,-86.424)(271.548,-74.425)(276.909,-74.479) \Bezier(276.909,-74.479)(282.269,-74.533)(282.228,-86.533)(271.019,-86.324) \Bezier(271.019,-86.324)(259.811,-86.115)(260.268,-74.124)(265.601,-74.27) \Bezier(265.601,-74.27)(270.934,-74.415)(270.736,-86.414)(258.968,-85.913) \Bezier(258.968,-85.913)(247.2,-85.413)(248.138,-73.45)(253.408,-73.757) \Bezier(253.408,-73.757)(258.677,-74.064)(258.22,-86.055)(246.313,-84.989) \Bezier(246.313,-84.989)(234.405,-83.923)(236.366,-72.084)(241.466,-72.705) \Bezier(241.466,-72.705)(246.566,-73.326)(245.627,-85.29)(233.185,-82.733) \Bezier(233.185,-82.733)(220.743,-80.177)(225.474,-69.149)(229.927,-70.452) \Bezier(229.927,-70.452)(234.379,-71.755)(232.418,-83.593)(220.213,-76.78) \Bezier(220.213,-76.78)(208.008,-69.966)(218.351,-63.881)(220.798,-66.037) \Bezier(220.798,-66.037)(223.246,-68.193)(218.515,-79.221)(210.573,-67.052) \Bezier(210.573,-67.052)(202.631,-54.882)(214.599,-54.011)(215.758,-57.728) \Bezier(215.758,-57.728)(216.917,-61.444)(206.574,-67.529)(204.295,-56.982) \Bezier(204.295,-56.982)(203.155,-51.709)(205.008,-46.218)(208,-46)
     \Bezier(208,82)(205.007,82.207)(203.008,87.557)(204.001,92.7) \Bezier(204.001,92.7)(205.987,102.985)(216.717,97.611)(215.631,93.64) \Bezier(215.631,93.64)(214.546,89.668)(202.575,90.498)(209.475,102.248) \Bezier(209.475,102.248)(216.375,113.998)(221.848,103.318)(219.55,100.999) \Bezier(219.55,100.999)(217.252,98.68)(206.523,104.054)(218.478,111.637) \Bezier(218.478,111.637)(230.434,119.221)(232.678,107.432)(228.43,105.973) \Bezier(228.43,105.973)(224.182,104.514)(218.709,115.194)(230.851,118.125) \Bezier(230.851,118.125)(242.993,121.056)(244.096,109.107)(239.04,108.394) \Bezier(239.04,108.394)(233.985,107.681)(231.741,119.469)(243.362,120.69) \Bezier(243.362,120.69)(254.983,121.911)(255.545,109.925)(250.295,109.559) \Bezier(250.295,109.559)(245.045,109.194)(243.942,121.143)(255.435,121.738) \Bezier(255.435,121.738)(266.928,122.332)(267.186,110.335)(261.865,110.153) \Bezier(261.865,110.153)(256.545,109.971)(255.983,121.958)(267.668,122.237) \Bezier(267.668,122.237)(279.352,122.516)(279.427,110.516)(274.073,110.442) \Bezier(274.073,110.442)(268.72,110.368)(268.462,122.365)(280.152,122.454) \Bezier(280.152,122.454)(291.841,122.543)(291.805,110.543)(286.432,110.535) \Bezier(286.432,110.535)(281.06,110.526)(280.986,122.526)(292.476,122.5) \Bezier(292.476,122.5)(303.966,122.474)(303.858,110.474)(298.476,110.507) \Bezier(298.476,110.507)(293.094,110.539)(293.13,122.539)(305.191,122.437) \Bezier(305.191,122.437)(317.252,122.335)(317.1,110.336)(311.712,110.394) \Bezier(311.712,110.394)(306.323,110.452)(306.43,122.452)(318.03,122.31) \Bezier(318.03,122.31)(329.629,122.168)(329.452,110.169)(324.059,110.243) \Bezier(324.059,110.243)(318.666,110.316)(318.817,122.315)(330.879,122.141) \Bezier(330.879,122.141)(342.94,121.967)(342.751,109.968)(337.355,110.05) \Bezier(337.355,110.05)(331.959,110.132)(332.135,122.131)(343.466,121.955) \Bezier(343.466,121.955)(354.797,121.778)(354.604,109.78)(349.206,109.866) \Bezier(349.206,109.866)(343.808,109.952)(343.997,121.95)(355.674,121.763) \Bezier(355.674,121.763)(367.352,121.577)(367.16,109.578)(361.76,109.665) \Bezier(361.76,109.665)(356.361,109.751)(356.554,121.75)(368.588,121.558) \Bezier(368.588,121.558)(380.623,121.367)(380.437,109.369)(375.036,109.454) \Bezier(375.036,109.454)(369.636,109.539)(369.827,121.537)(381.521,121.357) \Bezier(381.521,121.357)(393.215,121.176)(393.038,109.178)(387.636,109.259) \Bezier(387.636,109.259)(382.234,109.341)(382.42,121.339)(394.42,121.163) \Bezier(394.42,121.163)(406.42,120.987)(406.255,108.988)(400.852,109.065) \Bezier(400.852,109.065)(395.45,109.142)(395.627,121.141)(407.167,120.982) \Bezier(407.167,120.982)(418.707,120.823)(418.554,108.824)(413.151,108.896) \Bezier(413.151,108.896)(407.749,108.967)(407.914,120.966)(419.708,120.816) \Bezier(419.708,120.816)(431.502,120.666)(431.364,108.667)(425.96,108.733) \Bezier(425.96,108.733)(420.557,108.798)(420.71,120.797)(432.763,120.659) \Bezier(432.763,120.659)(444.817,120.52)(444.694,108.521)(439.29,108.579) \Bezier(439.29,108.579)(433.887,108.638)(434.025,120.637)(446.343,120.512) \Bezier(446.343,120.512)(458.661,120.386)(458.555,108.386)(453.151,108.438) \Bezier(453.151,108.438)(447.747,108.489)(447.87,120.489)(459.557,120.385) \Bezier(459.557,120.385)(471.245,120.281)(471.155,108.281)(465.751,108.326) \Bezier(465.751,108.326)(460.347,108.37)(460.454,120.37)(472.353,120.28) \Bezier(472.353,120.28)(484.252,120.19)(484.178,108.19)(478.774,108.228) \Bezier(478.774,108.228)(473.369,108.265)(473.46,120.264)(485.574,120.19) \Bezier(485.574,120.19)(497.687,120.115)(497.63,108.115)(492.226,108.145) \Bezier(492.226,108.145)(486.822,108.174)(486.896,120.174)(499.228,120.116) \Bezier(499.228,120.116)(511.559,120.057)(511.519,108.057)(506.115,108.079) \Bezier(506.115,108.079)(500.711,108.101)(500.768,120.101)(512.303,120.062) \Bezier(512.303,120.062)(523.838,120.023)(523.814,108.023)(518.41,108.038) \Bezier(518.41,108.038)(513.006,108.052)(513.046,120.052)(524.747,120.029) \Bezier(524.747,120.029)(536.449,120.005)(536.441,108.005)(531.037,108.012) \Bezier(531.037,108.012)(525.633,108.02)(525.657,120.02)(534.83,120.01) \Bezier(534.83,120.01)(539.417,120.005)(544.002,117)(544,114)
    \Gluon(336,114)(320,18){6}{7}
    \Gluon(320,18)(304,82){6}{4}
    \Gluon(416,-78)(416,82){6}{12}
    \Text(560,114)[l]{\fontsize{30}{2}\selectfont {$\xvec_1$}} 
    \Text(560,82)[l]{\fontsize{30}{2}\selectfont {$\xvec_2$}} 
    \Text(560,-46)[l]{\fontsize{30}{2}\selectfont {$\xvec_3$}} 
    \Text(560,-78)[l]{\fontsize{30}{2}\selectfont {$\xvec_4$}} 
  \end{axopicture}
}}}}

\def \csIO (#1,#2) {
\resizebox{#1 mm}{!}{\raisebox{#2 pt}{
\fcolorbox{white}{white}{
  \begin{axopicture}(420,258) (175,-111)
    \SetWidth{1.0}
    \SetColor{Black}
    \GBox(256,-110)(464,146){0.882}
    \SetWidth{2.0}
    \Line[arrow,arrowpos=0.58,arrowlength=10,arrowwidth=4,arrowinset=0.2](176,82)(544,82)
    \Line[arrow,arrowpos=0.42,arrowlength=10,arrowwidth=4,arrowinset=0.2](544,-46)(176,-46)
    \SetWidth{3.0}
    \Line(426,8)(406,28)
    \Line(330,8)(310,28)
    \Line(330,28)(310,8)
    \Line(426,28)(406,8)
    \SetWidth{1.0}
     \Bezier(544,-78)(544.191,-80.994)(539.577,-84.362)(534.772,-84.737) \Bezier(534.772,-84.737)(525.16,-85.486)(523.781,-73.565)(529.312,-73.069) \Bezier(529.312,-73.069)(534.842,-72.573)(535.607,-84.548)(522.982,-84.91) \Bezier(522.982,-84.91)(510.358,-85.272)(512.647,-73.493)(517.063,-73.662) \Bezier(517.063,-73.662)(521.478,-73.832)(522.857,-85.752)(509.31,-80.54) \Bezier(509.31,-80.54)(495.763,-75.328)(506.722,-70.441)(508.305,-71.699) \Bezier(508.305,-71.699)(509.887,-72.956)(507.598,-84.736)(499.179,-71.632) \Bezier(499.179,-71.632)(490.76,-58.529)(502.737,-57.792)(503.748,-61.915) \Bezier(503.748,-61.915)(504.759,-66.038)(493.799,-70.926)(491.905,-58.647) \Bezier(491.905,-58.647)(490.958,-52.508)(493.006,-46.184)(496,-46)
     \Bezier(208,82)(205.007,82.207)(203.008,87.557)(204.001,92.7) \Bezier(204.001,92.7)(205.987,102.985)(216.717,97.611)(215.631,93.64) \Bezier(215.631,93.64)(214.546,89.668)(202.575,90.498)(209.475,102.248) \Bezier(209.475,102.248)(216.375,113.998)(221.848,103.318)(219.55,100.999) \Bezier(219.55,100.999)(217.252,98.68)(206.523,104.054)(218.478,111.637) \Bezier(218.478,111.637)(230.434,119.221)(232.678,107.432)(228.43,105.973) \Bezier(228.43,105.973)(224.182,104.514)(218.709,115.194)(230.851,118.125) \Bezier(230.851,118.125)(242.993,121.056)(244.096,109.107)(239.04,108.394) \Bezier(239.04,108.394)(233.985,107.681)(231.741,119.469)(243.362,120.69) \Bezier(243.362,120.69)(254.983,121.911)(255.545,109.925)(250.295,109.559) \Bezier(250.295,109.559)(245.045,109.194)(243.942,121.143)(255.435,121.738) \Bezier(255.435,121.738)(266.928,122.332)(267.186,110.335)(261.865,110.153) \Bezier(261.865,110.153)(256.545,109.971)(255.983,121.958)(267.668,122.237) \Bezier(267.668,122.237)(279.352,122.516)(279.427,110.516)(274.073,110.442) \Bezier(274.073,110.442)(268.72,110.368)(268.462,122.365)(280.152,122.454) \Bezier(280.152,122.454)(291.841,122.543)(291.805,110.543)(286.432,110.535) \Bezier(286.432,110.535)(281.06,110.526)(280.986,122.526)(292.476,122.5) \Bezier(292.476,122.5)(303.966,122.474)(303.858,110.474)(298.476,110.507) \Bezier(298.476,110.507)(293.094,110.539)(293.13,122.539)(305.191,122.437) \Bezier(305.191,122.437)(317.252,122.335)(317.1,110.336)(311.712,110.394) \Bezier(311.712,110.394)(306.323,110.452)(306.43,122.452)(318.03,122.31) \Bezier(318.03,122.31)(329.629,122.168)(329.452,110.169)(324.059,110.243) \Bezier(324.059,110.243)(318.666,110.316)(318.817,122.315)(330.879,122.141) \Bezier(330.879,122.141)(342.94,121.967)(342.751,109.968)(337.355,110.05) \Bezier(337.355,110.05)(331.959,110.132)(332.135,122.131)(343.466,121.955) \Bezier(343.466,121.955)(354.797,121.778)(354.604,109.78)(349.206,109.866) \Bezier(349.206,109.866)(343.808,109.952)(343.997,121.95)(355.674,121.763) \Bezier(355.674,121.763)(367.352,121.577)(367.16,109.578)(361.76,109.665) \Bezier(361.76,109.665)(356.361,109.751)(356.554,121.75)(368.588,121.558) \Bezier(368.588,121.558)(380.623,121.367)(380.437,109.369)(375.036,109.454) \Bezier(375.036,109.454)(369.636,109.539)(369.827,121.537)(381.521,121.357) \Bezier(381.521,121.357)(393.215,121.176)(393.038,109.178)(387.636,109.259) \Bezier(387.636,109.259)(382.234,109.341)(382.42,121.339)(394.42,121.163) \Bezier(394.42,121.163)(406.42,120.987)(406.255,108.988)(400.852,109.065) \Bezier(400.852,109.065)(395.45,109.142)(395.627,121.141)(407.167,120.982) \Bezier(407.167,120.982)(418.707,120.823)(418.554,108.824)(413.151,108.896) \Bezier(413.151,108.896)(407.749,108.967)(407.914,120.966)(419.708,120.816) \Bezier(419.708,120.816)(431.502,120.666)(431.364,108.667)(425.96,108.733) \Bezier(425.96,108.733)(420.557,108.798)(420.71,120.797)(432.763,120.659) \Bezier(432.763,120.659)(444.817,120.52)(444.694,108.521)(439.29,108.579) \Bezier(439.29,108.579)(433.887,108.638)(434.025,120.637)(446.343,120.512) \Bezier(446.343,120.512)(458.661,120.386)(458.555,108.386)(453.151,108.438) \Bezier(453.151,108.438)(447.747,108.489)(447.87,120.489)(459.557,120.385) \Bezier(459.557,120.385)(471.245,120.281)(471.155,108.281)(465.751,108.326) \Bezier(465.751,108.326)(460.347,108.37)(460.454,120.37)(472.353,120.28) \Bezier(472.353,120.28)(484.252,120.19)(484.178,108.19)(478.774,108.228) \Bezier(478.774,108.228)(473.369,108.265)(473.46,120.264)(485.574,120.19) \Bezier(485.574,120.19)(497.687,120.115)(497.63,108.115)(492.226,108.145) \Bezier(492.226,108.145)(486.822,108.174)(486.896,120.174)(499.228,120.116) \Bezier(499.228,120.116)(511.559,120.057)(511.519,108.057)(506.115,108.079) \Bezier(506.115,108.079)(500.711,108.101)(500.768,120.101)(512.303,120.062) \Bezier(512.303,120.062)(523.838,120.023)(523.814,108.023)(518.41,108.038) \Bezier(518.41,108.038)(513.006,108.052)(513.046,120.052)(524.747,120.029) \Bezier(524.747,120.029)(536.449,120.005)(536.441,108.005)(531.037,108.012) \Bezier(531.037,108.012)(525.633,108.02)(525.657,120.02)(534.83,120.01) \Bezier(534.83,120.01)(539.417,120.005)(544.002,117)(544,114)
    \Gluon(336,114)(320,18){6}{7}
    \Gluon(320,18)(304,82){6}{4}
    \Gluon(416,-46)(416,82){6}{10}
  \end{axopicture}
}}}}

\def \csOI (#1,#2) {
\resizebox{#1 mm}{!}{\raisebox{#2 pt}{
\fcolorbox{white}{white}{
  \begin{picture}(436,258) (175,-111)
    \SetWidth{1.0}
    \SetColor{Black}
    \GBox(256,-110)(464,146){0.882}
    \SetWidth{2.0}
    \Line[arrow,arrowpos=0.58,arrowlength=10,arrowwidth=4,arrowinset=0.2](176,82)(544,82)
    \Line[arrow,arrowpos=0.42,arrowlength=10,arrowwidth=4,arrowinset=0.2](544,-46)(176,-46)
    \SetWidth{3.0}
    \Line(426,8)(406,28)
    \Line(330,8)(310,28)
    \Line(330,28)(310,8)
    \Line(426,28)(406,8)
    \SetWidth{1.0}
     \Bezier(544,-78)(544.002,-81)(539.389,-84.005)(534.775,-84.01) \Bezier(534.775,-84.01)(525.545,-84.02)(525.521,-72.02)(530.925,-72.013) \Bezier(530.925,-72.013)(536.329,-72.005)(536.337,-84.005)(524.583,-84.029) \Bezier(524.583,-84.029)(512.828,-84.053)(512.788,-72.053)(518.192,-72.038) \Bezier(518.192,-72.038)(523.596,-72.024)(523.62,-84.024)(512.035,-84.063) \Bezier(512.035,-84.063)(500.449,-84.102)(500.394,-72.102)(505.797,-72.08) \Bezier(505.797,-72.08)(511.201,-72.058)(511.242,-84.058)(499.823,-84.112) \Bezier(499.823,-84.112)(488.404,-84.165)(488.332,-72.165)(493.736,-72.136) \Bezier(493.736,-72.136)(499.14,-72.108)(499.196,-84.107)(487.008,-84.181) \Bezier(487.008,-84.181)(474.821,-84.254)(474.732,-72.255)(480.136,-72.218) \Bezier(480.136,-72.218)(485.54,-72.182)(485.612,-84.182)(473.643,-84.27) \Bezier(473.643,-84.27)(461.674,-84.359)(461.569,-72.36)(466.973,-72.316) \Bezier(466.973,-72.316)(472.377,-72.272)(472.466,-84.272)(460.711,-84.375) \Bezier(460.711,-84.375)(448.956,-84.477)(448.836,-72.478)(454.24,-72.427) \Bezier(454.24,-72.427)(459.643,-72.376)(459.748,-84.376)(448.205,-84.492) \Bezier(448.205,-84.492)(436.661,-84.608)(436.526,-72.608)(441.929,-72.551) \Bezier(441.929,-72.551)(447.333,-72.493)(447.453,-84.493)(435.305,-84.63) \Bezier(435.305,-84.63)(423.158,-84.767)(423.008,-72.768)(428.411,-72.703) \Bezier(428.411,-72.703)(433.814,-72.639)(433.95,-84.638)(422.066,-84.787) \Bezier(422.066,-84.787)(410.182,-84.935)(410.019,-72.936)(415.422,-72.866) \Bezier(415.422,-72.866)(420.825,-72.795)(420.975,-84.794)(409.349,-84.952) \Bezier(409.349,-84.952)(397.724,-85.11)(397.55,-73.111)(402.952,-73.035) \Bezier(402.952,-73.035)(408.354,-72.959)(408.517,-84.958)(397.145,-85.123) \Bezier(397.145,-85.123)(385.772,-85.288)(385.589,-73.289)(390.99,-73.209) \Bezier(390.99,-73.209)(396.392,-73.128)(396.567,-85.127)(384.758,-85.307) \Bezier(384.758,-85.307)(372.949,-85.487)(372.759,-73.489)(378.16,-73.405) \Bezier(378.16,-73.405)(383.561,-73.32)(383.745,-85.319)(371.593,-85.511) \Bezier(371.593,-85.511)(359.441,-85.703)(359.247,-73.705)(364.647,-73.619) \Bezier(364.647,-73.619)(370.047,-73.532)(370.238,-85.531)(358.449,-85.719) \Bezier(358.449,-85.719)(346.661,-85.908)(346.47,-73.909)(351.869,-73.823) \Bezier(351.869,-73.823)(357.268,-73.737)(357.461,-85.735)(346.025,-85.915) \Bezier(346.025,-85.915)(334.588,-86.095)(334.409,-74.096)(339.805,-74.013) \Bezier(339.805,-74.013)(345.202,-73.93)(345.392,-85.928)(333.218,-86.107) \Bezier(333.218,-86.107)(321.044,-86.287)(320.888,-74.288)(326.282,-74.212) \Bezier(326.282,-74.212)(331.675,-74.137)(331.855,-86.135)(320.151,-86.285) \Bezier(320.151,-86.285)(308.448,-86.434)(308.329,-74.435)(313.719,-74.373) \Bezier(313.719,-74.373)(319.109,-74.311)(319.266,-86.31)(308.013,-86.417) \Bezier(308.013,-86.417)(296.76,-86.523)(296.701,-74.523)(302.085,-74.483) \Bezier(302.085,-74.483)(307.47,-74.443)(307.59,-86.443)(295.92,-86.491) \Bezier(295.92,-86.491)(284.25,-86.54)(284.291,-74.54)(289.666,-74.536) \Bezier(289.666,-74.536)(295.041,-74.532)(295.1,-86.532)(283.225,-86.478) \Bezier(283.225,-86.478)(271.35,-86.424)(271.548,-74.425)(276.909,-74.479) \Bezier(276.909,-74.479)(282.269,-74.533)(282.228,-86.533)(271.019,-86.324) \Bezier(271.019,-86.324)(259.811,-86.115)(260.268,-74.124)(265.601,-74.27) \Bezier(265.601,-74.27)(270.934,-74.415)(270.736,-86.414)(258.968,-85.913) \Bezier(258.968,-85.913)(247.2,-85.413)(248.138,-73.45)(253.408,-73.757) \Bezier(253.408,-73.757)(258.677,-74.064)(258.22,-86.055)(246.313,-84.989) \Bezier(246.313,-84.989)(234.405,-83.923)(236.366,-72.084)(241.466,-72.705) \Bezier(241.466,-72.705)(246.566,-73.326)(245.627,-85.29)(233.185,-82.733) \Bezier(233.185,-82.733)(220.743,-80.177)(225.474,-69.149)(229.927,-70.452) \Bezier(229.927,-70.452)(234.379,-71.755)(232.418,-83.593)(220.213,-76.78) \Bezier(220.213,-76.78)(208.008,-69.966)(218.351,-63.881)(220.798,-66.037) \Bezier(220.798,-66.037)(223.246,-68.193)(218.515,-79.221)(210.573,-67.052) \Bezier(210.573,-67.052)(202.631,-54.882)(214.599,-54.011)(215.758,-57.728) \Bezier(215.758,-57.728)(216.917,-61.444)(206.574,-67.529)(204.295,-56.982) \Bezier(204.295,-56.982)(203.155,-51.709)(205.008,-46.218)(208,-46)
    \Gluon(336,82)(320,18){6}{5}
    \Gluon(320,18)(304,82){6}{4}
    \Gluon(416,-78)(416,82){6}{12}
     \Bezier(496,82)(493.004,82.148)(490.63,87.727)(491.252,93.157) \Bezier(491.252,93.157)(492.497,104.017)(504.02,100.665)(503.248,96.066) \Bezier(503.248,96.066)(502.476,91.466)(490.49,92.059)(496.654,105.269) \Bezier(496.654,105.269)(502.817,118.478)(508.207,107.756)(506.574,105.793) \Bezier(506.574,105.793)(504.94,103.83)(493.418,107.182)(505.834,114.609) \Bezier(505.834,114.609)(518.251,122.036)(517.172,110.085)(513.688,109.423) \Bezier(513.688,109.423)(510.204,108.76)(504.814,119.481)(519.157,120.122) \Bezier(519.157,120.122)(533.5,120.762)(532.648,108.792)(527.209,109.231) \Bezier(527.209,109.231)(521.77,109.67)(522.849,121.621)(533.637,120.803) \Bezier(533.637,120.803)(539.031,120.394)(544.213,116.992)(544,114)
  \end{picture}
}}}}

\def \csOO (#1,#2) {
\resizebox{#1 mm}{!}{\raisebox{#2 pt}{
\fcolorbox{white}{white}{
  \begin{axopicture}(436,258) (175,-111)
    \SetWidth{1.0}
    \SetColor{Black}
    \GBox(256,-110)(464,146){0.882}
    \SetWidth{2.0}
    \Line[arrow,arrowpos=0.58,arrowlength=10,arrowwidth=4,arrowinset=0.2](176,82)(544,82)
    \Line[arrow,arrowpos=0.42,arrowlength=10,arrowwidth=4,arrowinset=0.2](544,-46)(176,-46)
    \SetWidth{3.0}
    \Line(426,8)(406,28)
    \Line(330,8)(310,28)
    \Line(330,28)(310,8)
    \Line(426,28)(406,8)
    \SetWidth{1.0}
    \Gluon(336,82)(320,18){6}{5}
    \Gluon(320,18)(304,82){6}{4}
    \Gluon(416,-46)(416,82){6}{10}
     \Bezier(496,82)(492.255,82.185)(489.24,88.211)(489.972,94.052) \Bezier(489.972,94.052)(491.434,105.733)(505.837,101.543)(504.872,95.793) \Bezier(504.872,95.793)(503.907,90.044)(488.926,90.785)(496.137,105.605) \Bezier(496.137,105.605)(503.349,120.424)(510.087,107.023)(508.045,104.569) \Bezier(508.045,104.569)(506.003,102.115)(491.601,106.305)(505.665,114.857) \Bezier(505.665,114.857)(519.73,123.409)(518.382,108.47)(514.027,107.642) \Bezier(514.027,107.642)(509.671,106.814)(502.934,120.215)(518.943,121.189) \Bezier(518.943,121.189)(534.953,122.162)(533.888,107.2)(527.089,107.749) \Bezier(527.089,107.749)(520.29,108.297)(521.639,123.237)(533.086,122.359) \Bezier(533.086,122.359)(538.809,121.92)(544.266,117.741)(544,114)
     \Bezier(544,-78)(544.239,-81.742)(539.357,-85.896)(534.235,-86.308) \Bezier(534.235,-86.308)(523.992,-87.131)(522.268,-72.23)(529.181,-71.61) \Bezier(529.181,-71.61)(536.093,-70.99)(537.05,-85.959)(522.898,-86.223) \Bezier(522.898,-86.223)(508.747,-86.487)(511.608,-71.763)(517.127,-71.975) \Bezier(517.127,-71.975)(522.647,-72.187)(524.37,-87.087)(509.107,-80.897) \Bezier(509.107,-80.897)(493.843,-74.706)(507.543,-68.597)(509.521,-70.169) \Bezier(509.521,-70.169)(511.499,-71.741)(508.637,-86.466)(498.908,-71.87) \Bezier(498.908,-71.87)(489.18,-57.273)(504.151,-56.352)(505.415,-61.506) \Bezier(505.415,-61.506)(506.679,-66.66)(492.979,-72.769)(490.747,-59.615) \Bezier(490.747,-59.615)(489.63,-53.038)(492.257,-46.23)(496,-46)
  \end{axopicture}
}}}}

\def \CrossSectionGGG (#1,#2) {
\resizebox{#1 mm}{!}{\raisebox{#2 pt}{
\fcolorbox{white}{white}{
  \begin{axopicture}(516,322) (47,-79)
    \SetWidth{1.0}
    \SetColor{Black}
    \GBox(192,-78)(448,242){0.882}
    \SetWidth{3.0}
    \Line(426,72)(406,92)
    \Line(330,72)(310,92)
    \Line(330,92)(310,72)
    \Line(426,92)(406,72)
    \Line(234,72)(214,92)
    \Line(234,92)(214,72)
    \SetWidth{1.0}
    \Gluon(48,178)(496,178){5.5}{34}
     \Bezier(96,178)(93.041,178.492)(92.623,184.324)(95.164,189.663) \Bezier(95.164,189.663)(100.246,200.341)(106.033,189.829)(104.737,187.999) \Bezier(104.737,187.999)(103.441,186.17)(91.604,188.139)(105.535,198.399) \Bezier(105.535,198.399)(119.465,208.66)(121.69,196.868)(117.562,195.385) \Bezier(117.562,195.385)(113.434,193.902)(107.647,204.415)(119.462,207.456) \Bezier(119.462,207.456)(131.277,210.497)(132.687,198.58)(127.556,197.793) \Bezier(127.556,197.793)(122.425,197.007)(120.2,208.799)(132.358,210.328) \Bezier(132.358,210.328)(144.515,211.858)(145.483,199.897)(140.221,199.373) \Bezier(140.221,199.373)(134.959,198.849)(133.549,210.766)(145.34,211.761) \Bezier(145.34,211.761)(157.13,212.757)(157.834,200.778)(152.513,200.406) \Bezier(152.513,200.406)(147.193,200.035)(146.225,211.996)(158.337,212.729) \Bezier(158.337,212.729)(170.448,213.462)(170.982,201.474)(165.632,201.198) \Bezier(165.632,201.198)(160.282,200.922)(159.578,212.901)(171.033,213.423) \Bezier(171.033,213.423)(182.487,213.945)(182.903,201.952)(177.537,201.739) \Bezier(177.537,201.739)(172.17,201.527)(171.637,213.515)(183.727,213.941) \Bezier(183.727,213.941)(195.817,214.366)(196.143,202.37)(190.768,202.204) \Bezier(190.768,202.204)(185.393,202.038)(184.977,214.031)(196.791,214.358) \Bezier(196.791,214.358)(208.604,214.685)(208.867,202.688)(203.485,202.556) \Bezier(203.485,202.556)(198.102,202.424)(197.776,214.419)(209.109,214.671) \Bezier(209.109,214.671)(220.443,214.923)(220.657,202.925)(215.27,202.818) \Bezier(215.27,202.818)(209.884,202.71)(209.621,214.708)(221.368,214.919) \Bezier(221.368,214.919)(233.115,215.13)(233.288,203.131)(227.899,203.045) \Bezier(227.899,203.045)(222.51,202.958)(222.296,214.956)(234.471,215.133) \Bezier(234.471,215.133)(246.647,215.311)(246.788,203.312)(241.396,203.241) \Bezier(241.396,203.241)(236.005,203.171)(235.832,215.169)(247.224,215.305) \Bezier(247.224,215.305)(258.617,215.44)(258.733,203.441)(253.339,203.383) \Bezier(253.339,203.383)(247.946,203.325)(247.805,215.324)(259.514,215.438) \Bezier(259.514,215.438)(271.224,215.552)(271.318,203.553)(265.924,203.505) \Bezier(265.924,203.505)(260.529,203.458)(260.413,215.457)(272.449,215.553) \Bezier(272.449,215.553)(284.484,215.649)(284.561,203.649)(279.165,203.61) \Bezier(279.165,203.61)(273.77,203.572)(273.675,215.572)(285.338,215.647) \Bezier(285.338,215.647)(297,215.723)(297.063,203.723)(291.666,203.691) \Bezier(291.666,203.691)(286.27,203.66)(286.193,215.66)(298.132,215.723) \Bezier(298.132,215.723)(310.072,215.786)(310.123,203.786)(304.726,203.76) \Bezier(304.726,203.76)(299.329,203.735)(299.266,215.735)(310.722,215.784) \Bezier(310.722,215.784)(322.179,215.833)(322.22,203.833)(316.823,203.812) \Bezier(316.823,203.812)(311.425,203.791)(311.374,215.791)(323.059,215.832) \Bezier(323.059,215.832)(334.744,215.873)(334.777,203.873)(329.379,203.856) \Bezier(329.379,203.856)(323.982,203.839)(323.94,215.839)(335.858,215.872) \Bezier(335.858,215.872)(347.777,215.906)(347.803,203.906)(342.405,203.892) \Bezier(342.405,203.892)(337.006,203.879)(336.973,215.879)(349.129,215.906) \Bezier(349.129,215.906)(361.285,215.932)(361.305,203.932)(355.906,203.922) \Bezier(355.906,203.922)(350.508,203.911)(350.482,215.911)(361.998,215.931) \Bezier(361.998,215.931)(373.515,215.951)(373.531,203.951)(368.132,203.943) \Bezier(368.132,203.943)(362.733,203.935)(362.713,215.935)(374.417,215.95) \Bezier(374.417,215.95)(386.122,215.966)(386.134,203.966)(380.735,203.96) \Bezier(380.735,203.96)(375.336,203.954)(375.32,215.954)(387.217,215.966) \Bezier(387.217,215.966)(399.113,215.978)(399.122,203.978)(393.723,203.973) \Bezier(393.723,203.973)(388.323,203.968)(388.311,215.968)(400.402,215.977) \Bezier(400.402,215.977)(412.492,215.986)(412.499,203.986)(407.099,203.983) \Bezier(407.099,203.983)(401.7,203.979)(401.691,215.979)(412.992,215.985) \Bezier(412.992,215.985)(424.292,215.992)(424.296,203.992)(418.897,203.989) \Bezier(418.897,203.989)(413.497,203.987)(413.491,215.987)(424.938,215.991) \Bezier(424.938,215.991)(436.386,215.995)(436.389,203.995)(430.989,203.994) \Bezier(430.989,203.994)(425.59,203.992)(425.585,215.992)(437.182,215.995) \Bezier(437.182,215.995)(448.778,215.998)(448.78,203.998)(443.38,203.997) \Bezier(443.38,203.997)(437.98,203.996)(437.977,215.996)(449.725,215.997) \Bezier(449.725,215.997)(461.472,215.999)(461.473,203.999)(456.073,203.999) \Bezier(456.073,203.999)(450.673,203.998)(450.671,215.998)(462.571,215.999) \Bezier(462.571,215.999)(474.471,216)(474.471,204)(469.071,203.999) \Bezier(469.071,203.999)(463.672,203.999)(463.671,215.999)(475.725,216) \Bezier(475.725,216)(487.779,216)(487.779,204)(482.379,204) \Bezier(482.379,204)(476.979,204)(476.979,216)(486.49,216) \Bezier(486.49,216)(491.245,216)(496,213)(496,210)
     \Bezier(496,146)(496,143)(490.392,140)(484.785,140) \Bezier(484.785,140)(473.569,140)(473.57,152)(478.97,152) \Bezier(478.97,152)(484.37,152)(484.369,140)(470.733,140.001) \Bezier(470.733,140.001)(457.096,140.002)(457.098,152.002)(462.498,152.001) \Bezier(462.498,152.001)(467.898,152.001)(467.897,140.001)(454.535,140.003) \Bezier(454.535,140.003)(441.172,140.006)(441.176,152.006)(446.576,152.004) \Bezier(446.576,152.004)(451.975,152.003)(451.973,140.003)(438.88,140.008) \Bezier(438.88,140.008)(425.787,140.012)(425.794,152.012)(431.194,152.01) \Bezier(431.194,152.01)(436.593,152.007)(436.589,140.007)(423.761,140.015) \Bezier(423.761,140.015)(410.933,140.022)(410.943,152.022)(416.342,152.018) \Bezier(416.342,152.018)(421.742,152.014)(421.735,140.015)(409.166,140.026) \Bezier(409.166,140.026)(396.598,140.038)(396.614,152.038)(402.013,152.032) \Bezier(402.013,152.032)(407.412,152.026)(407.401,140.026)(394.198,140.044) \Bezier(394.198,140.044)(380.996,140.061)(381.019,152.061)(386.417,152.053) \Bezier(386.417,152.053)(391.815,152.044)(391.799,140.044)(378.917,140.069) \Bezier(378.917,140.069)(366.034,140.094)(366.066,152.094)(371.464,152.081) \Bezier(371.464,152.081)(376.861,152.069)(376.839,140.069)(364.269,140.102) \Bezier(364.269,140.102)(351.7,140.136)(351.742,152.136)(357.139,152.119) \Bezier(357.139,152.119)(362.537,152.103)(362.505,140.103)(350.242,140.146) \Bezier(350.242,140.146)(337.98,140.189)(338.034,152.189)(343.431,152.167) \Bezier(343.431,152.167)(348.828,152.146)(348.786,140.146)(336.08,140.205) \Bezier(336.08,140.205)(323.373,140.263)(323.445,152.263)(328.84,152.235) \Bezier(328.84,152.235)(334.236,152.206)(334.181,140.206)(321.839,140.281) \Bezier(321.839,140.281)(309.496,140.355)(309.587,152.354)(314.982,152.318) \Bezier(314.982,152.318)(320.377,152.281)(320.306,140.281)(308.317,140.374) \Bezier(308.317,140.374)(296.328,140.466)(296.443,152.465)(301.837,152.419) \Bezier(301.837,152.419)(307.231,152.372)(307.14,140.373)(295.495,140.486) \Bezier(295.495,140.486)(283.851,140.599)(283.995,152.598)(289.387,152.539) \Bezier(289.387,152.539)(294.78,152.481)(294.665,140.482)(282.759,140.627) \Bezier(282.759,140.627)(270.854,140.773)(271.037,152.771)(276.427,152.698) \Bezier(276.427,152.698)(281.817,152.624)(281.672,140.625)(269.606,140.811) \Bezier(269.606,140.811)(257.54,140.998)(257.772,152.996)(263.159,152.903) \Bezier(263.159,152.903)(268.546,152.809)(268.363,140.811)(256.753,141.038) \Bezier(256.753,141.038)(245.142,141.265)(245.434,153.262)(250.818,153.144) \Bezier(250.818,153.144)(256.202,153.026)(255.969,141.029)(244.796,141.304) \Bezier(244.796,141.304)(233.623,141.58)(233.993,153.575)(239.371,153.426) \Bezier(239.371,153.426)(244.75,153.278)(244.458,141.282)(232.829,141.646) \Bezier(232.829,141.646)(221.2,142.01)(221.673,154.001)(227.044,153.812) \Bezier(227.044,153.812)(232.415,153.623)(232.045,141.629)(220.963,142.075) \Bezier(220.963,142.075)(209.881,142.521)(210.495,154.505)(215.854,154.262) \Bezier(215.854,154.262)(221.213,154.019)(220.739,142.029)(209.468,142.618) \Bezier(209.468,142.618)(198.196,143.207)(199.002,155.18)(204.344,154.863) \Bezier(204.344,154.863)(209.687,154.547)(209.073,142.563)(198.438,143.295) \Bezier(198.438,143.295)(187.804,144.028)(188.875,155.98)(194.191,155.563) \Bezier(194.191,155.563)(199.508,155.146)(198.703,143.173)(188.11,144.16) \Bezier(188.11,144.16)(177.516,145.147)(179.013,157.053)(184.275,156.487) \Bezier(184.275,156.487)(189.536,155.92)(188.465,143.968)(177.634,145.408) \Bezier(177.634,145.408)(166.802,146.847)(169.044,158.636)(174.189,157.824) \Bezier(174.189,157.824)(179.335,157.012)(177.838,145.106)(167.182,147.297) \Bezier(167.182,147.297)(156.526,149.488)(160.134,160.932)(165.02,159.702) \Bezier(165.02,159.702)(169.906,158.472)(167.664,146.683)(157.521,150.523) \Bezier(157.521,150.523)(147.378,154.362)(154.257,164.195)(158.226,162.239) \Bezier(158.226,162.239)(162.196,160.283)(158.588,148.838)(149.458,157.35) \Bezier(149.458,157.35)(140.327,165.863)(152.09,168.235)(153.616,165.906) \Bezier(153.616,165.906)(155.141,163.577)(148.262,153.744)(143.19,165.279) \Bezier(143.19,165.279)(140.654,171.046)(141.059,177.407)(144,178)
    \Gluon(496,-14)(48,-14){6}{34}
     \Bezier(496,-46)(496,-49)(490.207,-52)(484.415,-52) \Bezier(484.415,-52)(472.83,-52)(472.83,-40)(478.23,-40) \Bezier(478.23,-40)(483.63,-40)(483.63,-52)(469.612,-51.999) \Bezier(469.612,-51.999)(455.594,-51.998)(455.595,-39.998)(460.995,-39.999) \Bezier(460.995,-39.999)(466.395,-39.999)(466.394,-51.999)(452.639,-51.998) \Bezier(452.639,-51.998)(438.883,-51.996)(438.886,-39.996)(444.286,-39.997) \Bezier(444.286,-39.997)(449.685,-39.998)(449.684,-51.998)(436.187,-51.994) \Bezier(436.187,-51.994)(422.691,-51.991)(422.696,-39.991)(428.095,-39.993) \Bezier(428.095,-39.993)(433.495,-39.994)(433.492,-51.994)(420.249,-51.989) \Bezier(420.249,-51.989)(407.007,-51.983)(407.015,-39.983)(412.414,-39.986) \Bezier(412.414,-39.986)(417.814,-39.989)(417.809,-51.989)(404.817,-51.98) \Bezier(404.817,-51.98)(391.825,-51.971)(391.836,-39.971)(397.236,-39.976) \Bezier(397.236,-39.976)(402.635,-39.98)(402.627,-51.98)(389.882,-51.968) \Bezier(389.882,-51.968)(377.136,-51.956)(377.152,-39.956)(382.551,-39.962) \Bezier(382.551,-39.962)(387.95,-39.968)(387.938,-51.968)(374.557,-51.95) \Bezier(374.557,-51.95)(361.177,-51.932)(361.198,-39.932)(366.597,-39.94) \Bezier(366.597,-39.94)(371.995,-39.949)(371.979,-51.949)(358.9,-51.925) \Bezier(358.9,-51.925)(345.82,-51.901)(345.849,-39.901)(351.247,-39.912) \Bezier(351.247,-39.912)(356.645,-39.924)(356.623,-51.924)(343.839,-51.893) \Bezier(343.839,-51.893)(331.055,-51.862)(331.092,-39.862)(336.49,-39.877) \Bezier(336.49,-39.877)(341.888,-39.891)(341.859,-51.891)(329.364,-51.852) \Bezier(329.364,-51.852)(316.869,-51.813)(316.917,-39.813)(322.314,-39.832) \Bezier(322.314,-39.832)(327.711,-39.851)(327.674,-51.851)(314.718,-51.799) \Bezier(314.718,-51.799)(301.762,-51.747)(301.823,-39.747)(307.22,-39.771) \Bezier(307.22,-39.771)(312.617,-39.796)(312.569,-51.796)(299.955,-51.731) \Bezier(299.955,-51.731)(287.342,-51.667)(287.418,-39.667)(292.814,-39.698) \Bezier(292.814,-39.698)(298.21,-39.729)(298.149,-51.729)(285.87,-51.65) \Bezier(285.87,-51.65)(273.591,-51.571)(273.685,-39.571)(279.081,-39.61) \Bezier(279.081,-39.61)(284.476,-39.648)(284.4,-51.648)(272.447,-51.553) \Bezier(272.447,-51.553)(260.494,-51.458)(260.609,-39.459)(266.004,-39.506) \Bezier(266.004,-39.506)(271.398,-39.553)(271.304,-51.553)(259.669,-51.44) \Bezier(259.669,-51.44)(248.033,-51.326)(248.175,-39.327)(253.568,-39.385) \Bezier(253.568,-39.385)(258.961,-39.443)(258.846,-51.442)(246.941,-51.3) \Bezier(246.941,-51.3)(235.037,-51.158)(235.211,-39.159)(240.603,-39.231) \Bezier(240.603,-39.231)(245.994,-39.302)(245.852,-51.301)(233.769,-51.123) \Bezier(233.769,-51.123)(221.686,-50.945)(221.902,-38.947)(227.291,-39.035) \Bezier(227.291,-39.035)(232.68,-39.123)(232.505,-51.121)(220.841,-50.91) \Bezier(220.841,-50.91)(209.177,-50.698)(209.442,-38.701)(214.828,-38.809) \Bezier(214.828,-38.809)(220.215,-38.917)(219.999,-50.915)(208.74,-50.664) \Bezier(208.74,-50.664)(197.481,-50.413)(197.806,-38.417)(203.189,-38.549) \Bezier(203.189,-38.549)(208.573,-38.682)(208.308,-50.679)(197,-50.368) \Bezier(197,-50.368)(185.691,-50.058)(186.094,-38.065)(191.472,-38.228) \Bezier(191.472,-38.228)(196.85,-38.391)(196.525,-50.387)(185.254,-50.003) \Bezier(185.254,-50.003)(173.982,-49.62)(174.484,-37.63)(179.855,-37.833) \Bezier(179.855,-37.833)(185.227,-38.036)(184.823,-50.029)(174.037,-49.57) \Bezier(174.037,-49.57)(163.25,-49.112)(163.882,-37.128)(169.243,-37.382) \Bezier(169.243,-37.382)(174.605,-37.635)(174.103,-49.625)(163.116,-49.036) \Bezier(163.116,-49.036)(152.128,-48.447)(152.932,-36.474)(158.279,-36.795) \Bezier(158.279,-36.795)(163.626,-37.115)(162.995,-49.098)(152.574,-48.384) \Bezier(152.574,-48.384)(142.153,-47.67)(143.191,-35.715)(148.516,-36.125) \Bezier(148.516,-36.125)(153.842,-36.535)(153.037,-48.508)(142.626,-47.579) \Bezier(142.626,-47.579)(132.215,-46.649)(133.595,-34.729)(138.881,-35.265) \Bezier(138.881,-35.265)(144.168,-35.8)(143.13,-47.755)(132.889,-46.528) \Bezier(132.889,-46.528)(122.649,-45.301)(124.543,-33.451)(129.763,-34.171) \Bezier(129.763,-34.171)(134.983,-34.89)(133.602,-46.81)(123.672,-45.141) \Bezier(123.672,-45.141)(113.741,-43.472)(116.474,-31.788)(121.56,-32.788) \Bezier(121.56,-32.788)(126.645,-33.788)(124.751,-45.637)(114.94,-43.132) \Bezier(114.94,-43.132)(105.13,-40.626)(109.462,-29.436)(114.209,-30.902) \Bezier(114.209,-30.902)(118.956,-32.368)(116.223,-44.053)(106.502,-39.574) \Bezier(106.502,-39.574)(96.782,-35.095)(104.597,-25.989)(108.293,-28.202) \Bezier(108.293,-28.202)(111.989,-30.414)(107.656,-41.604)(99.687,-33.009) \Bezier(99.687,-33.009)(91.718,-24.414)(103.541,-22.362)(105.05,-25.017) \Bezier(105.05,-25.017)(106.558,-27.672)(98.742,-36.778)(94.415,-25.902) \Bezier(94.415,-25.902)(92.252,-20.464)(93.044,-14.513)(96,-14)
     \Bezier(144,-14)(141.065,-13.38)(140.882,-6.969)(143.635,-1.178) \Bezier(143.635,-1.178)(149.141,10.404)(155.167,0.027)(153.703,-1.996) \Bezier(153.703,-1.996)(152.239,-4.019)(140.498,-1.54)(151.798,7.286) \Bezier(151.798,7.286)(163.097,16.113)(165.44,4.344)(161.378,2.809) \Bezier(161.378,2.809)(157.316,1.274)(151.29,11.651)(166.78,15.521) \Bezier(166.78,15.521)(182.271,19.391)(183.508,7.455)(178.451,6.691) \Bezier(178.451,6.691)(173.394,5.928)(171.05,17.697)(183.445,19.154) \Bezier(183.445,19.154)(195.84,20.611)(196.675,8.64)(191.396,8.183) \Bezier(191.396,8.183)(186.118,7.726)(184.881,19.662)(197.194,20.557) \Bezier(197.194,20.557)(209.508,21.452)(210.098,9.467)(204.769,9.15) \Bezier(204.769,9.15)(199.44,8.833)(198.605,20.804)(211.215,21.444) \Bezier(211.215,21.444)(223.825,22.085)(224.255,10.092)(218.9,9.864) \Bezier(218.9,9.864)(213.545,9.637)(212.955,21.622)(225.624,22.088) \Bezier(225.624,22.088)(238.293,22.554)(238.614,10.558)(233.244,10.39) \Bezier(233.244,10.39)(227.874,10.222)(227.444,22.214)(239.919,22.556) \Bezier(239.919,22.556)(252.394,22.897)(252.641,10.9)(247.261,10.773) \Bezier(247.261,10.773)(241.881,10.645)(241.56,22.641)(253.593,22.893) \Bezier(253.593,22.893)(265.626,23.144)(265.817,11.146)(260.432,11.048) \Bezier(260.432,11.048)(255.046,10.949)(254.8,22.947)(267.366,23.15) \Bezier(267.366,23.15)(279.931,23.354)(280.081,11.354)(274.692,11.278) \Bezier(274.692,11.278)(269.303,11.202)(269.112,23.2)(280.931,23.349) \Bezier(280.931,23.349)(292.75,23.499)(292.867,11.499)(287.476,11.439) \Bezier(287.476,11.439)(282.084,11.379)(281.935,23.378)(294.155,23.5) \Bezier(294.155,23.5)(306.375,23.621)(306.467,11.621)(301.074,11.574) \Bezier(301.074,11.574)(295.681,11.527)(295.563,23.527)(308.198,23.624) \Bezier(308.198,23.624)(320.834,23.722)(320.904,11.722)(315.51,11.685) \Bezier(315.51,11.685)(310.115,11.649)(310.024,23.649)(322.316,23.722) \Bezier(322.316,23.722)(334.609,23.796)(334.664,11.796)(329.268,11.768) \Bezier(329.268,11.768)(323.872,11.739)(323.801,23.739)(335.645,23.794) \Bezier(335.645,23.794)(347.489,23.85)(347.532,11.85)(342.135,11.828) \Bezier(342.135,11.828)(336.738,11.806)(336.683,23.805)(348.817,23.849) \Bezier(348.817,23.849)(360.951,23.893)(360.984,11.893)(355.586,11.876) \Bezier(355.586,11.876)(350.189,11.859)(350.146,23.859)(362.578,23.893) \Bezier(362.578,23.893)(375.009,23.927)(375.033,11.927)(369.635,11.914) \Bezier(369.635,11.914)(364.237,11.902)(364.205,23.902)(376.939,23.927) \Bezier(376.939,23.927)(389.674,23.953)(389.691,11.953)(384.293,11.944) \Bezier(384.293,11.944)(378.895,11.935)(378.871,23.935)(390.966,23.952) \Bezier(390.966,23.952)(403.061,23.97)(403.073,11.97)(397.675,11.963) \Bezier(397.675,11.963)(392.276,11.957)(392.259,23.957)(404.597,23.97) \Bezier(404.597,23.97)(416.936,23.982)(416.944,11.982)(411.545,11.978) \Bezier(411.545,11.978)(406.146,11.973)(406.134,23.973)(418.72,23.982) \Bezier(418.72,23.982)(431.307,23.991)(431.312,11.991)(425.912,11.988) \Bezier(425.912,11.988)(420.513,11.985)(420.505,23.985)(433.344,23.99) \Bezier(433.344,23.99)(446.182,23.996)(446.185,11.996)(440.786,11.994) \Bezier(440.786,11.994)(435.386,11.992)(435.381,23.992)(447.391,23.995) \Bezier(447.391,23.995)(459.401,23.998)(459.403,11.998)(454.003,11.997) \Bezier(454.003,11.997)(448.604,11.996)(448.601,23.996)(460.804,23.998) \Bezier(460.804,23.998)(473.007,24)(473.008,12)(467.608,11.999) \Bezier(467.608,11.999)(462.208,11.999)(462.207,23.999)(474.606,23.999) \Bezier(474.606,23.999)(487.005,24)(487.005,12)(481.605,12) \Bezier(481.605,12)(476.205,12)(476.204,24)(486.102,24) \Bezier(486.102,24)(491.051,24)(496,21)(496,18)
    \Text(510,200)[lb]{\fontsize{28}{2}\selectfont {$\pvec_1$}} 
    \Text(510,168)[lb]{\fontsize{28}{2}\selectfont {$\pvec_2$}} 
    \Text(510,136)[lb]{\fontsize{28}{2}\selectfont {$\pvec_3$}} 
    \Gluon(224,82)(208,178){5}{8}
    \Gluon(240,210)(224,82){5}{10}
    \Gluon(320,146)(320,-46){5}{15}
    \Gluon(416,146)(416,18){5}{10}
  \end{axopicture}}}}}

  \def \KetR (#1,#2) {
\resizebox{#1 mm}{!}{\raisebox{#2 pt}{
\fcolorbox{white}{white}{
  \begin{axopicture}(132,196) (46,-46)
    \SetWidth{3.0}
    \SetColor{Black}
    \Line(48,148)(130,148)
    \Line(48,-44)(130,-44)
    \Line(130,-44)(130,148)
    \Text(70,130)[l]{\fontsize{40}{2}\selectfont $R$}
  \end{axopicture}
}}}}

\def \FIGgedanken (#1,#2) {
\resizebox{#1 mm}{!}{\raisebox{#2 pt}{
\fcolorbox{white}{white}{
  \begin{axopicture}(451,155) (125,-125)
    \SetWidth{1.0}
    \SetColor{White}
    \GBox(307.847,-124.188)(363.819,29.735){0.882}
    \SetWidth{2.0}
    \SetColor{Black}
    \Line[arrow,arrowpos=0.1,arrowlength=6.667,arrowwidth=2.667,arrowinset=0.2](153.923,-40.23)(335.833,-40.23)
    \Line[arrow,arrowpos=0.1,arrowlength=6.667,arrowwidth=2.667,arrowinset=0.2,flip](153.923,-54.223)(335.833,-54.223)
    \Line(335.833,-40.23)(545.729,1.749)
    \Line(335.833,-54.223)(545.729,-96.202)
    \SetWidth{1.0}
    \Line[dash,dashsize=8.746](125.937,-47.227)(573.715,-47.227)
    \Vertex(321.84,-82.209){4.459}
    \Vertex(348.077,-83.084){4.459}
    \Vertex(348.077,-82.209){4.459}
    \Photon(321.84,-40.23)(321.84,-82.209){2.624}{6}
    \Photon(348.077,-57.721)(348.951,-82.209){2.624}{4}
    \Text(547,-12.244)[l]{\mbox{\fontsize{24}{2}\selectfont \Black{$\pvec_1$}}}
    \Text(547,-82.209)[l]{\mbox{\fontsize{24}{2}\selectfont \Black{$\pvec_2$}}}
    \Text(148.676,-20.115)[l]{\mbox{\fontsize{24}{2}\selectfont \Black{$e^-$}}}
    \Text(147.801,-76.962)[l]{\mbox{\fontsize{24}{2}\selectfont \Black{$e^+$}}}
    \Text(321.84,-117.192)[l]{\mbox{\fontsize{24}{2}\selectfont \Black{$L$}}}
    \Line[arrow,arrowpos=1,arrowlength=5,arrowwidth=4,arrowinset=0.2](545.729,-47.227)(545.729,1.749)
    \Line[arrow,arrowpos=0,arrowlength=5,arrowwidth=4,arrowinset=0.2,flip](545.729,-96.202)(545.729,-47.227)
    \Text(321.84,-103.199)[l]{\mbox{\fontsize{48}{2}\selectfont \Black{$\leftrightarrow$}}}\end{axopicture}}}}}

\def \FIGdijetpA (#1,#2) {
\resizebox{#1 mm}{!}{\raisebox{#2 pt}{
\fcolorbox{white}{white}{
  \begin{axopicture}(451,119) (37,-90)
    \SetWidth{2.0}
    \SetColor{Black}
    \Line(56.832,-42.624)(151.553,-42.624)
    \Line(57.509,-50.067)(73.747,-54.126)
    \Line(56.832,-40.595)(73.747,-34.505)
    \Line(56.156,-36.535)(71.717,-25.71)
    \Line(37.888,-46.007)(56.156,-46.007)
    \Line(37.888,-39.918)(56.156,-39.918)
    \SetWidth{1.5}
    \Gluon(320.697,-77.806)(320.697,-18.944){3.045}{8}
    \SetWidth{1.0}
    \GOval(56.156,-43.977)(12.855,6.089)(0){0.882}
    \SetWidth{1.5}
    \Gluon(151.553,-42.624)(421.506,-3.383){3.045}{29}
    \SetWidth{2.0}
    \Line(151.553,-42.624)(422.183,-53.449)
    \SetWidth{1.5}
    \Gluon(383.618,-75.1)(382.941,-52.773){3.045}{3}
    \Gluon(203.649,-79.159)(204.326,-37.212){3.045}{6}
    \SetWidth{2.0}
    \Line(202.296,-75.1)(216.504,-73.747)
    \Line(188.088,-76.453)(202.973,-76.453)
    \Line(188.088,-79.159)(202.973,-79.159)
    \Line(202.973,-77.806)(217.857,-77.806)
    \Line(202.296,-79.836)(215.827,-81.866)
    \SetWidth{1.0}
    \GOval(202.973,-78.483)(4.736,6.089)(0){0.882}
    \SetWidth{2.0}
    \Line(382.265,-73.747)(396.473,-72.394)
    \Line(368.057,-75.1)(382.941,-75.1)
    \Line(368.057,-77.806)(382.941,-77.806)
    \Line(382.941,-76.453)(397.826,-76.453)
    \Line(382.265,-78.483)(395.796,-80.512)
    \SetWidth{1.0}
    \GOval(382.941,-77.13)(4.736,6.089)(0){0.882}
    \SetWidth{2.0}
    \Line(320.02,-74.423)(334.228,-73.07)
    \Line(305.812,-75.776)(320.697,-75.776)
    \Line(305.812,-78.483)(320.697,-78.483)
    \Line(320.697,-77.13)(335.581,-77.13)
    \Line(320.02,-79.159)(333.552,-81.189)
    \SetWidth{1.0}
    \GOval(320.697,-77.806)(4.736,6.089)(0){0.882}
    \SetWidth{1.5}
    \Gluon(261.835,-77.13)(251.686,-46.684){3.045}{5}
    \Gluon(277.396,-27.74)(261.835,-77.806){3.045}{8}
    \SetWidth{2.0}
    \Line(246.273,-75.776)(261.158,-75.776)
    \Line(246.273,-78.483)(261.158,-78.483)
    \Line(261.835,-75.776)(276.719,-75.776)
    \Line(261.835,-78.483)(276.719,-78.483)
    \SetWidth{1.0}
    \GOval(261.158,-77.806)(4.736,6.089)(0){0.882}
    \SetWidth{2.0}
    \Arc[clock](470.558,-353.781)(304.096,99.024,87.131)
    \Arc(538.992,366.232)(435.634,-105.553,-96.837)
    \Arc[clock](478.11,-104.839)(115.258,119.028,93.926)
    \Arc(453.945,82.245)(91.963,-110.205,-77.226)
    \Line(444.51,-1.353)(470.22,4.736)
    \Line(449.246,-4.736)(472.926,-2.706)
    \Line(459.394,-54.126)(483.751,-55.479)
    \Line(462.101,-58.862)(483.751,-60.892)
    \SetColor{Red}
    \Line[dash,dashsize=6.766](230.712,11.502)(230.712,-96.75)
    \Text(160,15)[lb]{\mbox{\fontsize{24}{2}\selectfont \Red{${\bf 3 \otimes 8 = 3 \oplus \bar{6} \oplus 15}$}}}    
   \end{axopicture}}}}}

\def \FIGpotential (#1,#2) {
\resizebox{#1 mm}{!}{\raisebox{#2 pt}{
\fcolorbox{white}{white}{
  \begin{axopicture}(169,108) (207,-342)
    \SetWidth{2.0}
    \SetColor{Black}
    \Line(208,-260)(368,-260)
    \SetWidth{1.5}
    \Gluon(288,-324)(289,-262){4.5}{6}
    \Text(206,-245)[l]{\Large{\Black{$p$}}}
    \Text(341,-244)[l]{\Large{\Black{$p+q$}}}
    \Text(239,-330)[l]{\Large{\Black{$P$}}}
    \SetWidth{1.0}
    \Line[arrow,arrowpos=1,arrowlength=5,arrowwidth=2,arrowinset=0.2](304,-308)(304,-284)
    \Text(307,-297)[l]{\Large{\Black{$q$}}}
    \SetWidth{2.0}
    \Line(287,-325)(312,-316)
    \Line(266,-327)(288,-327)
    \Line(266,-331)(288,-331)
    \Line(288,-329)(310,-329)
    \Line(287,-332)(312,-340)
    \SetWidth{1.0}
    \GOval(288,-330)(7,9)(0){0.882}
    \end{axopicture}}}}}

\def \FIGkinetic (#1,#2) {
\resizebox{#1 mm}{!}{\raisebox{#2 pt}{
\fcolorbox{white}{white}{
  \begin{axopicture}(451,276) (76,-135)
    \SetWidth{1.0}
    \SetColor{White}
    \GBox(77.838,-132.51)(341.005,95.444){0.882}
    \SetWidth{2.0}
    \SetColor{Black}
    \Line(265.948,-24.091)(286.331,-3.708)\Line(265.948,-3.708)(286.331,-24.091)
    \Line(198.303,-22.238)(218.686,-1.855)\Line(198.303,-1.855)(218.686,-22.238)
    \SetWidth{1.0}
    \Gluon(212.201,-13.9)(211.275,52.819){5.56}{5}
    \Gluon(209.422,-105.637)(210.348,-14.826){5.56}{6}
    \Gluon(279.846,-12.046)(251.12,69.498){5.56}{6}
    \Gluon(280.773,-13.9)(301.159,51.892){5.56}{5}
    \SetWidth{2.0}
    \EBox(119.537,-87.105)(505.021,51.892)
    \SetWidth{1.0}
    \Line[arrow,arrowpos=0.5,arrowlength=13,arrowwidth=5.2,arrowinset=0.2](350.271,51.892)(394.75,51.892)
    \Line[arrow,arrowpos=0.5,arrowlength=13,arrowwidth=5.2,arrowinset=0.2](384.557,-87.105)(354.905,-87.105)
    \SetWidth{2.0}
    \Line(392.899,-25.018)(413.282,-4.635)\Line(392.899,-4.635)(413.282,-25.018)
    \SetWidth{1.0}
    \Gluon(409.577,-86.178)(408.65,-13.9){5.56}{5}
    \Gluon(409.577,-14.826)(408.65,70.425){5.56}{6}
    \Gluon(502.241,75.058)(122.317,75.058){5.56}{28}
    \Gluon(121.39,-110.271)(502.241,-110.271){5.56}{29}
    \GOval(117.684,-15.753)(101.004,23.166)(0){0.8}
    \COval(504.094,-16.68)(101.004,23.166)(0){Black}{White}
    \Text(455,94)[l]{\mbox{\fontsize{24}{2}\selectfont \Black{$\pvec_1$}}}
    \Text(455,35)[l]{\mbox{\fontsize{24}{2}\selectfont \Black{$\pvec_2$}}}
    \Text(455,-72)[l]{\mbox{\fontsize{24}{2}\selectfont \Black{$\pvec_3$}}}
    \Text(455,-129)[l]{\mbox{\fontsize{24}{2}\selectfont \Black{$\pvec_4$}}}
    \Text(108,-10.193)[l]{\mbox{\fontsize{32}{2}\selectfont \Black{$\alpha$}}}
    \Text(492,-12.046)[l]{\mbox{\fontsize{32}{2}\selectfont \Black{$\beta$}}}
    \Text(423.476,21.313)[l]{\mbox{\fontsize{24}{2}\selectfont \Black{$\ellvec$}}}
    \Text(410,122.317)[l]{\mbox{\fontsize{24}{2}\selectfont \Black{$t+\dd t$}}}
    \Text(337,131.583)[l]{\mbox{\fontsize{24}{2}\selectfont \Black{$t$}}}
    \Text(336,110.271)[l]{\mbox{\fontsize{24}{2}\selectfont \Black{$\downarrow$}}}
    \Text(427,96.371)[l]{\mbox{\fontsize{24}{2}\selectfont \Black{$\downarrow$}}}
  \end{axopicture}
  }}}}

\title{Nuclear ${\pvec}_\perp$-broadening of an energetic parton pair} 

\author[a]{Florian Cougoulic,}
\author[a]{St\'ephane Peign\'e}

\affiliation[a]{SUBATECH UMR 6457 (IMT-Atlantique, Universit\'e de Nantes, CNRS-IN2P3) F-44307 Nantes}

\emailAdd{Florian.Cougoulic@subatech.in2p3.fr}\emailAdd{peigne@subatech.in2p3.fr}

\abstract{We revisit the transverse momentum broadening of a fast parton pair crossing a nuclear medium, putting emphasis on the pair global color state, for any number of colors $N$ and within the eikonal limit for parton propagation and the Gaussian approximation for the gluon field of the target. The pair transverse momentum probability distribution is derived in a kinetic equation approach, and is determined by an operator $\B$ describing the possible transitions between the pair color states when crossing the medium. The exponential of $\B$ encompasses the 4-point correlators of Wilson lines in the saturation formalism. We emphasize the relation of $\B$ with the anomalous dimension matrices appearing in the study of soft gluon radiation associated to hard $2 \to 2$ partonic processes. In a well-chosen, orthonormal basis of the pair color states, we rederive $\B$ for any type of parton pair, making maximal use of ${\rm SU}(N)$ invariants and using `birdtrack' color pictorial notations, providing a quite economical derivation of all previously known 4-point correlators (or equivalently, anomalous dimension matrices for $2 \to 2$ parton scattering). We discuss some general features of the pair transverse momentum distribution. The latter simplifies in the `compact pair expansion' which singles out the {global charges} (Casimirs) of the pair color states. This study should provide the necessary tools to address nuclear broadening of $n$-parton systems in phenomenology while highlighting the color structure of the process.}  


\begin{document}
\maketitle
\setcounter{footnote}{0}
\renewcommand{\thefootnote}{\arabic{footnote}} 	

\section{Introduction}
\label{sec:intro}

In proton-nucleus collisions at sufficiently high energies, hadron production arises from the production of partons which live long enough to propagate through the whole nuclear target. In this situation, not only the underlying hard partonic process, but the in-medium parton propagation itself is commonly addressed in perturbative QCD (pQCD). Within the perturbative framework different nuclear effects, responsible for the modification of production rates in proton-nucleus (\pA) when compared to proton-proton (\pp) collisions, have been addressed using various formalisms. 

Some of those effects can be encoded, within the collinear factorization formalism~\cite{Collins:1989gx}, in the {\it leading-twist} nuclear parton distribution functions (nPDFs, obtained from global fits based on DGLAP evolution), such as the depletion of the gluon PDF in the target nucleus at small gluon longitudinal momentum fraction $x \lesssim 10^{-2}$, namely, leading-twist shadowing~\cite{Nikolaev:1990ja}. Several other nuclear effects, not accounted for by leading-twist QCD factorization, have been widely discussed. In the saturation formalism~\cite{Mueller:1999wm,Mueller:2001fv,Iancu:2003xm,Gelis:2010nm,Albacete:2014fwa}, gluon shadowing (determined from non-linear QCD evolution) incorporates `higher-twist' effects induced by gluon rescattering in nuclear matter, and thus differs from the shadowing included in leading-twist nPDFs. (See Ref.~\cite{Armesto:2006ph} for a topical review on shadowing.) Other effects such as transverse momentum nuclear broadening and medium-induced energy loss have been addressed theoretically in different formalisms, for instance in a path-integral approach in Refs.~\cite{Zakharov:1996fv,Zakharov:1997uu}, in a Glauber picture for multiple soft scattering~\cite{Baier:1996kr,Baier:1996sk}, in the opacity expansion~\cite{Gyulassy:1999zd,Gyulassy:2000er,Kovner:2003zj,Arleo:2010rb,Peigne:2014uha,Peigne:2014rka,Armesto:2012qa,Armesto:2013fca}, in the higher-twist approach~\cite{Qiu:1990xxa,Qiu:1990xy,Qiu:2001hj,Luo:1993ui,Wang:2001ifa,Majumder:2007hx,Majumder:2007ne}, and in the saturation formalism (see \eg~\cite{Mueller:2012bn,Liou:2014rha,Munier:2016oih}). Those effects have been implemented in numerous phenomenological studies of various observables in \pA\ collisions (see \eg~\cite{Albacete:2014fwa,Accardi:2009qv} for reviews). 

Maybe the simplest of the above nuclear effects is nuclear $p_\perp$-broadening, usually defined in hadron (jet) \pA\ vs \pp\ production as $\Delta \ave{p_\perp^2} \equiv \ave{p_\perp^2}_{\rm pA} - \ave{p_\perp^2}_{\rm pp}$, where $\ave{p_\perp^2}$ is the average of the hadron (jet) $p_\perp^2$. At high-energy, hadron $p_\perp$-broadening is inherited from the parent parton (up to a rescaling by the hadron longitudinal momentum fraction w.r.t.~the parton), which dominantly arises from parton diffusion in transverse momentum space due to in-medium rescatterings. The main features of $p_\perp$-broadening are those of a classical high-energy particle undergoing a random walk in transverse space, namely, $\Delta \ave{p_\perp^2} = C_R {\hat{q}} L$, where $L$ is the longitudinal size of the nucleus (assumed to be much larger than the proton size), $C_R$ the squared color charge (Casimir) of the parton in color representation $R$ crossing the medium, and $\hat{q}$ the parton transverse momentum squared acquired per unit path length, defined here for a parton of unit color charge. With the latter convention $\hat{q} = \bar{Q}^2/L$, where $\bar{Q}$ is the `color-stripped' saturation scale defined by ${\bar Q}^2 \equiv Q_s^2 / N$, with $Q_s$ the gluon saturation scale. At high energy, $\hat{q}$ is related to the small-$x$ gluon distribution, $\hat{q} = \hat{q}(x) = \bar{Q}^2(x)/L \sim xG(x)$~\cite{Baier:1996sk,Mueller:1999wm}. In addition to the latter $x$-dependence, $\hat{q}$ also acquires some logarithmic $L$-dependence arising from radiative corrections~\cite{Blaizot:2013vha,Liou:2013qya}.

In spite of its apparent simplicity, $p_\perp$-broadening is an interesting observable which can bring valuable information. For some processes, the observed magnitude of $p_\perp$-broadening may help disentangling various production mechanisms. For instance, for $\jpsi$ production in \pA\ collisions, $p_\perp$-broadening may differ by a factor of 2 depending whether the $c \bar{c}$ pair is assumed to be produced in a color octet or color singlet state~\cite{Kang:2008us,Johnson:2006wi}. The actual magnitude of $\hat{q}(x)$ (equivalently, of the saturation scale $\bar{Q}(x)$) may be accessed through various phenomenological studies, for instance (see Ref.~\cite{Albacete:2014fwa} for a review) of small-$x$ DIS data~\cite{Albacete:2010sy}, saturation effects in single and double inclusive hadron production~\cite{Albacete:2010bs,Albacete:2010pg}, broadening in Drell-Yan or quarkonium production~\cite{Kang:2016ron,Kopeliovich:2010aa}, parton energy loss effects in quarkonium production~\cite{Arleo:2012rs,Arleo:2013zua}, shadowing/saturation effects in Drell-Yan production~\cite{Basso:2016ulb}. 

In most of studies on dihadron/dijet production (see \eg\ Refs.~\cite{Gelis:2001da,JalilianMarian:2004da,Blaizot:2004wv,Qiu:2004da,Kharzeev:2004bw,Baier:2005dv,Marquet:2007vb,Tuchin:2009nf,Albacete:2010pg,Stasto:2011ru,Kang:2011bp,Lappi:2012nh}), the accent is put on the small-$x$ saturation physics rather than on the color structure of the underlying process, and a sum over final color indices is performed. However, addressing the non-trivial color structure of parton pair production has proven to be quite enlightening~\cite{Nikolaev:2005qs,Nikolaev:2005dd,Nikolaev:2005zj}. In the present study, we revisit the $p_\perp$-broadening of a parton pair, putting emphasis on the pair SU($N$) irreducible representation (irrep) or {\it color state} (with a different viewpoint from Refs.~\cite{Nikolaev:2005qs,Nikolaev:2005dd,Nikolaev:2005zj}). Expressing the dijet cross section as an explicit sum over the parton pair color states may give an additional handle to probe small-$x$ saturation effects (see section~\ref{sec:discussion} for a discussion). 

For a parton pair in a given color state $\alpha$, the broadening of transverse momentum imbalance $\qvec \equiv \pvec_1+\pvec_2$\footnote{We will denote a transverse vector $\vec{p}_\perp$ as $\pvec$, and its modulus by either $p_\perp$ or $|\pvec|$.}  in general depends not only on the Casimirs of the individual partons ($C_F=(N^2-1)/(2N)$ and $C_A =N$ for quark and gluon, respectively), but also on the pair global charge $C_\alpha$. In the particular case of a {\it pointlike} pair, the broadening depends solely on $C_\alpha$, $\ave{\qvec^2}_{_\alpha} \propto C_\alpha$. Thus, dihadron (dijet) production may probe `unusual' Casimirs larger than $C_F$ or $C_A$, simply because of the presence of higher dimensional color states in the inclusive pair production cross section. As an illustration, consider a quark-gluon pair. For SU($N$) with $N\geq 3$, the latter can appear in three different color states,\footnote{We label SU($N$) irreps according to their dimensions in the case $N =3$.}
\be
\label{quark-glu}
{\bf 3 \otimes 8 = 3 \oplus \bar{6} \oplus 15} \, ,
\ee
whose dimensions $K_{\alpha}$ and Casimirs $C_{\alpha}$ are given by 
\bea
K_{\alpha} &=& \{K_3, K_6, K_{15} \} = \left\{\mbox{\fontsize{12}{2}\selectfont ${N,  \frac{N (N -2) (N+1)}{2},\frac{N (N +2) (N-1)}{2}}$}\right\} \, , \label{quark-glu-dim} \\
C_{\alpha} &=&\{C_3, C_6, C_{15} \} =  \left\{\mbox{\fontsize{12}{2}\selectfont  ${\frac{N^2-1}{2 N}, \frac{(N -1) (3 N+1)}{2N}, \frac{(N +1) (3 N-1)}{2N}}$}\right\} \, . \label{quark-glu-cas}
\eea
For $N=3$ we have $C_3 = C_F= \frac{4}{3}$, $C_6= \frac{10}{3}$, $C_{15}= \frac{16}{3}$. Since the broadening of an individual quark (gluon) scales as $C_F$ ($C_A$), a pointlike $qg$ pair in color state ${\bf \bar{6}}$ or ${\bf 15}$ suffers more broadening than individual partons. For instance, for a pointlike $qg$ pair in color state ${\bf 15}$, $\ave{q_\perp^2}_{qg}^{\bf 15} = 4 \, \ave{p_\perp^2}_{q} = \frac{16}{9} \, \ave{p_\perp^2}_{g}$. In practice, a parton pair produced in \pA\ collisions has a finite relative momentum $|\pvec_1 - \pvec_2|$ and thus cannot be truly pointlike. However, one may approach the pointlike pair limit by considering the domain of large relative momenta, where the pair effective transverse size $\sim 1/|\pvec_1 - \pvec_2|$ is smaller than the resolution power $\sim 1/\bar{Q}$ of the medium. We will call such a pair a {\it compact} pair and the kinematical domain $|\pvec_1 - \pvec_2| \gg \bar{Q}$ the {\it compact pair limit}. 

In this study we review the $p_\perp$-broadening of an energetic parton pair in a self-contained theoretical model, within the eikonal limit for parton propagation, the Gaussian approximation for the gluon field of the target~\cite{McLerran:1993ni} (where the saturation scale $\bar{Q}$ is independent of $x$), and putting emphasis on the pair color state. We work at finite number of colors $N$.\footnote{In view of phenomenological applications, working at finite $N$ is preferable to using the large-$N$ limit. In the case of a $qg$ pair for instance, in the large-$N$ limit the dimensions (as well as the Casimirs) of the representations ${\bf \bar{6}}$ and ${\bf 15}$ coincide (see \eq{quark-glu-dim}--\eq{quark-glu-cas}), which is clearly not a good approximation for $N=3$.} We first review in section~\ref{sec:single-parton} the case of a single parton crossing a nuclear medium, which will be useful in the rest of the study. We recall the derivation of the broadening distribution $f(\pvec;L)$ (see \eq{f-density}) from a simple kinetic equation, as well as its limiting behaviors and the distinction between typical and average $p_\perp$-broadening. In section~\ref{sec:asym-pair}, the main part of the study, we review the transverse momentum broadening of an `asymptotic parton pair' ($qq$, $q\bar{q}$, $qg$, or $gg$), using the same kinetic equation approach. We derive the pair transverse momentum probability distribution $f_{\alpha \rightarrow \beta}(\pvec_1,\pvec_2;L)$ (see \eq{master-eq-asym-pair-2}) which is fully determined by an operator $\B$ describing the possible transitions between the pair color states $\alpha$ when crossing the medium. The exponential of $\B$ encompasses the 4-point correlators of Wilson lines in the saturation formalism calculated previously, and $\B$ is also directly related to the anomalous dimension matrices appearing in the study of multiple soft gluon radiation associated to hard $2 \to 2$ partonic processes. In a well-chosen, orthonormal basis of the pair color states, $\B$ can be easily rederived for any type of parton pair, making maximal use of ${\rm SU}(N)$ invariants and using `birdtrack' color pictorial notations, thus providing a quite economical derivation of all previously known 4-point correlators (or equivalently, anomalous dimension matrices for $2 \to 2$ parton scattering). We discuss some general features of $f_{\alpha \rightarrow \beta}(\pvec_1,\pvec_2;L)$, and introduce the `compact pair expansion'. The study of an asymptotic pair is somewhat academic but contains all ingredients necessary to address the more realistic case of a parton pair produced in a hard process briefly reviewed in section~\ref{sec:hard-prod}, as well as to substantiate the expected features of dijet (or more generally $n$-jet) nuclear broadening, as illustrated in the final discussion of section~\ref{sec:discussion}. 

\section{Transverse momentum broadening of a single energetic parton}
\label{sec:single-parton}

Here we briefly review the derivation of $p_\perp$-broadening for a single parton, based on a simple kinetic equation, as done in Ref.~\cite{Baier:1996sk}. This introduces the notations and basic ingredients used in section~\ref{sec:asym-pair} to study the $p_\perp$-broadening of a parton pair. 

\subsection{Transverse momentum probability distribution $f(\pvec;L)$}
\label{sec:fpL}

Consider a massless parton $a$ of Casimir $C_R$ (with $C_R=C_F$ for $a=q$, and $C_R=C_A$ for $a=g$) prepared in the far past with $\pvec = \vec{0}_\perp$ and longitudinal momentum $p^z >0$, and traversing a nuclear medium of length $L$.  For simplicity we consider the medium to be homogeneous and invariant under translations in the transverse plane. We work in the {\it eikonal limit} where $p^z$ can be arbitrarily large, the other scales (in particular, transverse momenta) being upper bounded.

\begin{figure}[t]
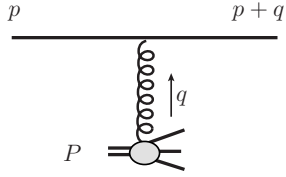

\bc
\FIGpotential(40,-20)
\ec
\caption{Parton-nucleon scattering amplitude from which the potential \eq{eq:potential} is obtained.}
\label{fig:potential}
\end{figure}

When studying $p_\perp$-broadening within the Glauber model where successive parton scatterings are assumed to be independent, the central quantity is the parton `scattering potential' $V(\qvec)$, defined as the normalized probability distribution for the parton to absorb $q_\perp$ in an individual parton-nucleon scattering~\cite{Baier:1996sk}. Treating the parton-nucleon scattering amplitude in the single gluon exchange approximation (see Fig.~\ref{fig:potential}), $V(\qvec)$ takes the form of a screened Coulomb potential,
\be
\label{eq:potential}
V(\qvec) = \frac{\mu^2}{\pi (q_\perp^2 + \mu^2)^2} \  ;  \ \ \  \int \!  \dd^2 \qvec \, V(\qvec) = 1 \, ,
\ee
where the parameter $\mu$ plays the role of the inverse screening length of the medium. In cold nuclear matter, the Coulomb interaction is effectively screened by color neutrality at distances larger than the nucleon size $\sim \lqcd^{-1}$, and thus $\mu \sim \lqcd$. One easily checks that $\mu$ is the magnitude of the {\it typical} transverse momentum transfer in parton-nucleon scattering. (The domain $q_\perp \lesssim \mu$ saturates half of the integral $\int \! \dd^2 \qvec \, V(\qvec)$.)

Within the above setup, the probability density $f(\pvec; t)$ for the parton to have transverse momentum $\pvec$ at time $t$ satisfies the kinetic equation~\cite{Baier:1996sk} 
\be
\label{kinetic}
f(\pvec; t + \dd t) -  f(\pvec; t) = \frac{\dd t}{\lambda_R} \int \! \dd^2 \ellvec \,V(\ellvec) \,f(\pvec - \ellvec; t) -\frac{\dd t}{\lambda_R} \, f(\pvec; t) \, ,
\ee
where $\lambda_R$ is the average time (or parton mean free path) between two parton-nucleon scatterings. The first term in the r.h.s. of \eq{kinetic} is the `gain term' between $t$ and $t + \dd t$, given by the probability ${\dd t}/{\lambda_R}$ to have one scattering in time $\dd t$, times the probability that the parton transverse momentum is $\pvec$ {\it after} that scattering. The second term of \eq{kinetic} is the loss term, given by the product of ${\dd t}/{\lambda_R}$ and the probability that the transverse momentum is $\pvec$ {\it before} the scattering.\footnote{In a diagrammatic formulation, the gain (loss) term is given by diagrams where between $t$ and $t + \dd t$, there is one gluon exchange in the parton-nucleon scattering amplitude and one in the conjugate amplitude (two gluon exchanges in the amplitude or in the conjugate). The gain and loss terms thus correspond to the parton-nucleon scattering between $t$ and $t + \dd t$ being inelastic or elastic, respectively, when viewed from the nucleon's side.} Note that in the single gluon exchange approximation for parton-nucleon scattering (Fig.~\ref{fig:potential}), $\lambda_R \propto C_R^{-1}$. This follows from the relation $\lambda_R = 1/(\rho \sigma_R)$, where $\rho$ is the nuclear density and $\sigma_R$ the parton-nucleon cross section, and the fact that $\sigma_R \propto C_R$ in this approximation. In the following, it will be convenient to use the `color-stripped' mean free path $\lambda_0 \equiv C_R \lambda_R = C_F \lambda_q = N \lambda_g$.

The kinetic equation \eq{kinetic} is an integral-differential equation, 
\be
\label{partialf}
\frac{\partial f (\pvec; t)}{\partial t} = \frac{C_R}{\lambda_0} \int \dd^2 \ellvec \,V(\ellvec) \, \left[f(\pvec - \ellvec; t) -f(\pvec; t) \right] \, ,
\ee
which is easily solved by going to transverse coordinate space~\cite{Baier:1996sk}. Defining
\be
\label{Fourier-single}
f(\pvec; t) = \int \! \!\frac{\dd^2 \xvec}{(2\pi)^2} \ e^{i \pvec \cdot \xvec}  \tilde{f}(\xvec; t) \ ; \ \ \ V(\ellvec) = \int \! \!\frac{\dd^2 \xvec}{(2\pi)^2} \ e^{i \ellvec \cdot \xvec} \, \tilde{V}(\xvec)  \, ,
\ee
we obtain from \eq{partialf}
\be
\label{eq:diff}
\frac{\partial \tilde{f} (\xvec; t)}{\partial t} = \frac{C_R}{\lambda_0} \tilde{f} (\xvec; t) \, [\tilde{V}(\xvec) - 1 ] \ \ \Longrightarrow \ \  \tilde{f} (\xvec; t) =  e^{- C_R \frac{t}{\lambda_0}  [1-\tilde{V}(\xvec)]} \, , 
\ee
where we used the initial condition $\tilde{f} (\xvec; 0) = 1$, corresponding to the initial distribution $f(\pvec; 0) = \delta^{(2)}(\pvec)$ in transverse momentum space. The probability density for the parton to have transverse momentum $\pvec$ after crossing a medium of length $L$ thus reads
\bea
\label{f-density}
f(\pvec;L) &=& \int \! \! \frac{\dd^2 \xvec}{(2 \pi)^2} \ e^{i \pvec \cdot \xvec} \, e^{- C_R \hat \Gamma({\xvec})} \, , \\
\label{Gamma-hat}
\hat \Gamma({\xvec}) &=& \frac{L}{\lambda_0} [1-\tilde{V}(\xvec)] \, ,
\eea
with the Fourier transform $\tilde{V}(\xvec)$ of the scattering potential \eq{eq:potential} given by
\be
\label{Vtilde}
\tilde{V}(\xvec) = \int \! \dd^2 \qvec  \ e^{-i \qvec \cdot \xvec} \, V(\qvec) = \mu x_\perp \, {\rm K}_1(\mu x_\perp) \, .
\ee

The expression \eq{f-density} is well-known and was derived in different formalisms (in addition to the kinetic equation approach of Ref.~\cite{Baier:1996sk}). Let us recall that production processes can be related to forward amplitudes \cite{Nikolaev:1990ja,Zakharov:1996fv}. In the present case, the transverse momentum {\it probability distribution} of a parton $a$ is related to the {\it scattering amplitude} (or $S$-matrix element) of an $a \bar a$ color dipole of transverse size $\xvec$ (namely, the factor $e^{- C_R \hat \Gamma({\xvec})}$ in \eq{f-density})~\cite{Mueller:1999wm}. Note that within the `dipole picture', the `single gluon exchange approximation' (for {\it parton}-nucleon scattering, see Fig.~\ref{fig:potential}) is also called the `two-gluon exchange approximation' (for {\it dipole}-nucleon scattering), as \eg\ in Refs.~\cite{Nikolaev:2005qs,Nikolaev:2005dd,Nikolaev:2005zj}. In the saturation formalism, the $a \bar a$ dipole scattering amplitude $e^{- C_R \hat \Gamma({\xvec})}$ results from evaluating the correlator of two Wilson lines (of parton $a$) in the McLerran-Venugopalan (MV) model~\cite{McLerran:1993ni}, where color sources in the nucleus have Gaussian correlations, and can be found for instance in Refs.~\cite{McLerran:1998nk,Gelis:2001da}. It depends on the `saturation scale' $Q_a$ of parton $a$ in the nucleus, related to the parameters $C_R$, $L$, $\mu$ and $\lambda_0$ as
\be
\label{Qsat}
Q_a^2 \equiv C_R \, {\bar Q}^2  \ ; \ \ \ {\bar Q}^2 = 2 \mu^2 \frac{L}{\lambda_0} \, ,
\ee
where ${\bar Q}$ is the `color-stripped' saturation scale.

\subsection{Limiting behaviors of $f(\pvec;L)$ and {typical} {\it vs} {average} $p_\perp$-broadening} 
\label{sec:limits-of-f}

The expression of the transverse momentum distribution \eq{f-density} will be recurrent in our study. (Up to the color factor $C_R$, similar expressions will describe the $p_\perp$-broadening of a compact parton pair, see section~\ref{sec:asym-pair}.) We briefly recall those features of $f(\pvec;L)$ which will be useful in the next sections.  

The expression \eq{f-density} depends on the three parameters $p_\perp$, $C_R L/\lambda_0$ and $\mu$, or equivalently (see \eq{Qsat}) on $p_\perp$, $Q_a$ and $\mu$. In the following we will always consider $p_\perp$ and $Q_a$ (or ${\bar Q}$) to be much larger than $\mu$. In the limit $p_\perp, {\bar Q} \gg \mu$, we readily see that the integration region $x_\perp \gsim 1/\mu \gg 1/p_\perp$ can be neglected in \eq{f-density}. Thus, \eq{f-density} arises dominantly from the region $x_\perp \ll 1/\mu$, where to logarithmic accuracy we have (use \eq{Gamma-hat}, \eq{Vtilde} and \eq{Qsat})
\be
\label{gammahat-appr}
\hat{\Gamma}({\xvec}) \mathop{\simeq}_{x_\perp \ll 1/\mu} \frac{\bar{Q}^2}{4} \, \xvec^2 \log\left(\frac{1}{\mu |\xvec|}\right) \, .
\ee
We obtain\footnote{For \eq{f-approx} to be mathematically well-defined, it should be remembered that there is an implicit cutoff $x_\perp \leq x_{\perp \,{\rm max}} \ll 1/\mu$ in the $\xvec$-integral.}
\be
\label{f-approx}
f(\pvec;L) \simeq \int \! \! \frac{\dd^2 \xvec}{(2 \pi)^2} \ e^{i \pvec \cdot \xvec} \, e^{- C_R \frac{\bar{Q}^2}{4} \, \xvec^2 \log(\frac{1}{\mu |\xvec|}) } \, .
\ee

The limiting behaviors of \eq{f-approx} when $p_\perp \lsim \bar{Q}$ and $p_\perp \gg \bar{Q}$ were obtained in~\cite{Baier:1996sk}. For completeness, a quick derivation of those limits is presented in Appendix~\ref{app:single-parton}. In the limit $p_\perp \lsim \bar{Q}$, $f(\pvec;L)$ is well approximated by a Gaussian distribution $f^{{\rm G}}(\pvec; L)$ of width $p_{\perp {\rm w}} \sim \morder{\bar{Q}}$,
\be
\label{f-gaussian}
f(\pvec;L) \mathop{\simeq}_{p_\perp \lsim \bar{Q}} \frac{e^{-\pvec^2/ p_{\perp {\rm w}}^2}}{\pi p_{\perp {\rm w}}^2} \equiv f^{{\rm G}}(\pvec; L)  \ ; \ \ \  p_{\perp {\rm w}}^2 \equiv C_R \, \bar{Q}^2 \log\left(\frac{\bar{Q}}{\mu}\right)  \, ,
\ee
whereas the behavior at $p_\perp \gg \bar{Q}$ is algebraic,  
\be
\label{f-tail}
f(\pvec;L) \mathop{\simeq}_{p_\perp \gg \bar{Q}} \frac{C_R \bar{Q}^2}{2 \pi |\pvec|^4} = \frac{L}{\lambda_R} \frac{\mu^2}{\pi |\pvec|^4} \simeq \frac{L}{\lambda_R} V(\pvec) \, .
\ee
Quite intuitively, at large $p_\perp$ the distribution $f(\pvec;L)$ arises dominantly from a {\it single},  {\it hard} Coulomb exchange, yielding the factor $V(\pvec) \simeq \mu^2/(\pi p_\perp^4)$ in \eq{f-tail}, the additional factor being the number of ways to choose that hard exchange among $L/\lambda_R$ scatterings. 

We emphasize that when $p_\perp^2 \sim p_{\perp {\rm w}}^2$, \eq{f-gaussian} is still larger than \eq{f-tail} by a factor $\log{(\bar{Q}/\mu)}$. Thus, within the logarithmic accuracy defined by
\be
\label{log-acc-single}
\log\left(\frac{\bar{Q}}{\mu}\right) \gg 1 \, ,
\ee
the Gaussian behavior \eq{f-gaussian} holds slightly {\it beyond} the value $p_\perp^2 \sim p_{\perp {\rm w}}^2$ corresponding to the width of the Gaussian. 

From the limiting behaviors \eq{f-gaussian} and \eq{f-tail} one can infer that:
\bi
\item[(i)] The integral $\int\dd^2 \pvec \, f(\pvec;L)$ (equal to unity) is dominated by the region $p_\perp \sim p_{\perp {\rm w}}$. In particular, the {\it typical} (or median) value $\bar{p}_\perp$ of $p_\perp$, defined by the implicit equation
\be
\label{median-def}
\int\dd^2 \pvec \, f(\pvec;L) \,  \Theta(\bar{p}_\perp - p_\perp) = \frac{1}{2} \, ,
\ee
can be found by replacing in \eq{median-def} the exact distribution $f(\pvec;L)$ by $f^{{\rm G}}(\pvec; L)$. This yields~\cite{Baier:1996sk} (see also Appendix A of \cite{Peigne:2008wu} for an alternative derivation of $\bar{p}_\perp$)
\be
\label{typical-p}
\bar{p}_\perp^2 \simeq (\log{2}) \, p_{\perp {\rm w}}^2 = (\log{2}) \, C_R  \bar{Q}^2 \log\left(\frac{\bar{Q}}{\mu}\right)  \, .
\ee 
\item[(ii)] The {\it average} $p_\perp^2$, defined by $\ave{p_\perp^2} \equiv \int\dd^2 \pvec \, \, \pvec^2 f(\pvec;L)$, is logarithmically divergent, due to the large-$p_\perp$ Coulomb behavior $f(\pvec;L) \sim 1/p_\perp^4$. In practice, an upper cut-off on $p_\perp$ is provided by experimental constraints. As long as $p_{\perp\, {\rm max}} \ll p^z$, so that the eikonal approximation remains valid, we expect the parametric dependence
\be
\label{average-p}
\ave{p_\perp^2} \simeq \int_{\bar{Q}^2}^{p_{\perp\, {\rm max}}^2} \dd^2 \pvec \, \, \pvec^2 \, \frac{C_R \bar{Q}^2}{2 \pi |\pvec|^4}  = C_R  \bar{Q}^2 \log\left(\frac{p_{\perp\, {\rm max}}}{\bar{Q}}\right) \, .
\ee
\ei

For a single parton, `$p_\perp$-broadening' (defined as being either typical or average) is trivially proportional to the parton Casimir $C_R$. We stress that this color dependence can be read directly from the $p_\perp$-broadening {distribution} \eq{f-density}. 
In the case of a parton pair, we will similarly read the color dependence of nuclear broadening on transverse momentum distributions. 

\section{Transverse momentum broadening of an asymptotic parton pair}
\label{sec:asym-pair}

In this section we study the $p_\perp$-broadening of an `asymptotic' pair of partons $a$ and $b$, with $a,b=g,q$ or $\bar{q}$. We assume that the pair enters the nuclear medium with some initial transverse momentum distribution and in a given color state $\alpha$. We derive the probability distribution $f_{\alpha \rightarrow \beta}(\pvec_1,\pvec_2;L)$ to find, after crossing the length $L$, the partons with transverse momenta $\pvec_1$ and $\pvec_2$ and the parton pair in the color state $\beta$, given the initial condition $f_{\alpha \rightarrow \beta}(\pvec_1,\pvec_2;0)$. 

In section~\ref{sec:color-structure} we discuss the color structure of the process and define color states, using {\it birdtrack} color pictorial notations~\cite{Cvitanovic:2008zz,Dokshitzer:1995fv,Keppeler:2017kwt}. We then derive in section~\ref{sec:fp1p2-kin} the distribution $f_{\alpha \rightarrow \beta}(\pvec_1,\pvec_2;L)$ using a kinetic equation approach, as in the case of the single parton transverse momentum distribution $f(\pvec;L)$ reviewed in section~\ref{sec:single-parton}. In sections~\ref{sec:relationBtoQ} and \ref{sec:relationBtoCorr} we emphasize the relation of the $\B$ `evolution operator' \eq{B-def} to soft anomalous dimension matrices and to correlators of Wilson lines, respectively. In section~\ref{sec:gen-features} we present some general properties of  $f_{\alpha \rightarrow \beta}(\pvec_1,\pvec_2;L)$.

\subsection{Color structure and pictorial notations} 
\label{sec:color-structure}

\begin{figure}[t]
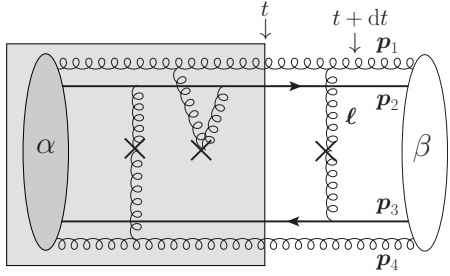
 
\bc
\FIGkinetic(60,-120)
\ec
\caption{Generic contribution to the distribution $f_{\alpha \rightarrow \beta}$ describing the evolution, in transverse momentum and color space, of an $ab$ parton pair (drawn here for $a=g$, $b=q$). The partons $a$, $b$ (in the amplitude) and $\bar{b}$, $\bar{a}$ (in the conjugate amplitude) are labelled by $i=1\ldots 4$, respectively. The scattering between the nucleus and the color singlet $ab\bar{b}\bar{a}$ system is treated in the two-gluon exchange approximation (nucleons are represented by crosses).}
\label{fig:kinetic-4}
\end{figure}

A generic contribution to transverse momentum broadening of the $ab$ parton pair is shown in Fig.~\ref{fig:kinetic-4}, where the upper half of the diagram represents the scattering amplitude of the pair in the medium (referred to as the {\it $s$-channel} scattering amplitude), and the lower half its conjugate. For illustration we take the $ab$ pair to be a quark-gluon pair. In the following, we will often take a quark-gluon pair as a generic pair when discussing general features applying to any type of pair, and otherwise explicitly mention the specific case under consideration ($qq$, $q\bar{q}$, $qg$, or $gg$). We will use the `two-gluon exchange approximation' where the scattering off the nucleus consists in two-gluon exchanges between the overall color singlet $ab\bar{b}\bar{a}$ system and any number of nucleons in the nucleus, see Fig.~\ref{fig:kinetic-4}.

At any time $t$, the color state of the $ab$ pair can be decomposed into a sum of SU($N$) irreps labelled by $\alpha$, namely, $a \otimes b = \sum_{\alpha} R_{\alpha}$. (Recall that an irrep $\alpha$ can be defined by its associated Young {\it tableau}.) We denote by ${\cal P}_\alpha$ the {\it $s$-channel projector} on the irrep $\alpha$, satisfying ${\cal P}_\alpha \, {\cal P}_\beta = \delta_{\alpha \beta} \, {\cal P}_\alpha$ and the ($s$-channel) completeness relation $\unit =  \sum_{\alpha} \, {\cal P}_\alpha$. The latter is written in pictorial form as
\be
\label{qgidentity-schannel}
\QGidentitystraight(16,-16) =  \sum_{\alpha} \proj(\alpha,22,-15) \ ; \ \ \   {\cal P}_\alpha \equiv  \proj(\alpha,22,-15) \, .
\ee
Inserting in Fig.~\ref{fig:kinetic-4}, at a given time $t$ between rescatterings, the completeness relation \eq{qgidentity-schannel} in both the amplitude and its conjugate, allows one to express the whole diagram as
\be
\label{double-sum}
\matchinga (20,-90) \ = \ \sum_{\beta, \gamma} \  \ \matchingb (\beta,\gamma,20,-90) \ \, ,
\ee
where the shaded areas include the whole dynamics occurring before or after time $t$. 

From color conservation, the $s$-channel irreps $\beta$ (in the amplitude) and $\gamma$ (in the conjugate amplitude) have to `match' (in other words, to be associated with the same Young {\it diagram}). For a $qg$ pair for instance, there are three different irreps, ${\bf 3 \otimes 8 = 3 \oplus \bar{6} \oplus 15}$,\footnote{We recall that SU($N$) representations are named according to their dimensions in the case $N =3$.}  $\beta$ and $\gamma$ can take three values ($\beta, \gamma = {\bf 3,\bar{6},15}$), and color conservation imposes $\beta = \gamma$ in the double sum of \eq{double-sum}. The same holds for $qq$ pairs (${\bf 3 \otimes 3 = \bar{3} \oplus 6}$) and $q \bar{q}$ pairs (${\bf 3 \otimes \bar{3} = 1 \oplus 8}$). In general, the `matching' of irreps is {\it not} equivalent to setting $\beta = \gamma$ in \eq{double-sum}. In the case of a $gg$ pair for instance,  ${\bf 8 \otimes 8 = 1 \oplus 8_a \oplus  8_s \oplus 10 \oplus \overline{10} \oplus 27 \oplus 0}$,\footnote{The irrep ${\bf 0}$ appears only for $N>3$, see Appendix~\ref{app:Qgggg}.} color conservation reduces the double sum of \eq{double-sum} to all allowed $s$-channel {\it transitions} $\beta \to \gamma$, including ${\bf 8_a \to 8_s}$ and ${\bf 8_s \to 8_a}$. However, the two latter transitions turn out to be irrelevant in our study, where the product of {\it symmetry signums} $\sigma_\beta \sigma_\gamma$ is always positive.\footnote{The signum $\sigma_{\alpha}$ is defined as the eigenvalue of the irrep $\alpha$ under permutation of the two gluons in the tensor product ${\bf 8 \otimes 8}$.}  This follows from the fact that $\sigma_\beta \sigma_\gamma = +1$ in the initial state (the incoming parton pair is in the same color state in both the amplitude and its conjugate) and from the conservation of the product $\sigma_\beta \sigma_\gamma$ during the evolution, easily inferred from pictorial equations of the type 
\be
\label{eq:sym-match}
\symmetrymatchinga (\beta,\gamma,\beta',\gamma',35,-90) = \symmetrymatchingb (\beta,\gamma,\beta',\gamma',35,-90) = \sigma_\beta \sigma_{\beta'} \sigma_\gamma \sigma_{\gamma'} \, \symmetrymatchinga (\beta,\gamma,\beta',\gamma',35,-90) \, ,
\ee
implying $\sigma_\beta \sigma_{\beta'} \sigma_\gamma \sigma_{\gamma'} = 1$ and thus $\sigma_\beta \sigma_\gamma = \sigma_{\beta'} \sigma_{\gamma'}$.\footnote{\label{foot:signum}In \eq{eq:sym-match} the first equality simply follows from the relation $T_1 T_3 = T_2 T_4$, where $T_i$ denotes the color generator of gluon $i$. Indeed, $2 T_1 T_3 = (T_1+T_3)^2 - T_1^2 -T_3^2 = (T_2+T_4)^2 - T_2^2 -T_4^2 = 2 T_2 T_4 $, since $T_i^2 = C_A$ for $i=1\ldots4$ and $(T_1+T_3)^2 = (T_2+T_4)^2$ from color conservation. The second equality in \eq{eq:sym-match} is obtained by permuting the two gluons of the pair {\it twice} in the amplitude (respectively, right before and right after the scattering, yielding the factor $\sigma_\beta \sigma_{\beta'}$), and {\it twice} in the conjugate amplitude (yielding the factor $\sigma_\gamma \sigma_{\gamma'}$). Note that the conservation of the product $\sigma_\beta \sigma_\gamma$ holds within the two-gluon exchange approximation.} Thus, `mixed transitions' ${\bf 8_a \to 8_s}$ and ${\bf 8_s \to 8_a}$ never appear in our study. In summary, for any type of parton pair we can set $\beta = \gamma$ in \eq{double-sum}. As a consequence, {\it we can label the pair color state at any time by the $s$-channel irreps $\alpha$}.

Similarly to the case of a single parton $a$ where the distribution $f(\pvec;L)$ is related to the $a \bar{a}$ dipole scattering amplitude (see section~\ref{sec:single-parton}), for an $ab$ parton pair the distribution $f_{\alpha \rightarrow \beta}(\pvec_1,\pvec_2;L)$ is related to the $ab\bar{b}\bar{a}$ {\it quadrupole} scattering amplitude~\cite{Nikolaev:2005qs,Nikolaev:2005dd,Nikolaev:2005zj}, which is a matrix in color space. From color conservation, the color singlet $ab\bar{b}\bar{a}$ states are in one-to-one correspondence with all possible color structures for $ab \to ab$ transitions, \ie \ $s$-channel $\beta \to \gamma$ transitions.\footnote{In general, for a given process $ab \to cde \ldots$, the number of possible color transitions equals the number of independent color singlet $\bar{a}\bar{b}cde \ldots$ states in the `color space' of the process~\cite{Keppeler:2012ih}.} Since for our purpose, the latter can be reduced to those with $\beta = \gamma$, we see that we can also label the relevant color singlet $ab\bar{b}\bar{a}$ states by the label $\alpha$ of $s$-channel irreps.\footnote{\label{footnote:mixed}In the $gg$ case, allowing for an initial pair in a `mixed' color state (\eg, ${\bf 8_a}$ in the amplitude and ${\bf 8_s}$ in the conjugate), would invalidate the statement that $\beta = \gamma$. In such a situation, one would need to label the $gg$ color state by $\beta$ {\it and} $\gamma$ at any time of the evolution, as for instance in Ref.~\cite{Nikolaev:2005zj}. For $N>3$, this enlarges the relevant color space from 7 to 9 dimensions, but does not entail any formal complication. Our study could be simply extended to such a situation by trading the `projector basis' \eq{ortho-basis} for the larger `vector basis' of all independent color singlet $gggg$ states~\cite{Keppeler:2012ih}.}

We thus choose the following orthonormal basis of color singlet $ab\bar{b}\bar{a}$ states,
\be
\label{ortho-basis}
\ket{\alpha} = \frac{1}{\sqrt{K_\alpha}} \ \OBKet(\alpha,10,-60) \, ,
\ee
where the arrow and symbol $\alpha$ in the blob indicate that the pair is projected onto the $s$-channel irrep $\alpha$, and $K_\alpha$ is the dimension of the irrep $\alpha$. Using color pictorial rules~\cite{Cvitanovic:2008zz,Dokshitzer:1995fv,Keppeler:2017kwt}, we directly check that the basis \eq{ortho-basis} is orthonormal, 
\be
\langle \alpha \vert \beta \rangle = \frac{1}{\sqrt{K_\alpha K_\beta}}  \ \OBBra(\alpha,10,-60) \! \!  \OBKet(\beta,10,-60) = \frac{1}{\sqrt{K_\alpha K_\beta}}\,  \delta_{\alpha \beta} \  \OBBra(\alpha,10,-60) \! \! \! \QGidentity(11,-62) = \delta_{\alpha \beta} \, ,
\ee
where the sum over color indices of the intermediate four-parton state is implicit, and we used
\be
\label{dim}
\OBBra(\alpha,10,-60) \! \! \! \! \QGidentity(11,-62) = \tr{{\cal P}_\alpha} = K_{\alpha}  \, .
\ee
For further use, let us mention the completeness relation\footnote{In a situation involving mixed color states, the sum in \eq{eq:completeness} should be extended to all independent color singlet four-parton states, see footnote~\ref{footnote:mixed}.}
\be
\label{eq:completeness}
\left(\! \GQQGidentity(9.5,-59) \right)_{\rm \! \! \! singlet} = \sum_{\gamma} \ket{\gamma} \bra{\gamma} =  \sum_{\gamma}  \frac{1}{K_\gamma} \ \OBKet(\gamma,10,-60) \ \OBBra(\gamma,10,-60) 
\, ,
\ee 
where $\ket{\gamma} \bra{\gamma}$ is the projector on the color singlet four-parton state $\ket{\gamma}$.

It will be convenient to view the distribution $f_{\alpha \rightarrow \beta}$ as the matrix element of a `color transition operator' $\hat{f}$ acting in the space of (orthonormal) color singlet states $\ket{\alpha}$,
\be
\label{def-hatf}
\hat{f} = \sum_{\alpha,\beta} \,  f_{\alpha \rightarrow \beta} \times \ket{\alpha} \bra{\beta} \ ; \ \ \ f_{\alpha \rightarrow \beta} = \bra{\alpha} \hat{f} \ket{\beta} \, .
\ee
Note that the initial and final color states $\alpha$ and $\beta$ are read from left to right in $\bra{\alpha} \hat{f} \ket{\beta}$, \ie, an initial (final) state appears as a bra (ket) in the transition matrix element. This convention is consistent with drawing Feynman diagrams with increasing time from left to right, which facilitates the reading of equations involving pictorial color factors.  

As a last color pictorial rule needed in this study, we choose the convention where a gluon coupling to a parton $i$ of color generator $T_i$ provides a factor $T_i$ when the coupling occurs from below, and a factor $-T_i$ when it occurs from above~\cite{Dokshitzer:1995fv}. Within this convention, conjugate partons have opposite color generators, and color conservation is made pictorially explicit. For instance, in an overall color singlet parton system, the individual color generators satisfy 
\vspace{-4mm}
\begin{equation}
\label{pict-color-cons}
\qgTa(12,-45)  + \qgTb(12,-45) + \qgTc(12,-45) + \qgTd(12,-45) = \ \sum_{i=1}^4 \, T_i \, = 0  \, .
\end{equation}
\vspace{2mm}

\subsection{Parton pair distribution $f_{\alpha \rightarrow \beta}(\pvec_1,\pvec_2;L)$ from a kinetic equation}  
\label{sec:fp1p2-kin}

We are now ready to derive the distribution $f_{\alpha \rightarrow \beta}$ using a kinetic equation approach, using the same setup as in section~\ref{sec:single-parton}. 

At a given time $t$ of the evolution of the $qg$ pair (chosen as a generic pair), the relevant quantity is the distribution $f_{\alpha \rightarrow \beta}(\{\pvec_i \}; t)$ to find the gluon and quark with momenta $\pvec_1$ and $\pvec_2$ in the $qg$ scattering amplitude, and the antiquark and gluon with momenta $\pvec_3$ and $\pvec_4$ in the conjugate amplitude, with $\sum_i \pvec_i = \zerovec$ (see Fig.~\ref{fig:kinetic-4}). The final distribution $f_{\alpha \rightarrow \beta}(\pvec_1,\pvec_2;L)$ to `tag' the gluon and quark with momenta $\pvec_1$ and $\pvec_2$ after crossing the medium will be obtained from $f_{\alpha \rightarrow \beta}(\{\pvec_i\};L)$ by setting $\pvec_4 = - \pvec_1$ and $\pvec_3 = - \pvec_2$. 

Let us consider the generic contribution to $f_{\alpha \rightarrow \beta}(\{\pvec_i\};t + \dd t)$ represented in Fig.~\ref{fig:kinetic-4}, where the last scattering (and only that one) happens between $t$ and $t + \dd t$. Let us moreover focus on those contributions where in the last scattering, the two exchanged gluons attach to {\it distinct} parton lines $i$ and $j$, $1 \leq i < j \leq 4$. This type of contribution modifies the $\pvec_i$'s between $t$ and $t + \dd t$, and thus contributes to the `gain term' of the evolution equation for $f_{\alpha \rightarrow \beta} (\{\pvec_i\};t)$ (in analogy with \eq{partialf}). Inserting, at time $t$ of the evolution, the completeness relation \eq{eq:completeness} in the generic diagram of Fig.~\ref{fig:kinetic-4}, we easily find for the gain term,
\bea
\frac{\partial f_{\alpha \rightarrow \beta}^{\rm gain} (\{\pvec_i\};t)}{\partial t} = \frac{1}{\lambda_0} \int \! \dd^2 \ellvec \,V(\ellvec) \left[\, f_{\alpha \rightarrow \gamma}(\pvec_1 - \ellvec, \pvec_2, \pvec_3 + \ellvec, \pvec_4; t) \, {\cal C}_{13}^{\gamma \beta} \right.  \hskip 2cm && \nn \\
\left. + f_{\alpha \rightarrow \gamma}(\pvec_1 - \ellvec, \pvec_2 + \ellvec, \pvec_3, \pvec_4; t) \, {\cal C}_{12}^{\gamma \beta} + \ldots \, \right] \, , \hskip 0.7cm && 
\label{partialf-pair}
\eea
where the sum over $\gamma$ is implicit, and the color factor ${\cal C}_{ij}^{\gamma \beta}$ corresponding to the pair $ij$ of partons reads\footnote{The contribution to the gain term where the two exchanged gluons (in the last scattering of Fig.~\ref{fig:kinetic-4}) attach to partons 1 and 2 has a color factor $T_1 T_2$ and an additional minus sign since the exchanged gluons both appear in the amplitude. (The relative weights of such contributions have been discussed previously, see for instance Ref.~\cite{Baier:1998kq}.) This is consistent with the color factor \eq{color-fact-ij}, where pictorially the two exchanged gluons couple to partons $i$ and $j$ from below and from above, respectively.} 
\be
\label{color-fact-ij}
{\cal C}_{ij}^{\gamma \beta} = - \bra{\gamma} \, T_i T_j \, \ket{\beta} = \bra{\gamma} \! \IJexchange(9.2,-55) \ket{\beta} =   \frac{1}{\sqrt{K_\beta K_\gamma}} \ \OBBra(\gamma,10,-60) \! \! \! \IJexchange(9.2,-55) \OBKet(\beta,10,-60) \, .
\ee

The calculation of the loss term, arising from contributions where the two gluons exchanged in the last scattering couple to the {\it same} parton line, presents no difficulty.\footnote{In addition to a minus sign, those contributions have a symmetry factor $1/2$~\cite{Baier:1998kq}.} Using color conservation, see \eq{pict-color-cons}, the color factors associated with such contributions can always be traded for color factors associated with the gain terms. For instance,
\be
\qgTdTd(12,-50)  = \qgTaTb(12,-50) + \qgTaTc(12,-50) + \qgTaTd(12,-50) \ \, .
\vspace{2mm}
\ee

Adding gain and loss terms we obtain
\be
\label{partialf-pair-tot}
\frac{\partial f_{\alpha \rightarrow \beta} (\{\pvec_i\};t)}{\partial t} = \frac{1}{\lambda_0} \int \! \dd^2 \ellvec \,V(\ellvec) \sum_{i < j} \left[f_{\alpha \rightarrow \gamma}(\{\pvec_i - \ellvec, \pvec_j + \ellvec \}; t)- f_{\alpha \rightarrow \gamma}(\{\pvec_i \}; t) \right] {\cal C}_{ij}^{\gamma \beta} \, ,
\ee
where in $f_{\alpha \rightarrow \gamma}(\{\pvec_i - \ellvec, \pvec_j + \ellvec \}; t)$ only the two momenta $\pvec_i$ and $\pvec_j$ are shifted by $\pm \ellvec$, the two other momenta being unchanged. Obviously, in absence of a medium (\ie, replacing formally $V(\ellvec) \to \delta^{(2)}(\ellvec)$) the loss term exactly compensates the gain term, similarly to the case of a single parton reviewed in section~\ref{sec:single-parton}, see \eq{partialf}. 

We now go to transverse coordinate space using
\be
\label{Fourier-pair}
f_{\alpha \rightarrow \beta} (\{\pvec_i\};t) = \left[ \, \prod_{i=1}^{4} \int \! \!\frac{\dd^2 \xvec_i}{(2\pi)^2} \ e^{i \pvec_i \cdot \xvec_i} \right]  \tilde{f}_{\alpha \rightarrow \beta}(\{\xvec_i\};t) \, ,
\ee
and obtain from \eq{partialf-pair-tot},
\be
\label{eq:diff-pair}
\frac{\partial \tilde{f}_{\alpha \rightarrow \beta} (\{\xvec_i \};t)}{\partial t} = \frac{1}{\lambda_0} \tilde{f}_{\alpha \rightarrow \gamma} (\{\xvec_i\};t) \sum_{i < j}  \, [\tilde{V}(\xvec_{ij}) - 1 ] \, {\cal C}_{ij}^{\gamma \beta}  \, ,
\ee
where $\xvec_{ij} \equiv \xvec_i - \xvec_j$ and $\tilde{V}$ is defined in \eq{Vtilde}. The solution of \eq{eq:diff-pair} is
\be
\label{solution-kineq}
\tilde{f}_{\alpha \rightarrow \beta} (\{\xvec_i \};L) = \tilde{f}_{\alpha \rightarrow \gamma} (\{\xvec_i \};0) \, \left[ e^{\B} \right]_{\gamma \beta} \, , 
\ee
with $\B$ a color matrix which can be expressed in terms of the function $\hat{\Gamma}(\xvec)$ defined in \eq{Gamma-hat}. Denoting $\hat{\Gamma}_{ij} \equiv \hat{\Gamma}(\xvec_{ij})$, the matrix elements of $\B$  in the orthonormal basis \eq{ortho-basis} read
\be
\label{color-matrix}
\B_{\gamma \beta} \equiv \bra{\gamma} \B \ket{\beta} = - \sum_{i < j} \, \hat{\Gamma}_{ij} \, {\cal C}_{ij}^{\gamma \beta} =  - \sum_{i < j} \hat{\Gamma}_{ij} \, \bra{\gamma} \! \IJexchange(9.2,-55) \ket{\beta}  \, ,
\ee
yielding the basis-independent expression of $\B$, 
\be
\label{B-def}
\B =  - \sum_{i<j} \hat{\Gamma}_{ij} \! \IJexchange(9.2,-55) \ = \sum_{i < j} \hat{\Gamma}_{ij} \, T_i T_j \, .
\ee

For a nuclear medium invariant under translations in the transverse plane, $\tilde{f}_{\alpha \rightarrow \beta}(\{\xvec_i \},t)$ depends only on three {\it relative} transverse positions, and one of the $\xvec_i$-integrals in \eq{Fourier-pair} produces a trivial factor $\delta^{(2)}(\sum_i \pvec_i)$. Redefining $f_{\alpha \rightarrow \beta} (\{\pvec_i\};t)$ by removing this factor (consistently with the definition of the distribution in the single parton case, see \eq{Fourier-single}), choosing the three independent relative coordinates to be $\{\xvec_{ij}\} \equiv (\xvec_{12}, \xvec_{23}, \xvec_{34})$,  using \eq{solution-kineq} and finally setting $\pvec_4 = - \pvec_1$ and $\pvec_3 = - \pvec_2$, we obtain from \eq{Fourier-pair}
\be
\label{master-eq-asym-pair}
f_{\alpha \rightarrow \beta}(\pvec_1,\pvec_2;L) = \int \! \frac{\dd^2 \xvec_{12} \, \dd^2 \xvec_{23} \, \dd^2 \xvec_{34}}{(2\pi)^6} \ e^{i \pvec_1 \cdot (\xvec_{12}+\xvec_{34}) + i (\pvec_1+\pvec_2) \cdot \xvec_{23}} \, \tilde{f}_{\alpha \rightarrow \gamma} (\{\xvec_{ij} \};0) \, \bra{\gamma} e^\B \ket{\beta}  \, .
\ee

Although we will not really need an explicit form of the initial condition $\tilde{f}_{\alpha \rightarrow \gamma} (\{\xvec_{ij} \};0)$ in the following, it is natural to consider a dependence of the form
\be
\label{initial-cond-cdn}
\tilde{f}_{\alpha \rightarrow \beta} (\{\xvec_{ij}\};0) = \delta_{\alpha \beta} \, \psi(\xvec_{12}) \, \psi^*(\xvec_{43}) \, , 
\ee
with $\psi(\xvec)$ the incoming pair wavefunction. Setting $L=0$ (implying $\B = 0$) in \eq{master-eq-asym-pair} and using \eq{initial-cond-cdn} we find the corresponding initial condition in momentum space,
\be
\label{initial-cond-mom}
f_{\alpha \rightarrow \beta}(\pvec_1,\pvec_2;L=0) =  \delta_{\alpha \beta}  \, |{\psi}(\pvec_1)|^2  \, \delta^{(2)}(\pvec_1 + \pvec_2)  \, , 
\ee
where ${\psi}(\pvec)$ is the Fourier transform of ${\psi}(\xvec)$. The factor $|{\psi}(\pvec_1)|^2$ in \eq{initial-cond-mom} is nothing but the initial distribution in the {\it relative} momentum $\rvec \equiv \halft (\pvec_1-\pvec_2)$. For a normalized wavefunction, $\int \!\dd^2 \pvec \, |{\psi}(\pvec)|^2= 1$, we have $\int \dd^2 \pvec_1 \dd^2 \pvec_2 \, f_{\alpha \rightarrow \beta}(\pvec_1,\pvec_2;L=0) = \delta_{\alpha \beta}$.

With the choice \eq{initial-cond-cdn} the result \eq{master-eq-asym-pair} can be rewritten as
\be
\label{master-eq-asym-pair-2}
f_{\alpha \rightarrow \beta}(\pvec_1,\pvec_2;L) = \int \! \frac{\dd^2 \xvec_{12} \, \dd^2 \xvec_{23} \, \dd^2 \xvec_{34}}{(2\pi)^6} \ e^{i \pvec_1 \cdot (\xvec_{12}+\xvec_{34}) + i (\pvec_1+\pvec_2) \cdot \xvec_{23}} \, \psi(\xvec_{12}) \, \psi^*(\xvec_{43}) \, \bra{\alpha} e^\B \ket{\beta}  \, .
\ee
Given an incoming wavefunction $\psi(\xvec)$, the parton pair transverse momentum distribution is fully determined by the knowledge of the color matrix $\B$.

\subsection{Relation of $\B$ to soft anomalous dimension matrix} 
\label{sec:relationBtoQ}

It is noteworthy that the expression \eq{B-def} is formally identical to the square of the soft gluon emission current $j^{\mu}(k)$ associated to $2\to 2$ hard parton scattering. Indeed, the latter reads~\cite{Dokshitzer:2005ek}
\be
\label{j2}
j^{\mu}(k) = \sum_{i=1}^{4} \omega \frac{p_i^\mu}{(k p_i)} \, T_i \ \ \Longrightarrow \ \   \half \, j^2(k) = \sum_{i<j} \,   w_{ij}(k) \, T_i T_j  \, ,
\ee
with $p_i$ the four-momentum of parton $i$ (assumed to be massless, $p_i^2=0$), $\omega$ and $k$ the energy and four-momentum of the soft radiated gluon, and $w_{ij}(k)$ the {\it dipole antenna} distribution
\be
w_{ij} \equiv \omega^2  \frac{(p_ip_j)}{(kp_i)(kp_j)} \, .
\ee 
Thus, the matrix $\B$ defined by \eq{B-def} is obtained from $\half \, j^2(k) $ given in \eq{j2} by formally replacing dipole antenna distributions by dipole `scattering cross sections', 
\be
w_{ij} \  \longleftrightarrow \ \hat{\Gamma}_{ij} \, .
\ee

Following Ref.~\cite{Dokshitzer:2005ek}, we can express \eq{B-def} in terms of squared color charges by using 
\be
\label{eq:TiTj}
T_i T_j = \half \left[ (T_i +  T_j)^2 - C_i - C_j \right]  \, ,
\ee
where $C_i= T_i^2$ is the Casimir operator of parton $i$, and by introducing the squared color charges exchanged in the $s$, $t$ and $u$-channels of the scattering process, namely,
\bea
\label{Tstu}
&& T_s^2 = (T_1+T_2)^2, \quad T_t^2 = (T_2+T_3)^2, \quad T_u^2= (T_1+T_3)^2  \, , \\ 
&& \hskip 2.5cm  T_s^2 + T_t^2 + T_u^2 =  \sum_{i=1}^{4} C_i  \, . 
\label{sumT} 
\eea
We obtain from \eq{B-def}, 
\be
\label{BofQ}
\B =  - \frac{1}{2} \left[ C_1 W^{(1)}_{34} + C_2 W^{(2)}_{34} + C_3 W^{(3)}_{12} + C_4 W^{(4)}_{12} \right] -  \frac{1}{2} N (X_t + X_u) \, {\cal Q} \, ,
\ee
where $W^{(i)}_{jk}$, $X_t$, $X_u$ are combinations of $\hat{\Gamma}_{ij}$'s which are the formal analogs of the combinations of $w_{ij}$'s defined in~\cite{Dokshitzer:2005ek}, 
\bea
W^{(i)}_{jk} &=& \hat{\Gamma}_{ij} + \hat{\Gamma}_{ik} - \hat{\Gamma}_{jk}  \, , \label{Wijk}  \\
X_t &=& \hat{\Gamma}_{12} + \hat{\Gamma}_{34} - \hat{\Gamma}_{14} - \hat{\Gamma}_{23}  \, , \label{Xt}  \\
X_u &=& \hat{\Gamma}_{12} + \hat{\Gamma}_{34} - \hat{\Gamma}_{13} - \hat{\Gamma}_{24}  \, , 
\label{Xu} 
\eea
and ${\cal Q}$ is the analog of the `soft anomalous dimension matrix' associated with $2 \to 2$ parton scattering~\cite{Dokshitzer:2005ek},
\be
\label{Q-def} 
{\cal Q} \equiv  \frac{1}{2N} \left[ \, T_t^2 +T_u^2  +  \frac{X_t - X_u}{X_t + X_u} \, (T_t^2 - T_u^2) \, \right]  \, .
\ee

Soft anomalous dimension matrices are a key ingredient in the study of QCD obser\-va\-bles sensitive to multiple soft gluon radiation~\cite{Dokshitzer:2005ek,Botts:1989kf,Sotiropoulos:1993rd,Contopanagos:1996nh,Kidonakis:1998nf, Oderda:1999kr,Bonciani:2003nt,Appleby:2003hp,Banfi:2004yd,Kyrieleis:2005dt,Sjodahl:2008fz,Forshaw:2008cq}. In general, such observables receive {\it double logarithmic} contributions, which exponentiate and can be expressed as the product of Sudakov form factors associated with each parton participating to the hard process. The form factor of a given parton is related to (the exponential of) the single gluon radiation probability off that parton, which is proportional to the parton Casimir. Form factors are thus diagonal in color space. Resumming {\it single logarithmic} contributions, in particular those arising from soft and non-collinear gluon radiation, is more complicated, because it involves in general mixing in the color space of the partonic process. For hard processes involving at least four partons (\eg, $2 \to n$ processes with $n \geq2$), such contributions cannot be expressed solely in terms of parton color charges, and resum to the exponential of a {\it non-diagonal} color matrix named {\it (soft) anomalous dimension matrix}~\cite{Dokshitzer:2005ek,Botts:1989kf,Sotiropoulos:1993rd,Contopanagos:1996nh,Kidonakis:1998nf, Oderda:1999kr,Bonciani:2003nt,Appleby:2003hp,Banfi:2004yd,Kyrieleis:2005dt,Sjodahl:2008fz,Forshaw:2008cq}.

In the present study of $p_\perp$-broadening of a parton pair, the first term of \eq{BofQ} provides (after exponentiation) the analog of the product of Sudakov form factors in the problem of soft gluon radiation. As for the second term of \eq{BofQ}, proportional to the anomalous dimension matrix ${\cal Q}$ associated with $2 \to 2$ scattering, it provides the analog of the `fifth form factor'~\cite{Dokshitzer:2005ek}. The analogy between the matrix encoding $p_\perp$-broadening of a parton pair and the anomalous dimension matrix of $2 \to 2$ processes was noted in Ref.~\cite{Nikolaev:2005zj}. 
 
Finding the explicit form of the matrix $\B$ in a given basis amounts to find the matrix ${\cal Q}$ in this basis. Anomalous dimension matrices have been derived in various bases~\cite{Dokshitzer:2005ek,Botts:1989kf,Sotiropoulos:1993rd,Contopanagos:1996nh,Kidonakis:1998nf, Oderda:1999kr,Bonciani:2003nt,Appleby:2003hp,Banfi:2004yd,Kyrieleis:2005dt,Sjodahl:2008fz,Forshaw:2008cq}, and their explicit form is known for all $2 \to 2$ parton processes (see for instance Ref.~\cite{Kidonakis:1998nf}). However, we found it useful to rederive the matrix ${\cal Q}$ associated with $ab \to ab$ scattering in a simple way, choosing the orthonormal basis \eq{ortho-basis}. The latter is more convenient for our purpose for two reasons. First, a basis emphasizing the {\it $s$-channel} irreps of the parton pair is obviously more appropriate to discuss $p_\perp$-broadening of the pair. In particular, the compact pair limit should single out the Casimirs of those irreps, as mentioned in the Introduction. Second, ${\cal Q}$ is always symmetric in an {\it orthonormal} basis~\cite{Seymour:2005ze,Seymour:2008xr}, simplifying substantially the derivation.\footnote{Strictly speaking, as proven in Ref.~\cite{Seymour:2008xr}, soft anomalous dimension matrices are symmetric in the subset of orthonormal bases built from color tensors involving delta functions, group generators and structure constants (which is the case of the basis \eq{ortho-basis} when the projectors ${\cal P}_\alpha$ are built from those elements).} 

A simple derivation of ${\cal Q}$ for $ab \to ab$ scattering is presented in Appendices~\ref{app:Qaqaq} and~\ref{app:Qgggg}. In Appendix~\ref{app:Qaqaq} we consider the cases where $b$ is a quark, $aq\to aq$, with $a = g, q$ or $\bar{q}$. Those three cases can be treated on the same footing. The associated ${\cal Q}$-matrix can be obtained with minimal effort with the help of the Fierz identity, in terms of ${\rm SU}(N)$ invariants (Casimirs $C_\alpha$ and dimensions $K_\alpha$) of $s$-channel irreps, resulting in the expression \eq{eq:Qaqaq}. In Appendix \ref{app:Qgggg} we consider $gg \to gg$, and present a simple derivation of ${\cal Q}$ which only uses the explicit form of the $s$-channel projector ${\cal P}_{\bf a}$ on the adjoint representation of a gluon pair. We find \eq{Tt2minusTu2-ortho-basis}--\eq{eq:Qgggg}, recovering the results of Refs.~\cite{Dokshitzer:2005ek,Nikolaev:2005zj} in that case.\footnote{Existing calculations, such as that in Ref.~\cite{Dokshitzer:2005ek}, are based on the following observation. In an $s$-channel basis such as \eq{ortho-basis}, the operator $T_s^2$ is the diagonal matrix of Casimirs of $s$-channel irreps. From \eq{sumT}, only one additional operator, \eg\  $T_t^2$, needs to be evaluated to determine ${\cal Q}$ (given by \eq{Q-def}) in this basis. Since $T_t^2$ is known (and diagonal) in a {\it $t$-channel} basis, one needs to determine the transition matrix $K_{st}$ between $t$-channel and $s$-channel bases. Calculating $K_{st}$ can be quite involved in general, since it requires knowing the explicit form of all $s$-channel and $t$-channel projectors. In Ref.~\cite{Dokshitzer:2005ek}, this procedure is carried out explicitly for $gg \to gg$. Using pictorial tools simplifies the color algebra, but the calculation remains quite elaborate. (See also the tensorial calculation of Ref.~\cite{Nikolaev:2005zj}.) The derivation of ${\cal Q}$ in Appendix \ref{app:Qgggg} bypasses the calculation of the matrix $K_{st}$.}   

For further use, we give below the matrix elements of $\B$ in the basis \eq{ortho-basis}, for the cases $aq\to aq$ and $gg \to gg$ respectively, which directly follow from \eq{BofQ} and the expressions of ${\cal Q}$ recalled in Appendices~\ref{app:Qaqaq} and~\ref{app:Qgggg}.  

\subsubsection*{\centerline{$\B$-matrix for $aq\to aq$}}

For $aq\to aq$, $C_2 = C_3 = C_F$ and $C_1 = C_4 = C_a$. Using \eq{BofQ} and \eq{eq:Qaqaq} we obtain
\bea
\bra{\alpha} \B \ket{\beta}= \frac{\sqrt{K_\alpha K_\beta}}{2 K_a} (X_t - X_u) - \frac{\delta_{\alpha \beta}}{2} \left[ C_a X_1 + C_F X_2 + N (X_t - X_u) - C_\alpha X_u \right] \, , \hskip 1cm && \label{B-aq} \\
X_1 = \hat{\Gamma}_{12} + \hat{\Gamma}_{34} - \hat{\Gamma}_{13} - \hat{\Gamma}_{24} + 2 \hat{\Gamma}_{14} \, , \label{X1} \hskip 4cm && \\
X_2 = \hat{\Gamma}_{12} + \hat{\Gamma}_{34} - \hat{\Gamma}_{13} - \hat{\Gamma}_{24} + 2 \hat{\Gamma}_{23} \, . \hskip 4cm && \label{X2} 
\eea
We stress that \eq{B-aq} encompasses the three cases $gq \to gq$, $qq \to qq$, $\bar{q}q \to \bar{q}q$, and is expressed only in terms of ${\rm SU}(N)$ invariants. 

\subsubsection*{\centerline{$\B$-matrix for $gg \to gg$}}

For $gg \to gg$, $C_i = N$ ($i=1\ldots 4$), and from \eq{BofQ} and \eq{eq:Qgggg} we get 
\bea 
\bra{\alpha} \B  \ket{\beta} = -\frac{1}{4} (X_t - X_u) \, \bra{\alpha} (T_t^2 - T_u^2) \ket{\beta}  - \frac{\delta_{\alpha \beta}}{2} \left[ N Z - C_\alpha  \frac{X_t + X_u}{2} \right] \, , \hskip 5mm && \label{B-gg} \\ 
Z = 2 (\hat{\Gamma}_{12} + \hat{\Gamma}_{34}) \, , \hskip 5cm &&
\eea
where $\bra{\alpha} (T_t^2 - T_u^2)  \ket{\beta}$ is given by \eq{Tt2minusTu2-mat-elements} or \eq{Tt2minusTu2-ortho-basis}.

\subsection{Relation of $\B$ to correlators of Wilson lines}
\label{sec:relationBtoCorr}

\subsubsection{$e^{\B}$ and 4-point correlators}
\label{sec:expBto4pt}

The matrix $e^\B$, specifying the distribution $f_{\alpha \rightarrow \beta}$ (see \eq{master-eq-asym-pair-2}) as shown from a kinetic equation approach, is directly related to 4-point correlators of Wilson lines.  

For a system of $n$ partons crossing a nucleus in the eikonal approximation, the $n$-point correlator of the associated product of Wilson lines reads (see for instance Ref.~\cite{Fukushima:2007dy})
\be
\label{FH-correlator}
\ave{ U_1(\xvec_1)_{\alpha_1 \beta_1} U_2(\xvec_2)_{\alpha_2 \beta_2} \cdots U_n(\xvec_n)_{\alpha_n \beta_n} } \, , 
\ee
where the Wilson line $U_i$ associated to parton $i$ of transverse position $\xvec_i$ and color generator $T_i$ is given by (in light-cone gauge $A^+=0$)
\be 
U_i(\xvec) = \mathcal{P} \exp \left[ ig \int dx^+ A_c^-(x^+, {\xvec}) \, T_i^c \right] \, .
\ee
The average in \eq{FH-correlator} is over the gluon field ($A^\mu = A_c^\mu \, T^c$) configurations in the nucleus, and $\alpha_i, \beta_i$ refer to the initial and final color indices of parton $i$ (our convention for positioning color indices is explained after \eq{def-hatf}). The $n$-point correlator \eq{FH-correlator} is the $S$-matrix element for the eikonal scattering of an $n$-parton system off the target nucleus.

The correlator \eq{FH-correlator} is derived in Ref.~\cite{Fukushima:2007dy} in the so-called Gaussian approximation (which should be valid for sufficiently large nuclei), as used in the MV model~\cite{McLerran:1993ni}. Namely, color sources in the nucleus are assumed to have Gaussian correlations if evaluated at the same space position, but are otherwise uncorrelated. Within this approximation, and in the particular case of a {\it color-singlet} parton system relevant to our study ($\sum_{i} T_i =0$, see \eq{pict-color-cons}), the $n$-point correlator \eq{FH-correlator} reduces to~\cite{Fukushima:2007dy}
\be
\label{FH-correlator-singlet}
\ave{ U_1(\xvec_1)_{\alpha_1 \beta_1} U_2(\xvec_2)_{\alpha_2 \beta_2} \cdots U_n(\xvec_n)_{\alpha_n \beta_n} } = \left[ \exp{ \left( \sum_{i<j} \hat{\Gamma}_{ij} \, T_i T_j  \right) } \right]_{\alpha_1\cdots\alpha_n ; \, \beta_1\cdots\beta_n} \, .
\ee

Recalling \eq{B-def}, we see that the matrix $e^\B$ obtained in section~\ref{sec:fp1p2-kin} within a kinetic approach coincides with the 4-point correlator given by \eq{FH-correlator-singlet} for $n=4$.\footnote{The saturation scales $\left. Q_s^2 \right|_{^{\rm FH}}$ and ${\bar Q}^2$ used respectively in \cite{Fukushima:2007dy} and in our study are related by $\left. Q_s^2 \right|_{^{\rm FH}} = {\pi(N^2-1){\bar Q}^2}/{(2N g^4)}$. Using this relation in \cite{Fukushima:2007dy} gives \eq{FH-correlator-singlet} with $\hat{\Gamma}$ defined by \eq{Gamma-hat}, where $L/\lambda_0$ is related to the gluon saturation scale $Q_g^2$ by \eq{Qsat}, $L/\lambda_0 = Q_g^2 /(2 N \mu^2)$, with $\mu \sim \lqcd$ the inverse screening length in cold nuclear matter. Our convention for the gluon saturation scale is the same as in Ref.~\cite{Blaizot:2004wv}.} This follows from the equivalence between the Gaussian approximation and the two-gluon exchange approximation used for scattering of the color-singlet parton system off a nucleon. In fact, $n$-point correlators are often defined as specific linear combinations of matrix elements of \eq{FH-correlator-singlet}. In this sense $e^\B$ encompasses all 4-point correlators calculated previously. 

The matrix $\B$ in the basis \eq{ortho-basis} being real and symmetric, it can be diagonalized and thus in principle easily exponentiated. This can be done analytically for ${\bar q}q \to {\bar q}q$ and $q q \to q q$, where the eigenvalues of the $\B$-matrix are roots of a quadratic polynomial. The expression \eq{B-aq} being given in terms of $s$-channel SU$(N)$ invariants, the ${\bar q}q {\bar q}q$ and $q q {\bar q} {\bar q}$ correlators can be derived on the same footing and in a rather economical way. This is illustrated in Appendix~\ref{app:4-point}, where we recover the ${\bar q}q {\bar q}q$ correlator calculated in Refs.~\cite{Blaizot:2004wv,Fukushima:2007dy}, see \eq{corr-BGV}, and give the result for the $q q {\bar q} {\bar q}$ correlator, see \eq{eB-qq}--\eq{lambda-qq}. The latter might be relevant when addressing higher-twist phenomena in \pA\ collisions and diquark states in the projectile proton. 

The $gq\bar{q}g$ correlator can also be expressed analytically (the eigenvalues of $\B$ associated to $gq \to gq$ being roots of a cubic polynomial), yielding however a cumbersome expression. It has been noted that some simplification arises in the large-$N$ limit, where one of the three eigenvectors of ${\B}$ decouples from the problem~\cite{Nikolaev:2005dd}. As for the $gggg$ correlator~\cite{Dokshitzer:2005ek,Nikolaev:2005zj},\footnote{See also Ref.~\cite{Kovner:2001vi} where a specific linear combination of matrix elements of $e^\B$ is evaluated in that case.} the eigenvalues of $\B$ (related to the eigenvalues of the ${\cal Q}$-matrix \eq{eq:Qgggg} by a simple shift, see \eq{BofQ}) are roots of a polynomial of sixth degree, but quite surprisingly they also turn out to be expressible analytically, with three of them being roots of a cubic polynomial~\cite{Dokshitzer:2005ek}. Thus, an analytical expression of the $gggg$ correlator $e^\B$ is in principle also available. 

A purpose of our study is to address the {\it compact pair limit} (see the Introduction). The expression of $e^{\B}$ in this limit simplifies, and can be found without needing the exact analytical expression of $e^\B$, see section~\ref{sec:color-transitions}. In the limit of an infinitely compact, \ie\ {\it pointlike} parton pair, $e^{\B}$ further simplifies and reduces to known 3-point and 2-point correlators, as recalled in the next section. 

\subsubsection{3-point and 2-point correlators as limits of $e^{\B}$}
\label{sec:32point-corr}

We quote here the limit of $e^\B$ when the pair is {pointlike}, which is easily found from the general expression \eq{B-def} of the operator $\B$. 

Let us first make the pair of partons 1 and 2 pointlike, $\xvec_1 = \xvec_2 = \uvec$, giving in \eq{B-def}, 
\begin{equation}
\label{Gamma-uj}
\hat{\Gamma}_{12} = 0 \, ; \  \hat{\Gamma}_{1j} = \hat{\Gamma}_{2j} \equiv \hat{\Gamma}_{uj} \ {\rm for} \ j= 3, 4  \, .
\end{equation}
We then introduce the operator $\mathcal{R}_\beta$ projecting from the four-particle state onto the three-particle state containing one particle in representation $\beta$ in the amplitude, and two particles in the conjugate amplitude, 
\begin{equation}
\mathcal{R}_\beta \equiv \OperatorReduction-22-12 (8,-75, \beta) \ \, .
\end{equation}
Multiplying \eq{B-def} on the right by this operator and using \eq{Gamma-uj} we obtain
\bea
\B(\uvec, \uvec ; \xvec_3, \xvec_4) \, \mathcal{R}_\beta \ &=& 
- \left[  \left( \Bfig (8,-80,148,20,6) + \Bfig (8,-80,84,20,4) \right) \hat{\Gamma}_{u3} 
+ \left( \Bfig (8,-80,148,-44,8) + \Bfig (8,-80,84,-44,6) \right)\hat{\Gamma}_{u4} 
+ \Bfig (8,-80,20,-44,4) \ \hat{\Gamma}_{34} \right] \OperatorReduction-22-12 (8,-80,\beta) \, \ . \nn \\ &&
\eea
The expressions in parenthesis can be simplified using color conservation, 
\begin{equation}
  \left( \Bfig (8,-80,148,20,6) + \Bfig (8,-80,84,20,4) \right) \OperatorReduction-22-12 (8,-80,\beta) = \OperatorReduction-22-12 (8,-80,\beta) \Bc1_2 (8,-48,84,-12,5) \, \, ,
\end{equation}
leading to 
\be
\label{Eq:BP_PB}
\B(\uvec, \uvec ; \xvec_3, \xvec_4) \, \mathcal{R}_\beta =  \mathcal{R}_\beta \, \B_{(1,2)}(\uvec; \xvec_3, \xvec_4) \, , 
\ee
\be
\B_{(1,2)}(\uvec; \xvec_3, \xvec_4) = - \left[ \Bc1_2 (8,-43,84,-12,5) \, \hat{\Gamma}_{u3} + \Bc1_2 (8,-43,84,-76,7) \, \hat{\Gamma}_{u4} + \Bc1_2 (8,-43,-12,-76,4) \, \hat{\Gamma}_{34} \right] \, ,
\label{eq:B12}  
\ee
where $\B_{(1,2)}$ denotes the operator $\B$ (see \eq{B-def}) in the case of one parton in the amplitude and two partons in the conjugate amplitude. 

An immediate consequence of \eq{Eq:BP_PB} is
\begin{equation}
\label{eq:expB2expB1}
e^{\B(\uvec, \uvec ; \xvec_3, \xvec_4)} \, \mathcal{R}_\beta = \mathcal{R}_\beta \, e^{\B_{(1,2)}(\uvec; \xvec_3, \xvec_4)} \, .
\end{equation}
This shows that making the pair pointlike in the amplitude, $\xvec_1 = \xvec_2 = \uvec$, and reducing the number of particles from 2 to 1 in the amplitude with the help of the operator $\mathcal{R}_\beta$, turns the 4-point correlator $e^{\B}$ into the 3-point correlator $e^{\B_{(1,2)}}$. 
 
Using \eq{eq:TiTj}, we easily show that $e^{\B_{(1,2)}}$ reads\footnote{Strictly speaking, $e^{\B_{(1,2)}}$ is proportional to the 3-particle identity operator. The latter can be associated with $\mathcal{R}_\beta$ in \eq{eq:expB2expB1}, and $e^{\B_{(1,2)}}$ thus understood as a $c$-number. This is sometimes named `color triviality'~\cite{Dokshitzer:2005ek}. The possibility of mixing between non-equivalent color states requires at least four partons.} 
\bea
e^{\B_{(1,2)}(\uvec; \xvec_3, \xvec_4)} = \exp{\mbox{\fontsize{12}{2}\selectfont $ \left[ -\left( \frac{C_\beta + C_3-C_4}{2} \, \hat{\Gamma}_{u3} + \frac{C_\beta + C_4 - C_3}{2} \, \hat{\Gamma}_{u4} + \frac{C_3+C_4-C_\beta}{2} \, \hat{\Gamma}_{34} \right) \right] $}} \, .  \hskip 5mm \label{3pt-b}
\eea
Inserting now the complete sum over four-particle states, $\sum_{\gamma} \ket{\gamma} \bra{\gamma}$, in the l.h.s.~of \eq{eq:expB2expB1}, we see that only the term $\gamma = \beta$ contributes, and multiplying \eq{eq:expB2expB1} by $\bra{\alpha}$ on the left we then obtain 
\be
\label{eq:4to3-corr}
\bra{\alpha} e^{\B(\uvec, \uvec ; \xvec_3, \xvec_4)} \ket{\beta} = \delta_{\alpha \beta} \, 
e^{\B_{(1,2)}(\uvec; \xvec_3, \xvec_4)}  \, ,
\ee
which relates the matrix elements of $e^{\B}$ in the pointlike limit to the 3-point correlator of partons in representations $\{\beta,R_3,R_4\}$ and at transverse positions $\{ \uvec, \xvec_3, \xvec_4 \}$.

The expression \eq{3pt-b} encompasses all 3-point correlators addressed previously. Setting $C_{\beta} = C_3 = C_R$ and $C_4 = N$ in \eq{3pt-b} we get~\cite{Marquet:2010cf}
\be
e^{\B_{(R,\bar{R} g)}(\uvec; \xvec_3, \xvec_4)} =  \exp{\left\{-\frac{1}{2} \left[ N \, \hat{\Gamma}_{34} +N \, \hat{\Gamma}_{u4} + (2 C_R-N) \, \hat{\Gamma}_{u3} \right] \right\}} \, ,
\label{aabarg-corr}
\ee
which reproduces the 3-point $(q, \bar{q} g)$ correlator~\cite{Kovner:2001vi,Marquet:2010cf,Blaizot:2004wv} and $(g,gg)$ correlator~\cite{Kovner:2001vi,Marquet:2010cf} for $C_R = C_F$ and $C_R = N$, respectively. Setting $C_\beta = N$ and $C_3 = C_4 = C_F$ in \eq{3pt-b} provides the 3-point $(g, \bar{q} q)$ correlator, viewed as the limit of the 4-point ${\bar q}q {\bar q}q$ correlator where the ${\bar q}q$ pair in the amplitude is replaced by a gluon. The $(g,\bar{q} q)$ and $(q, \bar{q} g)$ correlators are obviously related by the permutation $\uvec \leftrightarrow \xvec_4$. 

Finally, making the pair of partons 3 and 4 also pointlike, \ie, setting $\xvec_3 = \xvec_4 = \vvec$ in \eq{3pt-b}--\eq{eq:4to3-corr}, yields the relation between $e^{\B}$ and the two-point correlator $e^{-  C_\alpha  \hat{\Gamma}_{uv}}$ of a parton of charge $C_\alpha$, 
\be
\label{2pt-ab}
\bra{\alpha} e^{\B(\uvec, \uvec ; \vvec, \vvec)} \ket{\beta} = \delta_{\alpha \beta} \, e^{-  C_\alpha  \hat{\Gamma}_{uv}} \, .
\ee

Let us note that the pictorial procedure presented in this section can be generalized to infer correlators of $n$ partons knowing those of $n+1$ partons, for any $n$. 

\subsection{General features of $f_{\alpha \rightarrow \beta}(\pvec_1,\pvec_2;L)$}
\label{sec:gen-features}

In this section we verify that the transverse momentum distribution $f_{\alpha \rightarrow \beta}(\pvec_1,\pvec_2;L)$ given by \eq{master-eq-asym-pair-2} satisfies simple properties expected from common sense. These properties follow from the general expression of the operator $\B$ (discussed in detail in sections~\ref{sec:relationBtoQ} and~\ref{sec:relationBtoCorr}), independently of the type of the parton pair. We then introduce the `compact pair expansion' in section~\ref{sec:color-transitions}.

\subsubsection{Transverse momentum broadening of a pointlike pair} 
\label{sec:pointlike-pair} 

Here we consider the drastic limit of an incoming pointlike parton pair. 

When $|\xvec_{12}| = |\xvec_{34}| = 0$, we insert the limit \eq{2pt-ab} of $e^\B$ in \eq{master-eq-asym-pair-2} to obtain
\be
f_{\alpha \rightarrow \beta}(\pvec_1,\pvec_2;L) = \delta_{\alpha \beta} \, |{\psi}(\pvec_1)|^2 \int \! \frac{\dd^2 \xvec_{23}}{(2\pi)^2} \ e^{i \qvec \cdot \xvec_{23}} \, e^{- C_{\alpha} \hat{\Gamma}_{23}}  \, .
\label{dist-approx}
\ee

The pointlike pair distribution \eq{dist-approx} after crossing the medium has the same $\pvec_1$-dependence as the initial distribution, see \eq{initial-cond-mom}, and the $\qvec$-dependence 
\be
f_{\alpha \rightarrow \beta}(\qvec;L) \equiv \int \!  \dd^2 \pvec_1 \, f_{\alpha \rightarrow \beta}(\pvec_1,\pvec_2;L)  =  \delta_{\alpha \beta} \int \! \frac{\dd^2 \xvec}{(2\pi)^2} \ e^{i \qvec \cdot \xvec} \, e^{- C_{\alpha} \hat{\Gamma}(\xvec)}  \, .
\label{fqL-pointlike}
\ee

Obviously,  a pointlike pair cannot be probed by the medium, and thus remains in the same color state (factor $\delta_{\alpha \beta}$) and suffers the same broadening as a single parton of color charge $C_\alpha$. In the particular case of a {\it color singlet} ($C_{\alpha} = 0$) {\it pointlike} pair, there is no broadening, and the final $\qvec$-distribution equals the initial one $\propto \delta^{(2)}(\qvec)$.

\subsubsection{Tagging a single parton}
\label{sec:1-out-of-2}

Here we verify that when only parton 1 is tagged, \ie, when summing over the color of parton $2$ and integrating the distribution $f_{\alpha \rightarrow \beta}(\pvec_1,\pvec_2;L)$ over $\pvec_2$, one recovers the distribution for the single parton 1 defined in \eq{f-density}, convoluted with the initial momentum distribution of parton 1 within the pair. 

Let us consider parton $1$ to be in color representation $R$. Being inclusive in the color of parton $2$ amounts to replace the $2$-particle final state $\ket{\beta}$ by the (normalized) one-particle state $\ket{R}$ defined by
\be
\label{2to1-replacement}
\ket{R} \equiv   \frac{1}{\sqrt{K_R}}  \, \KetR (10,-80) \, .
\ee
In order to emphasize this reduction of the number of particles, we introduce the operator $\unit^{(2 \rightarrow 1)}$ which amounts to trace over the color of parton $2$, namely
\begin{equation}
\label{Eq:Operator_Reduction_2_1}
\unit^{(2 \rightarrow 1)} = \Operator_Reduction_2_1 (10, -80, R)  \, .
\end{equation}

Recalling \eq{def-hatf} and integrating \eq{master-eq-asym-pair-2} over $\pvec_2$ at fixed $\pvec_1$ then yields
\begin{equation}
\label{dist-tag1}
\int_{\pvec_2} \bra{\alpha} \hat{f}(\pvec_1,\pvec_2;L) \unit^{(2 \rightarrow 1)} \ket{R} =  \int \! \frac{\dd^2 \xvec_{12}\,\dd^2 \xvec_{34}}{(2\pi)^4} \ e^{i \pvec_1 \cdot (\xvec_{12} + \xvec_{34})} \, \psi(\xvec_{12}) \, \psi^*(\xvec_{43}) \, \bra{\alpha}  e^{\B} \unit^{(2 \rightarrow 1)} \ket{R} \, ,
\end{equation}
where $\B$ must be evaluated at $\xvec_{23} = \zerovec$. When $\xvec_{2} = \xvec_{3}$ we have (recalling \eq{B-def}) 
\begin{equation}
\label{Eq:B_2_2}
\B = - \hat{\Gamma}_{12} \left( \,\Btwo (10,-80,148,84,4)  + \Btwo (10,-80,148,20,6) \right) - \hat{\Gamma}_{24} \left( \Btwo (10,-80,84,-44,6) + \Btwo (10,-80,20,-44,4) \right) - \hat{\Gamma}_{14} \, \Btwo (10,-80,148,-44,10) \, ,
\end{equation}
with $\hat{\Gamma}_{14} = \hat{\Gamma}(\xvec_{12}+\xvec_{34})$. Applying the operator $\unit^{(2 \rightarrow 1)}$ to the right of the latter equation, we find that the first two terms vanish, yielding
\begin{equation}
\label{comBunit}
 \B \, \unit^{(2 \rightarrow 1)} = - \hat{\Gamma}_{14} \, \Btwo (10,-80,148,-44,10) \Operator_Reduction_2_1 (10,-80,\ ) =    - \hat{\Gamma}_{14} \, \Operator_Reduction_2_1 (10,-80, \ )  \Btwobis(10,-80,148,-44,10)  \, .
\end{equation}
In the r.h.s.~we recognize the expression of $\B$ for the broadening of a single parton (see section~\ref{sec:single-parton} and \eq{f-density}). Denoting by $\B_1$ and $\B_{2}$ the color matrices for the broadening of a single parton and a parton pair, respectively, we rewrite \eq{comBunit} as 
\begin{equation}
  \label{Eq:Reduction_2_1}
  \B_{2} \, \unit^{(2\rightarrow 1)} =  \unit^{(2\rightarrow 1)} \, \B_{1} \, ,
\end{equation}
with the immediate consequence\footnote{The fact that the operator $\unit^{(2\rightarrow 1)}$ can be moved to the left of in-medium rescatterings as in \eq{comBunit2} actually follows from the eikonal approximation, where the transverse positions $\xvec_i$ are frozen. In particular, $\xvec_{23} = \zerovec$ at any time along the path as a result of integrating over $\pvec_2$.}
\begin{equation}
\label{comBunit2}
  e^{\B_{2}} \, \unit^{(2\rightarrow 1)} = \unit^{(2\rightarrow 1)} \, e^{\B_{1}} \, .
\end{equation}

Inserting \eq{comBunit2} in \eq{dist-tag1} we arrive at\footnote{Both sides have been divided by the Clebsch-Gordan coefficient $\bra{\alpha} \unit^{2 \rightarrow 1} \ket{R} = \sqrt{{K_\alpha}/{K_R}}$, in order to obtain a distribution normalized to unity after integration over $\pvec_1$, as can be checked from \eq{gp1-0} or \eq{gp1}.}
\be
\label{gp1-0}
\int_{\pvec_2} \! \frac{\bra{\alpha} \, \hat{f}(\pvec_1,\pvec_2;L) \unit^{2 \rightarrow 1} \ket{R}}{\bra{\alpha} \unit^{2 \rightarrow 1} \ket{R}}
= \int \! \frac{\dd^2 \xvec_{12}\,\dd^2 \xvec_{34}}{(2\pi)^4} \ e^{i \pvec_1 \cdot (\xvec_{12} + \xvec_{34})} \, \psi(\xvec_{12}) \, \psi^*(\xvec_{43}) \, e^{- C_R \hat{\Gamma}_{14}} \, .
\ee
Finally, writing $e^{- C_R \hat{\Gamma}_{14}}$ as the Fourier transform of the one-particle distribution \eq{f-density}, 
\be
\label{f-density-cdn}
e^{- C_R \hat{\Gamma}(\xvec_{12}+\xvec_{34})} =  \int \! \dd^2 \pvec \ e^{-i \pvec \cdot (\xvec_{12}+\xvec_{34})} f(\pvec;L) \, ,
\ee
the expression \eq{gp1-0} can be put in the form
\be
\label{gp1}
\int_{\pvec_2} \! \frac{\bra{\alpha} \, \hat{f}(\pvec_1,\pvec_2;L) \unit^{2 \rightarrow 1} \ket{R}}{\bra{\alpha} \unit^{2 \rightarrow 1} \ket{R}}
 = \int \! \dd^2 \pvec \, \, |{\psi}(\pvec_1- \pvec)|^2  f(\pvec;L)  \, .
\ee

Thus, the inclusive distribution in $\pvec_1$ is the convolution between the initial distribution $|{\psi}(\pvec)|^2$ (of parton 1 within the pair) and the one-particle broadening distribution $f(\pvec;L)$ (for a parton of charge $C_R$). This demonstrates that when only one parton of the pair is tagged, this parton suffers the same broadening as if it were travelling in isolation. Quite remarkably, this result holds independently of the precise form of the initial pair wavefunction. For instance, it holds even in the compact pair limit where the parton travels with a nearby companion.

\subsubsection{Tagging $n$ out of $m$ partons}
\label{sec:n-out-of-m}

It is instructive to generalize \eq{gp1} to the case of $n$ tagged partons picked out from a system of $m>n$ partons. Similarly to the case studied in the previous section, we show that the matrix $\B_m$ associated to the broadening of an $m$-parton system reduces to $\B_n$. 

Let us consider an initial $m$-parton state in color representation $\alpha$, denoted as $\ket{\alpha^{m}}$. In coordinate space, the probability distribution $\tilde{f}_{\alpha^m \rightarrow \beta^m}$ to evolve to a final state $\ket{\beta^{m}}$ can be easily inferred from the case $m=2$ (see \eq{solution-kineq} and \eq{initial-cond-cdn}), 
\be
\tilde{f}_{\alpha^m \rightarrow \beta^m} \left(\{ \xvec \}_m, \{ \xvec' \}_m, \Delta_m^-  \right) = \psi(\{\xvec \}_m) \,{\psi}^*( \{\xvec' \}_m) \, \bra{\alpha^m} e^{\B_m\left(\{ \xvec \}_m, \{ \xvec' \}_m, \Delta_m^-  \right)} \ket{\beta^m} \, , 
\ee
where $\{ \xvec \}_m$ denotes the set of relative positions $\xvec_{im} \equiv \xvec_i - \xvec_m$ (for $i = 1 \ldots m-1$) in the amplitude, a prime denotes a position in the conjugate amplitude, $\Delta_m^- \equiv \xvec_m - \xvec_m'$, $\psi(\{\xvec \}_m)$ is the initial wavefunction of the $m$-parton system, and $\B_m$ is the generalization of \eq{B-def} to a (color singlet) system of $2m$ partons. Going to momentum space (see \eq{Fourier-pair}), removing a factor $\delta^{(2)}(\sum_i \pvec_i)$, and setting final-state momenta as $\pvec_i'= - \pvec_i$ (for $i = 1 \ldots m$), we obtain the generalization of \eq{master-eq-asym-pair}--\eq{initial-cond-cdn} to any $m \geq 2$, namely,
\bea
\bra{\alpha^m} \hat{f}(\pvec_1 \ldots \pvec_m;L) \ket{\beta^m} = \int \left[ \prod_{i=1}^{m-1} \frac{\dd^2 \xvec_{im}}{(2\pi)^2}\frac{\dd^2 \xvec_{im}'}{(2\pi)^2} \right] \frac{\dd^2 \Delta_m^-}{(2\pi)^2} \, \psi(\{\xvec \}_m) \,{\psi}^*( \{\xvec' \}_m) \, \hskip 10mm &&  \nn \\ 
\times e^{i \sum_{i=1}^{m-1} \pvec_i \cdot (\xvec_{im} - \xvec_{im}') + i \Pvec_m \cdot \Delta_m^-} \,  \bra{\alpha^m} e^{\B_m\left(\{ \xvec \}_m, \{ \xvec' \}_m, \Delta_m^-  \right)} \ket{\beta^m} \, , \hskip 10mm && 
\label{master-eq-asym-pair-m}
\eea
where $\Pvec_m \equiv \sum_{i=1}^{m} \pvec_i$.

Tagging partons $i$ from $i=1$ up to $i=n < m$ implies that we integrate over the remaining momenta and replace the final color state $\ket{\beta^m}$ by $\ket{ \beta^n }$, yielding 
\bea
&& \int_{\pvec_{n+1} \ldots \pvec_m} \bra{\alpha^m} \hat{f}(\pvec_1 \ldots \pvec_m;L) \unit^{(m \rightarrow n)} \ket{\beta^n} =  \int \! \prod_{j=n+1}^m \! \! \dd^2 \pvec_j \int \left[ \prod_{i=1}^{m-1} \frac{\dd^2 \xvec_{im}}{(2\pi)^2}\frac{\dd^2 \xvec_{im}'}{(2\pi)^2} \right] \frac{\dd^2 \Delta_m^-}{(2\pi)^2}  \nn \\ 
&& \times  \psi(\{\xvec \}_m) \,{\psi}^*( \{\xvec' \}_m) \, e^{i \sum_{i=1}^{m-1} \pvec_i \cdot (\xvec_{im} - \xvec_{im}') + i \Pvec_m \cdot \Delta_m^-} \, \bra{\alpha^m} e^{\B_m\left(\{ \xvec \}_m, \{ \xvec' \}_m, \Delta_m^-  \right)} \unit^{(m \rightarrow n)} \ket{\beta^n} \, , \nn \\ 
\label{dist-tag-n}
\eea
where the reduction operator $\unit^{(m \rightarrow n)}$ is defined analogously to \eq{Eq:Operator_Reduction_2_1} (color trace over unobserved partons and Kronecker's in color indices for tagged partons). In the case $n=1$, $m=2$, the expression \eq{dist-tag-n} reproduces \eq{dist-tag1}. 

Expressing the wavefunction in terms of its Fourier transform,
\bea
\psi(\{\xvec \}_m) &=& \int \left[ \prod_{i=1}^{m-1} \dd^2 \rvec_i \, e^{-i \rvec_i \cdot  \xvec_{im}} \right]  \tilde{\psi}(\rvec_1 \ldots \rvec_{m-1}) \, , \\ 
\psi^*(\{\xvec' \}_m) &=& \int \left[ \prod_{i=1}^{m-1} \dd^2 \rvec_i' \, e^{i \rvec_i' \cdot  \xvec_{im}'} \right]  \tilde{\psi}^*(\rvec_1' \ldots \rvec_{m-1}') \, ,
\eea
allows one to evaluate \eq{dist-tag-n} as follows. Integrating over $\pvec_j$ sets $\xvec'_i = \xvec_i$ for $i=n+1 \ldots m$ (leaving $(m+n-1)$ $\xvec$-integrals in \eq{dist-tag-n}), and \eq{comBunit2} generalizes to
\begin{equation}
\label{Eq:Exp_Unit_m_n}
e^{\B_m} \unit^{(m \rightarrow n)} =  \unit^{(m \rightarrow n)} e^{\B_n} \, ,
\end{equation}
where the operator $\B_n$ depends on the $(2n-1)$ variables $\{ \xvec \}_n$, $\{ \xvec' \}_n$, and $\Delta_n^-$. The remaining integration variables are thus conveniently traded for the latter, together with $\xvec_{im}$ for $i=n \ldots m-1$. The integrals over $\xvec_{im}$ for $i=n+1 \ldots m-1$ then fix $\rvec_i' = \rvec_i$ for those $i$-values, and the integral over $\xvec_{nm}$ yields a factor $\delta^{(2)}(\Rvec_n - \Rvec_n')$, where $\Rvec_n  \equiv \sum_{i=1}^n \rvec_i$, $\Rvec_n'  \equiv \sum_{i=1}^n \rvec_i'$. We arrive at
\bea
\int_{\pvec_{n+1} \ldots \pvec_m} \bra{\alpha^m} \hat{f}(\pvec_1 \ldots \pvec_m;L) \unit^{(m \rightarrow n)} \ket{\beta^n} =  \int \prod_{i=1}^{n} \left[ \dd^2 \rvec_i \, \dd^2 \rvec_i' \right] \times  \hskip 1.5cm && \nn \\
\times \ \delta^{(2)}(\Rvec_n - \Rvec_n')  \int \! \!\prod_{j=n+1}^{m-1} \! \! \dd^2 \rvec_j \ \tilde{\psi}(\rvec_1 \ldots \rvec_n , \rvec_{n+1} \ldots \rvec_{m-1}) \, \tilde{\psi}^*(\rvec_1' \ldots \rvec_n' , \rvec_{n+1} \ldots \rvec_{m-1}) && \nn \\
\times \int \prod_{i=1}^{n-1} \left[ \frac{\dd^2 \xvec_{in}}{(2\pi)^2}\frac{\dd^2 \xvec_{in}'}{(2\pi)^2} \right] \int \frac{\dd^2 \Delta_n^-}{(2\pi)^2} \, e^{i \phi} \, \bra{\alpha^m}  \unit^{(m \rightarrow n)} e^{\B_n\left(\{ \xvec \}_n, \{ \xvec' \}_n, \Delta_n^-  \right)} \ket{\beta^n} \, , \hskip 1cm && 
\label{tag-n-in-m}
\eea
where the phase reads 
\be
\label{tag-n-in-m-phase}
\phi = \sum_{i=1}^{n-1} \left[ (\pvec_i - \rvec_i) \cdot \xvec_{in} - (\pvec_i - \rvec_i') \cdot\xvec_{in}' \right]  + (\Pvec_n - \Rvec_n) \cdot \Delta_n^-  \, .
\ee

The probability density \eq{tag-n-in-m} to tag $n$ partons with transverse momenta $\pvec_1 \ldots \pvec_n$ in a system of $m>n$ partons has the form of a convolution and generalizes \eq{gp1} to any $n \geq 1$ and $m>n$ (we readily check that \eq{tag-n-in-m} coincides with \eq{gp1} in the particular case $n=1$, $m=2$). When  $n>1$, novel features appear. The first factor of the convolution (second line of \eq{tag-n-in-m}) is a `skewed' initial momentum distribution, where there is no matching between parton momenta in the amplitude and its conjugate. (Those momenta are constrained by $\delta^{(2)}(\Rvec_n - \Rvec_n')$ and match only for $n=1$, where the skewed distribution coincides with the true probability density $\sim |\tilde{\psi}(\rvec_1 \ldots \rvec_{m-1})|^2$.) The second factor of the convolution (third line of \eq{tag-n-in-m}) can be readily interpreted by comparing it to the expression \eq{master-eq-asym-pair-m}. It plays the role of an evolution operator, in transverse momentum space, of a color singlet system of $2n$ partons. In particular, the system evolves according to the operator $e^{\B_n}$, and thus suffers the same broadening as if the untagged partons were not present. 

\subsubsection{Color transitions: off-diagonal elements of $f_{\alpha \rightarrow \beta}$}
\label{sec:color-transitions}

In this section we discuss the probability density for the parton pair to change its color state, \ie \ we consider $f_{\alpha \rightarrow \beta}$ in the case $\alpha \neq \beta$. Since a pointlike pair remains in the same color state (see section~\ref{sec:pointlike-pair}), changing color state is possible only for pairs which have a nonzero size. We will thus now address a less drastic limit than in the previous section, and consider the {\it compact pair limit} where the pair has a small but finite transverse size compared to the resolution $\sim 1/{\bar Q}$ of the medium. (We assume that the pair wavefunction $\psi(\xvec)$ selects values $|\xvec| \ll 1/{\bar Q}$.) 

The pair color transitions are encoded in the off-diagonal elements ($\alpha \neq \beta$) of the $\B$-matrix. It will be convenient to single out those elements by writing   
\be
\B = {\cal D} + X {\cal F} \, , 
\ee
where ${\cal D}$ is a diagonal matrix, ${\cal F}$ is a symmetric matrix with only off-diagonal elements, and $X \equiv X_t -X_u$ (with $X_t$ and $X_u$ given in \eq{Xt}--\eq{Xu}). In the present section we do not need to specify the precise form of the matrix ${\cal F}$. The specific cases, $aq \to aq$ (with $a=\bar{q}, q, g$) or $gg \to gg$, can be recovered using (see \eq{B-aq} and \eq{B-gg})  
\begin{numcases} {{\cal F}_{\alpha \beta} = (1-\delta_{\alpha \beta}) \, \times }
  \frac{\sqrt{K_\alpha K_\beta}}{2 K_a} & \text{for $aq \to aq$,} \label{case-1} \\
-\frac{1}{4} \,  \bra{\alpha} (T_t^2 - T_u^2) \ket{\beta} & \text{for $gg \to gg$,} \label{case-2}
\end{numcases}
where in the $gg \to gg$ case, the matrix element $\bra{\alpha} (T_t^2 - T_u^2) \ket{\beta}$ is given by \eq{Tt2minusTu2-ortho-basis}.
 
We will use the identity
\be
e^{{\cal D} + X {\cal F}} = e^{\cal D} + \int [\dd \delta]_{_2} \, e^{\delta_1 {\cal D}} \, X {\cal F} \, e^{\delta_2 {\cal D}} +  \int [\dd \delta]_{_3} \, e^{\delta_1 {\cal D}} \, X {\cal F} \, e^{\delta_2 {\cal D}} \, X {\cal F} \, e^{\delta_3 {\cal D}} + \ldots \, , 
\label{eq:compact-pair-expansion} 
\ee
where we introduced the notation
\be
\int [\dd \delta]_{_n} \equiv \left[ \prod_{i=1}^{n}  \int_0^1 \! \! \dd \delta_i \right] \delta \left( \sum_{j=1}^{n} \delta_j  -1 \right) \, .
\ee

The identity \eq{eq:compact-pair-expansion} can be shown to hold for any ${\cal D}$ and ${\cal F}$, but we will use it only for diagonal ${\cal D}$ and off-diagonal ${\cal F}$. It can be viewed as an expansion in powers of $X {\cal F}$ or equivalently in the number of color transitions. Independently of the type of parton pair, the off-diagonal matrix ${\cal F}$ comes along with a factor $X$, which becomes small in the compact pair limit. Each additional color transition thus brings a suppression factor $X$. In the following we will need the approximation of $X$ when $\xvec_{12}$ and $\xvec_{34}$ are small compared to $\xvec_{23}$, namely, 
\bea
X \equiv X_t -X_u &=& \hat{\Gamma}(\xvec_{23}+\xvec_{12}) + \hat{\Gamma}(\xvec_{23}+\xvec_{34}) -  \hat{\Gamma}(\xvec_{23}+\xvec_{12}+\xvec_{34}) -  \hat{\Gamma}(\xvec_{23}) \hskip 10mm  \nn \\
&\simeq& - (\xvec_{12} \cdot \nablavec) \, (\xvec_{34} \cdot \nablavec) \, \hat{\Gamma}(\xvec_{23}) \nn \\ 
&\simeq& - \frac{\bar{Q}^2}{4} \,  \xvec_{12} \cdot \xvec_{34} \,  \left[ \log\left(\frac{1}{\mu^2 \xvec_{23}^2}\right) + \morder{1} \right] \, , \label{X-compact}
\eea
where the third line is obtained by approximating $\hat{\Gamma}(\xvec_{23})$ using \eq{gammahat-appr}.\footnote{We thus assume $|\xvec_{23}| \ll 1/\mu$ and $\log(\frac{1}{\mu |\xvec_{23}|}) \gg 1$, which can be easily verified a posteriori. Indeed, the typical value of $|\xvec_{23}|$ contributing to the final expression \eq{fqL-compact} is $\sim {\rm min}(1/q_\perp, 1/\bar{Q}) \ll 1/\mu$, similarly to the single parton case discussed in section~\ref{sec:limits-of-f} and Appendix~\ref{app:single-parton}. Thus, the present derivation holds within the logarithmic accuracy \eq{log-acc-single}.}

We are now ready to evaluate $\bra{\alpha} e^\B \ket{\beta}$ for non-diagonal elements in the compact pair limit, to be then implemented in \eq{master-eq-asym-pair-2} to obtain the resulting probability density. When $\alpha \neq \beta$, the first term of the expansion \eq{eq:compact-pair-expansion} does not contribute, and the second term dominates, yielding
\be
\bra{\alpha} e^\B \ket{\beta} \mathop{\simeq}_{\alpha \neq \beta} X \int [\dd \delta]_{_2} \, e^{\delta_1 {\cal D}_{\alpha \alpha}}  \, {\cal F}_{\alpha \beta} \, e^{ \delta_2 {\cal D}_{\beta \beta}} + \morder{X^2} \, . 
\label{compact-linear0}
\ee
In the compact pair limit, $|\xvec_{12}|, |\xvec_{34}| \ll 1/\bar{Q}$, we have $X \ll 1$. Neglecting terms $\sim \morder{X^2}$, the matrix ${\cal D}$ in \eq{compact-linear0} can be evaluated at $\xvec_{12}= \xvec_{34} = \zerovec$, \ie, in the {\it pointlike} limit. In the latter limit $X=0$, $\B$ and ${\cal D}$ coincide, and using \eq{2pt-ab} we can rewrite \eq{compact-linear0} as
\be
\bra{\alpha} e^\B \ket{\beta} \mathop{\simeq}_{\alpha \neq \beta}  X \int [\dd \delta]_{_2} \, e^{-\delta_1 C_{\alpha} \hat{\Gamma}_{23}}  \, {\cal F}_{\alpha \beta} \, e^{- \delta_2 C_{\beta} \hat{\Gamma}_{23}} + \morder{X^2} \, .
\label{compact-linear1} 
\ee
The latter expression has a simple interpretation. On a path length $\delta_1 L$, the pair is effectively {\it pointlike} and of color charge equal to the initial charge $C_{\alpha}$. At the longitudinal position $\delta_1 L$, the $\alpha \to \beta$ transition associated to the color factor ${\cal F}_{\alpha \beta}$ occurs, which requires probing the {compact} pair and thus costs a factor $X$. Then the pair is again effectively pointlike, of charge $C_{\beta}$, on the path length $\delta_2 L$ (with $\delta_1+ \delta_2 =1$). 

It is interesting to quote the distribution $f_{\alpha \rightarrow \beta}(\qvec;L)$. Inserting \eq{compact-linear1} in \eq{master-eq-asym-pair-2}, using \eq{X-compact}, and integrating over $\pvec_1$ at fixed $\pvec_1 +\pvec_2 = \qvec$ we get\footnote{The integral over $\delta$ in \eq{fqL-compact} could be trivially performed, leading however to no real simplification in the following discussion.}  
\be
f_{\alpha \rightarrow \beta}(\qvec;L) \mathop{\simeq}_{\alpha \neq \beta} {\cal F}_{\alpha \beta} \, \frac{\bar{Q}^2 \ave{\xvec_{12}^2}}{4} \int_0^1 \! \! \dd \delta \int \! \frac{\dd^2 \xvec_{23}}{(2 \pi)^2} \log\left(\frac{1}{\mu^2 \xvec_{23}^2}\right) e^{i \qvec \xvec_{23}} \, e^{-[\delta C_{\alpha} + (1-\delta)  C_{\beta}] \, \hat{\Gamma}_{23}}  \, ,
\label{fqL-compact}
\ee
where $\ave{\xvec_{12}^2}$ is the average size of the parton pair of wavefunction $\psi(\xvec)$,
\be
\ave{\xvec_{12}^2} \equiv  \int \frac{\dd^2 \xvec_{12}}{(2 \pi)^2}  \, \xvec_{12}^2 \, |\psi(\xvec_{12})|^2 \, .
\ee
In the compact pair limit, $f_{\alpha \rightarrow \beta}(\qvec;L)$ given by \eq{fqL-compact} is the product of a suppression factor $\sim {\cal F}_{\alpha \beta} \, \bar{Q}^2 \ave{\xvec_{12}^2}$ to have color transition, and a {\it linear superposition} of distributions corresponding (up to the logarithmic factor in the integrand of \eq{fqL-compact} which is however harmless) to pointlike color charges 
\be
\bar{C}_{\delta} = \delta C_{\alpha} + (1-\delta)  C_{\beta} 
\label{C-delta}
\ee
given by the average of the initial and final Casimirs weighted according to the time (in units of $L$) spent in the initial or final color state. 

This simple feature also shows up in the limiting behaviors of $f_{\alpha \rightarrow \beta}(\qvec;L)$ when $q_\perp \lsim \bar{Q}$ and $q_\perp \gg \bar{Q}$, which can be derived along the same lines as in section~\ref{sec:limits-of-f} and Appendix~\ref{app:single-parton} for the broadening of a single parton. For $q_\perp \lsim \bar{Q}$ we have
\bea
\label{fqL-gaussian}
f_{\alpha \rightarrow \beta}(\qvec;L)  \mathop{\simeq}_{q_\perp \lsim \bar{Q}} {\cal F}_{\alpha \beta} \, \frac{\bar{Q}^2 \ave{\xvec_{12}^2}}{4} \log\left(\frac{\bar{Q}^2}{\mu^2}\right)  \, h^{{\rm G}}(\qvec; L) \, , \hskip 2cm && \\ 
{\rm where \ \ } h^{{\rm G}}(\qvec; L) = \int_0^1 \! \! \dd \delta \, \, \frac{e^{-\qvec^2/ q_{\perp {\rm w}}^2(\delta)}}{\pi q_{\perp {\rm w}}^2(\delta)} \ \ ; \ \ q_{\perp {\rm w}}^2(\delta) \equiv \bar{C}_{\delta} \, \bar{Q}^2 \log\left(\frac{\bar{Q}}{\mu}\right)  \, , && 
\label{gaussian-superposition}
\eea
and in the limit $q_\perp \gg \bar{Q}$ we find
\be
\label{fqL-tail}
 f_{\alpha \rightarrow \beta}(\qvec;L) \mathop{\simeq}_{q_\perp \gg \bar{Q}} {\cal F}_{\alpha \beta} \, \frac{\bar{Q}^2 \ave{\xvec_{12}^2}}{4\pi \qvec^2} \left[ 1+ \frac{\bar{Q}^2}{\qvec^2} \cdot \frac{C_\alpha+C_\beta}{2} \left(\log{\frac{\qvec^2}{\mu^2}} + \morder{1} \right) \right] \, .
\ee
This allows one to emphasize the following points: 
\bi
\item[(i)] The amount of transverse broadening in the events under consideration ($\alpha \neq \beta$)  may be estimated by the average value $q_{\perp {\rm w}}^2$ of $q_\perp^2$ associated to the linear superposition of Gaussians $h^{{\rm G}}(\qvec; L)$ appearing in \eq{fqL-gaussian},
\be
q_{\perp {\rm w}}^2 \equiv \int \dd^2 \qvec  \, \, \qvec^2 \, h^{{\rm G}}(\qvec; L) = \int_0^1 \! \! \dd \delta \, \, q_{\perp {\rm w}}^2(\delta) = \frac{C_\alpha+C_\beta}{2} \, \bar{Q}^2 \log\left(\frac{\bar{Q}}{\mu}\right)   \, .
\label{broad-small-q}
\ee
Since in average the color transition occurs at $\delta = \frac{1}{2}$, the effective Casimir equals $\bar{C}_{\delta=1/2} = (C_\alpha+C_\beta)/2$. 
\item[(ii)] The function $h^{{\rm G}}(\qvec; L)$ has the same zeroth and first moments (in the variable $q_\perp^2$) as the Gaussian distribution of width $q_{\perp {\rm w}}^2$. As a consequence, the latter is a very good approximation to $h^{{\rm G}}(\qvec; L)$, at least when $C_\alpha$ and $C_\beta$ are both non-zero, as can be checked numerically.\footnote{In the particular case $C_\alpha = C_\beta \neq 0$, $h^{{\rm G}}(\qvec; L)$ obviously {\it coincides} with the Gaussian of width $q_{\perp {\rm w}}$, see \eq{C-delta} and \eq{gaussian-superposition}. This case occurs for the ${\bf 8_a \to 8_s}$ and ${\bf 8_s \to 8_a}$ transitions of a $gg$ pair, where $C_\alpha = C_\beta = N$ and ${\cal F}_{\alpha \beta} = N/4$.} When either $C_\alpha$ or $C_\beta$ vanishes, the exact expression \eq{gaussian-superposition} of $h^{{\rm G}}(\qvec; L)$ should be preferred.
\item[(iii)] Similarly to the case of a single parton (section~\ref{sec:limits-of-f}), to the logarithmic accuracy \eq{log-acc-single} the approximation \eq{fqL-gaussian} holds in a region which extends slightly beyond $q_\perp \sim q_{\perp {\rm w}}$, before being overcome by the large-$q_\perp$ asymptotics \eq{fqL-tail}. 
\item[(iv)] For a target of small size $L_{\rm p} \to 0$ (associated with a saturation scale $\bar{Q}_{\rm p} \to 0$), the distribution \eq{fqL-compact} reads\footnote{For the calculation of the $\xvec_{23}$-integral in \eq{fqL-compact-Lp}, see \eg\ Appendix A2 of Ref.~\cite{Munier:2016oih}.}
\be
f_{\alpha \rightarrow \beta}(\qvec;L_{\rm p}) \mathop{\simeq}_{\alpha \neq \beta} {\cal F}_{\alpha \beta} \, \frac{\bar{Q}_{\rm p}^2 \ave{\xvec_{12}^2}}{4}  \! \int \! \frac{\dd^2 \xvec_{23}}{(2 \pi)^2} \log\left(\frac{1}{\mu^2 \xvec_{23}^2}\right) e^{i \qvec \xvec_{23}} = {\cal F}_{\alpha \beta} \, \frac{\bar{Q}_{\rm p}^2 \ave{\xvec_{12}^2}}{4 \pi \qvec^2}  \, .
\label{fqL-compact-Lp}
\ee
Normalizing the distribution in a target of size $L$ by \eq{fqL-compact-Lp}, and then removing a trivial factor $L/L_{\rm p}$ (corresponding to a $q_\perp$-distribution being additive in the number of nucleons encountered along the path), \eq{fqL-gaussian} and \eq{fqL-tail} give
\begin{numcases} {\frac{L_{\rm p}}{L}\frac{f_{\alpha \rightarrow \beta}(\qvec; L)}{f_{\alpha \rightarrow \beta}(\qvec; L_{\rm p})} \ \simeq \, }
 \pi \qvec^2 \log\left(\frac{\bar{Q}^2}{\mu^2}\right)  \, h^{{\rm G}}(\qvec; L) \, & \hskip -3mm \text{for $q_\perp \lsim \bar{Q}$,} \label{small-q-case-1} \\ 1 +  \frac{\bar{Q}^2}{\qvec^2} \cdot \frac{C_\alpha + C_\beta}{2} \left( \text{log}\ \frac{\qvec^2}{\mu^2} + \mathcal{O}(1) \right) & \hskip -3mm \text{for $q_\perp \gg \bar{Q}$.} \hskip 10mm \label{large-q-case-2}
\end{numcases}
The latter limiting behaviors mimic the well-known Cronin effect. In a large nucleus w.r.t.~a proton, the $q_\perp$-distribution is reduced (enhanced) at small (large) $q_\perp$, as a result of nuclear broadening.  
\ei
In summary, a simple picture emerges in the compact pair limit. The hard exchange needed for color transition occurs in average at the position $L/2$. Before (after), the pair behaves as a {\it pointlike} object of color charge $C_\alpha$ ($C_\beta$). This results in $q_\perp$-broadening proportional to $(C_\alpha+C_\beta)/2$, which can be read off from the average $q_\perp^2$ associated with the small-$q_\perp$ approximation \eq{small-q-case-1}, or from the deviation to unity of the large-$q_\perp$ asymptotics \eq{large-q-case-2}. 

\section{Parton pair produced in a hard process}
\label{sec:hard-prod}

Until now we have focussed on the transverse momentum {\it probability density} of an `asymptotic parton pair', putting emphasis on the pair color state. In QCD, colorful parton pairs cannot be truly asymptotic, but can be produced in hard QCD processes and propagate through a nuclear medium before later hadronizing. Based on the study of an asymptotic pair, in this section we sketch how to infer the {\it production cross section} of a colorful pair in parton-nucleus scattering. We recover the expression derived previously in Refs.~\cite{Nikolaev:2005qs,Nikolaev:2005dd,Nikolaev:2005zj}, which after a proper sum over color indices reproduces the dijet forward production cross section derived elsewhere, see \eg\ Refs.~\cite{Marquet:2007vb,Dominguez:2011wm,Iancu:2013dta}. 

\begin{figure}[t]
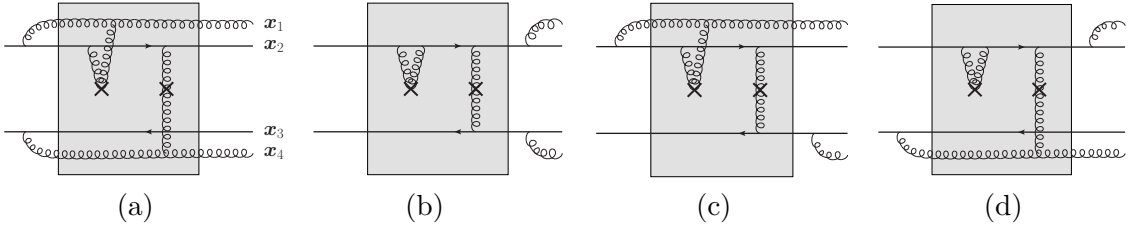
  \hskip 3mm
 \makebox[.22\textwidth]{\csII (40,0) } \hskip 6mm 
 \makebox[.22\textwidth]{\csOO (40,0) } \hskip 2mm
 \makebox[.22\textwidth]{\csIO (39,-3) } \hskip 3mm
 \makebox[.22\textwidth]{\csOI (40,0) }
 \makebox[.245\textwidth]{\centering (a)}
 \makebox[.245\textwidth]{\centering (b)}
 \makebox[.245\textwidth]{\centering (c)}
 \makebox[.245\textwidth]{\centering (d)}
 \caption{Contributions to the cross section for $q+ {\rm A} \to qg + {\rm X}$: (a) `initial-state', (b) `final-state', and (c,d) `interference' contributions (in the convention where the lower half of a diagram represents a conjugate amplitude).}
 \label{fig:3cont}
 \end{figure}

Let us consider forward dijet production in high-energy \pA\ collisions. When viewed in the nucleus rest frame, such a process typically arises from the forward scattering of an incoming parton (from the projectile proton) to an outgoing parton pair (further hadronizing into a dijet), mediated by gluon exchanges with the nuclear medium. 
Consider the generic case of $q \to qg$ scattering, see Fig.~\ref{fig:3cont}. (The discussion applies to any type of produced parton pair.) As is well-known, at high energy the typical lifetime of the $qg$ fluctuation in the incoming quark is large, and the $q \to qg$ splitting occurs, at the amplitude level, either long before or long after the nucleus. In light-cone $A^+=0$ gauge, the $qg$ production cross section is dominated by contributions which may be interpreted as `initial-state' (`final-state'), when the splitting occurs before (after) the nucleus in the amplitude and its conjugate (Figs.~\ref{fig:3cont}a and~\ref{fig:3cont}b), and as an interference when the splitting occurs before in the amplitude and after in its conjugate, or vice-versa (Figs.~\ref{fig:3cont}c and~\ref{fig:3cont}d). 

Those contributions are formally similar to the probability density $f_{\alpha \rightarrow \beta}(\pvec_1,\pvec_2;L) $ of an asymptotic pair and can be derived using the same kinetic equation approach as that of section~\ref{sec:fp1p2-kin}, up to the replacement of the asymptotic pair wavefunction $\psi(\xvec)$ by the $q \to q g$ splitting amplitude,\footnote{In the massless quark limit, any $1 \to 2$ splitting amplitude ($q \to q g$, $g \to gg$, or $g \to q \bar{q}$) is proportional to ${\xvec}/{\xvec^2}$ (with $\xvec$ the parton pair transverse separation), times a factor depending on spins (quark helicities and gluon polarizations)~\cite{Lepage:1980fj}. The square of the latter factor gives rise, after summing over spins, to the DGLAP splitting function. For simplicity we do not write the spin-dependent factor in the r.h.s.~of \eq{eq:psi-change}, and directly add the DGLAP splitting function in the final result \eq{a2ag-cross}.}
\be
\psi(\xvec) \longrightarrow \frac{\xvec}{\xvec^2} \, \, .
\label{eq:psi-change}
\ee

The initial-state contribution of Fig.~\ref{fig:3cont}a involves the 4-point $gq\bar{q}g$ correlator (similarly to \eq{master-eq-asym-pair-2}). The other contributions involve 3-point and 2-point correlators, which are limits of the 4-point correlator when the pair is pointlike (in the amplitude or/and in its conjugate), as recalled in section~\ref{sec:32point-corr}, see \eq{eq:4to3-corr} and \eq{2pt-ab}. Using \eq{eq:psi-change} and adding all contributions, the $qg$ production cross section in quark-nucleus scattering reads (compare to the probability density of an asymptotic pair \eq{master-eq-asym-pair-2})
\bea
\mbox{\fontsize{14}{2}\selectfont $ \frac{\dd \sigma_{\alpha \to \beta}(q +{\rm A}\to qg + {\rm X})}{ S_{\perp} \dd z_1 \dd^2 \pvec_1 \dd^2 \pvec_2} $}= \alpha_s \Phi_q^{g}(z_1) \! \int \!  \mbox{\fontsize{12}{2}\selectfont $ \frac{\dd^2 \xvec_{12} \, \dd^2 \xvec_{23} \, \dd^2 \xvec_{34}}{(2\pi)^6} \, e^{i \pvec_1 \cdot (\xvec_{12}+\xvec_{34}) + i (\pvec_1+\pvec_2) \cdot \xvec_{23}} \, \frac{\xvec_{12} \cdot \xvec_{43}}{\xvec_{12}^2 \xvec_{43}^2} $}  \,  && \nonumber \\
\times  \bra{\alpha} \left[ e^{\B(\xvec_1,\xvec_2;\xvec_3,\xvec_4)} - e^{\B(\vvec,\vvec;\xvec_3,\xvec_4)} - e^{\B(\xvec_1,\xvec_2;\vvec',\vvec')} + e^{\B(\vvec,\vvec;\vvec',\vvec')} \right] \ket{\beta}  \, , \hskip 1cm &&
\label{a2ag-cross}
\eea
where $S_\perp$ is the transverse area of the nuclear target, $\Phi_q^g(z) = C_F \cdot 2\left[1 + (1-z)^2 \right]/z$ the $q\to g$ DGLAP splitting function~\cite{Dokshitzer:1991wu}, $\ket{\alpha}$ the color state coinciding with the incoming parton representation ($\ket{\beta}$ the color state of the final parton pair), and the transverse positions ${\vvec}$ and ${\vvec}'$ are defined by ${\vvec}=z_1{\xvec_1} +z_2{\xvec_2}$ and ${\vvec'}=z_1 \xvec_4 +z_2 \xvec_3$, with $z_1$ and $z_2=1-z_1$ the longitudinal momentum fractions of partons 1 and 2 w.r.t.~the incoming quark.\footnote{When evaluating the various contributions in the kinetic equation approach of section~\ref{sec:fp1p2-kin}, ${\vvec}$ and ${\vvec}'$ arise as follows. Consider the contribution of Fig.~\ref{fig:3cont}d. First, note that if in the amplitude the parent quark of longitudinal momentum $p^z$ undergoes a transverse kick $\ellvec$, the $qg$ pair undergoes a global {\it rotation} of `angle' $\ellvec/p^z$, hence the individual partons acquire the transverse momenta $z_1 \ellvec$ and $z_2 \ellvec$. Thus, the probability to have final parton momenta $\pvec_1$ and $\pvec_2$ in the presence of a rescattering $\ellvec$ between times $t$ and $t + \dd t$, is related to the probability to have parton momenta $\pvec_1-z_1 \ellvec$ and $\pvec_2 - z_2 \ellvec$ in the absence of rescattering. The Fourier transform \eq{Fourier-pair} thus induces a phase shift $\sim i \ellvec \cdot (z_1{\xvec_1} +z_2{\xvec_2})$, fixing the transverse position of the parent quark at ${\vvec}=z_1{\xvec_1} +z_2{\xvec_2}$ for this contribution.} Up to proper replacements of the splitting function and $\B$ operator, the expression \eq{a2ag-cross} also holds in the $g\to gg$ and $g \to q \bar{q}$ cases. 

The production cross section \eq{a2ag-cross}, written for a final parton pair in a given color state and depending on specific matrix elements of the evolution operator $e^{\B}$, was derived previously in Refs.~\cite{Nikolaev:2005qs,Nikolaev:2005dd,Nikolaev:2005zj}. Summing over final color indices,\footnote{This amounts to consider a specific linear combination of matrix elements of $e^{\B}$, obtained by the replacement $\ket{\beta} \to \sum_{\beta} \sqrt{K_{\beta}} \, \ket{\beta}$ in \eq{a2ag-cross}, see \eq{app:color-sum} for the specific case of a final $q {\bar q}$ pair.} one recovers the dijet production cross section expressed in terms of usual (color-averaged) correlators~\cite{Marquet:2007vb,Dominguez:2011wm,Iancu:2013dta}. 
 
\section{Discussion: probing unusual Casimirs with compact pairs}
\label{sec:discussion} 
 
Our review of parton pair $p_\perp$-broadening may help addressing dijet broadening in phenomenology by keeping track of the color structure of the process. In particular, the `compact pair expansion' (introduced in section~\ref{sec:color-transitions}, see \eq{eq:compact-pair-expansion}) should allow one to probe Casimir charges of `unusual' ${\rm SU}(N)$ representations. 

In the case of an `asymptotic pair' studied in section~\ref{sec:asym-pair}, the parton pair can be made compact by choosing a wavefunction $\psi(\xvec)$ selecting values $|\xvec| \ll 1/{\bar Q}$. In the case of a parton pair produced in a hard process (section~\ref{sec:hard-prod}), the pair becomes effectively compact in the kinematical domain $|\pvec_1| \gg |\qvec| \sim {\bar Q}$. (In this limit the cross section \eq{a2ag-cross} is dominated by the region $|\xvec_{12}|, |\xvec_{34}| \sim 1/|\pvec_1| \ll |\xvec_{23}| \sim 1/|\qvec|$.) Thus, at large relative transverse momentum $\pvec_1 -\pvec_2$, dijet production should be sensitive to the global color charge of the pair, and the latter could be singled out by measuring the broadening of dijet transverse momentum imbalance $\qvec = \pvec_1 + \pvec_2$. 

As an illustration, let us consider the $g +{\rm A}\to gg + {\rm X}$ cross section. The latter is an incoherent sum of contributions corresponding to a given color state $\beta$ of the produced pair, \ie, a sum of terms of the type \eq{a2ag-cross} (with proper weights, see the comments in the end of section~\ref{sec:hard-prod}).~In the compact pair limit $|\pvec_1| \gg |\qvec| \sim {\bar Q}$, each contribution is associated with a $q_\perp$-broadening depending explicitly on $C_\beta$. As a consequence, the observed $q_\perp$-distribution should be sizable over a range given by the {\it largest} of the individual broadening widths.~In the $g \to gg$ case and for $N=3$, the largest broadening occurs when the $gg$ pair is produced in the color state $\beta = {\bf 27}$, and the associated $q_\perp$-broadening width scales as $(C_A+C_{27})/2 = 11/2$. The calculation is briefly summarized in Appendix~\ref{app:broadening-examples}, see the result \eq{app:g2gg-cross-27}. (The calculation is the same as for an asymptotic compact pair undergoing a color transition from $\alpha = {\bf 8_a}$ to $\beta = {\bf 27}$, up to the change \eq{eq:psi-change} which however does not lead to any difference for $q_\perp$-broadening.) We thus expect $q_\perp$-broadening of a dijet arising from a compact $gg$ pair to be nearly twice as wide as the broadening of a gluon. Let us mention that when the $gg$ pair is produced in a color octet state, $\beta = \alpha = {\bf 8_a}$, at any time of the evolution the parton system propagating in the medium is a compact color octet, and $q_\perp$-broadening scales as $C_A$. For completeness, we also summarize the calculation corresponding to this case in Appendix~\ref{app:broadening-examples}, see the result \eq{app:g2gg-cross-8-final}. 

\begin{figure}[t]
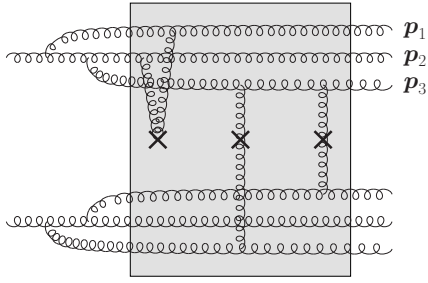
 
\bc
\CrossSectionGGG (60,0) 
\ec
\caption{Some contribution to the `3-jet' cross section $g+ {\rm A} \to ggg + {\rm X}$.}
\label{fig:3jet-example}
\end{figure} 
 
In the above discussion we considered forward dijet production. In \pA\ collisions at very high energies, the production cross section of $n_{\rm jet}>2$ forward jets becomes sizable, and the broadening of a tagged {\it pair of jets} in inclusive events may differ from the situation where only two jets are produced. However, such broadening could in principle be addressed using the tools presented in this study. As an illustration, consider `$3$-jet' production $g \to g(\pvec_1)g(\pvec_2)g(\pvec_3)$, and assume each gluon pair to be compact, \ie, $|\pvec_i - \pvec_j| \gg {\bar Q}$. In Fig.~\ref{fig:3jet-example} we show one particular contribution to the associated cross section, for which we expect the following features. The tagged pair of jets may arise from any gluon pair $(ij)$.  Similarly to the case of $n$ asymptotic partons picked out of $m$ studied in sections~\ref{sec:1-out-of-2} and~\ref{sec:n-out-of-m}, we expect this pair to broaden independently of the presence of the third gluon. In Fig.~\ref{fig:3jet-example} the gluon pair $(23)$ enters the medium as an octet, and we thus expect the associated broadening (of momentum imbalance $\pvec_2 + \pvec_3$) to scale at most as $(C_A+C_{27})/2$, similarly to dijet production.~In contrast, the gluon pair $(12)$ can already enter in a color state ${\bf 27}$, and can remain in this state. Since the pair is compact the associated broadening should depend solely on $C_{27}$. This suggests that the presence of $3$-jet events tends to increase the nuclear broadening of a tagged pair of jets. Finally, let us note that the $ggg$ system can be produced in higher dimensional color states than a $gg$ pair, namely ${\bf 35}$, ${\bf \overline{35}}$ and ${\bf 64}$. For instance, the state ${\bf 64}$ can be produced via two successive color transitions, ${\bf 8_a} \to {\bf 27} \to {\bf 64}$. Thus, if the three produced gluon jets are tagged, we expect the broadening distribution of $\pvec_1 + \pvec_2 + \pvec_3$ to receive a Gaussian contribution of width $(C_A+C_{27}+C_{64})/3$. The latter will however be suppressed by an additional factor $X \sim {\bar Q}^2/\pvec_{ij}^2$ (with $\pvec_{ij}$ a typical relative transverse momentum between final state partons), compared to contributions associated to a single color transition. 
 
In this study we assumed the saturation scale ${\bar Q}$ to be independent of the collision energy $\sqrt{s}$, or equivalently of $x \sim {\pvec_{ij}^2/s}$. At increasing energy, the presence of soft gluons in the incoming parton wavefunction usually promotes ${\bar Q}$ to an $x$-dependent function $\bar{Q}(x)$. It would be interesting to study whether this statement holds for dijet production when the produced parton pair is in a given color state. Indeed, on one hand the enhancement of $p_\perp$-broadening with energy is often attributed to the increase of ${\bar Q}(x)$, but on the other hand the color structure of the broadening process itself could explain (at least partly) such an enhancement. First, as argued above, the typical number of forward (hard) partons in inclusive events increases with energy, leading to the presence of higher dimensional color representations with larger Casimirs. Second, the latter Casimirs can be singled out in the compact pair limit, which is also easier to access at higher collision energies. We believe that the tools presented in this study should help quantifying the relative roles of small-$x$ evolution and color structure in the increase of $p_\perp$-broadening with energy. 
 
\acknowledgments

We would like to thank Fran\c{c}ois Arleo, Jamal Jalilian-Marian, Al Mueller, and St\'ephane Munier for fruitful discussions. Feynman diagrams have been drawn with the JaxoDraw software~\cite{Binosi:2008ig}.

\appendix

\section{Limits of the single parton distribution $f(\pvec;L)$}
\label{app:single-parton}

Here we derive the limits, stated in section~\ref{sec:limits-of-f}, of the single parton transverse momentum distribution $f(\pvec;L)$ given by \eq{f-approx}. 

Due to the exponential factors in the integrand of \eq{f-approx}, the integral \eq{f-approx} is dominated by $x_\perp \sim {\rm min}(1/p_\perp, 1/\bar{Q}) \ll 1/\mu$ (recall that $p_\perp, {\bar Q} \gg \mu$ is used in section~\ref{sec:limits-of-f} to obtain \eq{f-approx} from the more exact expression \eq{f-density}). We can thus distinguish two limiting cases.

\subsubsection*{Limit $p_\perp \lsim \bar{Q}$}

In this case, following Ref.~\cite{Baier:1996sk}, we have $x_\perp \sim 1/ \bar{Q}$, and to logarithmic accuracy we can replace $x_\perp \to 1/ \bar{Q}$ in $\log(\frac{1}{\mu x_\perp})$, yielding
\be
\label{eqapp:f-gaussian}
f(\pvec;L) \mathop{\simeq}_{p_\perp \lsim \bar{Q}} \int \! \! \frac{\dd^2 \xvec}{(2 \pi)^2} \ e^{i \pvec \cdot \xvec} \, e^{- C_R \frac{\bar{Q}^2}{8}\, \xvec^2 \log(\frac{\bar{Q}^2}{\mu^2}) } \, , 
\ee
which results in \eq{f-gaussian}. For $p_\perp \lsim \bar{Q}$, $f(\pvec;L)$ can be approximated by the Gaussian distribution $f^{{\rm G}}(\pvec,L)$ of width $p_{\perp {\rm w}}$. 

\subsubsection*{Limit $p_\perp \gg \bar{Q}$}

When $p_\perp \gg \bar{Q}$, we have $x_\perp \sim 1/ p_\perp \ll 1/\bar{Q}$, and the second exponential factor in the integrand of \eq{f-approx} can be Taylor-expanded,
\be
\label{f-large-p}
f(\pvec;L) \mathop{\simeq}_{p_\perp \gg \bar{Q}} \int \! \! \frac{\dd^2 \xvec}{(2 \pi)^2} \ e^{i \pvec \cdot \xvec} \left[ 1 + C_R \frac{\bar{Q}^2}{4} \,  \xvec^2 \log(\mu |\xvec|)  + \ldots \right] \, .
\ee
Here, the $x_\perp$-dependence of the non-analytic factor $\log(\mu x_\perp)$ is crucial to keep. Indeed, if one would replace $x_\perp \to 1/ p_\perp$ in this logarithm, then \eq{f-large-p} would become a series of terms of the type $\left({\bf \nabla}_\pvec^2 \right)^n \delta^{(2)}(\pvec)$, with a support only at $\pvec = \zerovec$, in contradiction with the assumption $p_\perp \gg \bar{Q}$. For $p_\perp \gg \bar{Q}$, the first term in the bracket of \eq{f-large-p} (which contributes to $\delta^{(2)}(\pvec)$) can be dropped, and the second term can be evaluated using the identity\footnote{For the calculation of integrals of the type \eq{int-log}, see Appendix A2 of Ref.~\cite{Munier:2016oih}.}
\be
\label{int-log}
\int \!\! \frac{\dd^2 \xvec}{(2 \pi)^2} \, e^{i \pvec \cdot \xvec}  \, \xvec^2  \log(\mu |\xvec|)  = \frac{2}{\pi |\pvec|^4} \, ,
\ee 
leading to the result stated in \eq{f-tail}.

\section{Anomalous dimension matrix for $aq \to aq$}
\label{app:Qaqaq}

Here we derive the matrix ${\cal Q}$ defined by \eq{Q-def} in the orthonormal basis \eq{ortho-basis}, for the partonic processes $aq\to aq$ (with $a=g$, $q$, or $\bar{q}$). Since parton $2$ is a quark, we have $C_2 = C_3 = C_F$, and we denote by $C_a= C_1 = C_4$ the Casimir operator of parton $a$. 

The matrix elements of $T_t^2=(T_2+T_3)^2$ read
\be
\bra{\alpha} T_t^2 \ket{\beta} = 2 C_F \delta_{\alpha \beta} +  2\, \bra{\alpha} T_2 T_3 \ket{\beta}  = 2 C_F \delta_{\alpha \beta} - \frac{2}{ \sqrt{K_\alpha K_\beta}} \qaTt(20,-30)  \, , 
\label{fierz-1}
\ee
where the generic parton $a=g$, $q$ or $\bar{q}$ is represented by the dashed line. Using the Fierz identity, 
\be
\label{fierz-identity}
2 \FierzOctet(10,-20) 
 = \FierzIdentity(10,-20) -\frac{1}{N} \FierzSinglet(10,-20)  \, ,
\ee
the color graph appearing in \eq{fierz-1} becomes
\bea
\label{fierz-2}
2 \qaTt(20,-30) \ 
 &=& \ \qaTtId(20,-30) \ 
-\frac{1}{N} \ \qaTtSinglet(20,-30) \ \, .
\eea
In the rhs of \eq{fierz-2}, the second graph equals $\delta_{\alpha \beta} K_\alpha$ (see \eq{dim}), and the first graph can be simplified by noting that in this graph the intermediate $a\bar{a}$ state is a fortiori color singlet, thus the value of the graph is unchanged by projecting $a\bar{a}$ on its color singlet part using
\be
\left( \! \aabar(15,-20)  \right)_{\rm \! \! singlet} 
= \frac{1}{K_a} \ \aabarSinglet(15,-20) \   \, ,
\ee
where $K_a$ is the dimension of the representation of parton $a$. As a result, the first graph in the rhs of \eq{fierz-2} equals $K_\alpha K_\beta/K_a$. Inserting then \eq{fierz-2} in \eq{fierz-1} gives
\be
\bra{\alpha} T_t^2 \ket{\beta} = N \delta_{\alpha \beta} - \frac{\sqrt{K_\alpha K_\beta}}{K_a} \, .
\ee
The matrix elements of ${\cal Q}$ given by \eq{Q-def} are then easily obtained with the help of \eq{sumT} and using $\bra{\alpha} T_s^2  \ket{\beta} = \delta_{\alpha \beta} C_\alpha$, 
\be
\label{eq:Qaqaq}
\bra{\alpha} {\cal Q} \ket{\beta} =  \delta_{\alpha \beta} \left[ (1-b) \, \frac{2C_F+2C_a-C_\alpha}{2N} + b \right] - b \,\frac{\sqrt{K_\alpha K_\beta}}{N K_a} \ ; \ \ \ b = \frac{X_t - X_u}{X_t + X_u} \, .
\ee

\section{Anomalous dimension matrix for $gg \to gg$}
\label{app:Qgggg}

In this section we derive the matrix ${\cal Q}$ (defined by \eq{Q-def}) in the basis \eq{ortho-basis}, for $gg \to gg$ scattering. The derivation of Appendix~\ref{app:Qaqaq} for $aq \to aq$, based on the Fierz identity \eq{fierz-identity}, does not apply when the four participating partons are gluons. However, it turns out that obtaining the matrix elements of ${\cal Q}$ is not more complicated for $gg \to gg$ than for $aq \to aq$. 

First, let us recall that for $N \geq 3$, a two-gluon system can be in the following irreducible (and self-conjugate) representations,
\be
\label{gluglu}
{\bf 8 \otimes 8 = 8_a  \oplus (10 \oplus \overline{10}) \oplus 1 \oplus 8_s \oplus 27 \oplus 0} \, ,
\ee
where as in \eq{quark-glu} the representations are labelled according to their dimensions in the case $N =3$. In particular ${\bf 0}$ is a symmetric representation which is absent when $N =3$. (The self-conjugate representation ${\bf 10 \oplus \overline{10}}$, of dimension 20 for $N=3$, will be simply denoted as ${\bf 10}$ in the following.) As in~\cite{Dokshitzer:2005ek}, we order the $s$-channel representations $\alpha$ (and thus the orthonormal basis vectors $\ket{\alpha}$ defined in section~\ref{sec:color-structure}) as in the rhs of \eq{gluglu}, namely $\alpha = \{ {\bf 8_a,10,1,8_s,27,0} \}$. Those representations are characterized by their dimensions $K_{\alpha}$, Casimirs $C_{\alpha}$ and symmetry signums $\sigma_{\alpha}=\pm 1$,
\bea
K_{\alpha} &=& \{ \mbox{\fontsize{10}{2}\selectfont ${N^2-1, \frac{(N^2-1)(N^2-4)}{2}, 1, N^2-1,  \frac{N^2(N-1)(N+3)}{4}, \frac{N^2(N+1)(N-3)}{4}} $} \} \, , \label{gluglu-cas} \label{invariant-dimension} \\ 
C_{\alpha} &=& \{N,\>2N,\>0,\>N,\>2(N\!+\!1), \>2(N\!-\!1)\} \, ,  \label{invariant-casimir} \\
\sigma_{\alpha} &=& \{ -1,\>-1,\>1,\>1,\>1,\>1 \} \, . \label{invariant-sigma}  
\eea

We start by evaluating the matrix elements of $T_t^2-T_u^2$. Using \eq{Tstu} and the fact that in the present case $T_i^2 = C_i = N$ ($i=1\ldots4$), we have
\be
\label{Tt2minusTu2}
\bra{\alpha} (T_t^2 - T_u^2) \ket{\beta} =  2\, \bra{\alpha} (T_2 T_3 - T_1 T_3) \ket{\beta}  = \frac{-2}{\sqrt{K_\alpha K_\beta}} 
\left\{ \ggTtbc(20,-40) - \ggTtac(20,-40) \right\} \, .
\ee
The two graphs appearing in \eq{Tt2minusTu2} are related by permutation of the two upper gluon lines. Since such a permutation introduces a factor $\sigma_{\alpha}  \sigma_{\beta}$, we obtain
\bea
\label{Tt2minusTu2-2}
\bra{\alpha} (T_t^2 - T_u^2) \ket{\beta} &=&  \frac{-2}{\sqrt{K_\alpha K_\beta}} \, (1-\sigma_{\alpha}  \sigma_{\beta}) \, I_{\alpha \beta} \, , \\ 
I_{\alpha \beta} &\equiv& \ggTtbc(20,-40)  \, .
\label{Ialphabeta}
\eea

We see that $\bra{\alpha} (T_t^2 - T_u^2) \ket{\beta}$ is symmetric under $\alpha \leftrightarrow \beta$, and non-zero only when the representations $\alpha$ and $\beta$ have different signums, $\sigma_{\alpha}  \sigma_{\beta} = -1$. Thus, it is sufficient to evaluate the graph $I_{\alpha \beta}$ when $\alpha$ is antisymmetric ($\alpha = {\bf 8_a,10}$) and $\beta$ is symmetric ($\beta= {\bf 1,8_s,27,0}$).

\bi
\item{} $I_{\alpha \beta}$ for $\alpha = {\bf 8_a}$
\ei 
When $\alpha = {\bf 8_a}$, we replace the left blob of the graph $I_{\alpha \beta}$ by the corresponding $s$-channel projector ${\cal P}_{\bf a}$, 
\be
{\cal P}_{\bf a} = \frac{1}{N} \GGoctet(7,-40)  \ \, ,
\ee
leading to
\be
\label{I-ab-octet}
\alpha = {\bf 8_a} \ \Longrightarrow \  I_{\alpha \beta} = \frac{1}{N} \ \, \ggIOctet(\beta,26,-90) = \frac{K_\beta}{4 N} \, ( 2 N - C_\beta)^2 \, ,
\ee    
where we used the fact that each gluon between blue and red ones contributes a factor $(N-\frac{C_\beta}{2})$.

\bi
\item{} $I_{\alpha \beta}$ for $\alpha = {\bf 10}$
\ei 
For $\alpha = {\bf 10}$, we use completeness in the subspace of antisymmetric $s$-channel irreps,
\be
\label{proj10}
{\cal P}_{\bf 10} = \half \left( \GGidentity(9,-24)  - \GGcross(9,-24)  \right) - {\cal P}_{\bf a} \ \, .
\ee 
Replacing the left blob of the graph $I_{\alpha \beta}$ by \eq{proj10} we obtain
\bea
\alpha = {\bf 10} \ \Longrightarrow \  I_{\alpha \beta} &=& 
\frac{1}{2} \left( \ggIdBeta(\beta,26,-90) - \ggXBeta(\beta,26,-90) \right) - \frac{K_\beta}{4 N} \, ( 2 N - C_\beta)^2 \nn \\ 
&=& \frac{1}{2} \left( N K_\beta + (N-\frac{C_\beta}{2}) \, \sigma_{\beta} K_\beta \right) - \frac{K_\beta}{4 N} \, ( 2 N - C_\beta)^2 \nn \\ 
&=& \frac{K_\beta}{4 N} \, C_\beta \, (3 N - C_\beta) \, ,
\label{I-ab-decuplet}
\eea
where we used $\sigma_\beta = +1$ to go from the second to third line. 

Using \eq{I-ab-octet} and \eq{I-ab-decuplet} in \eq{Tt2minusTu2-2} we arrive at 
\be
\label{Tt2minusTu2-mat-elements-0}
\bra{\alpha} (T_t^2 - T_u^2) \ket{\beta} = - \frac{2 (1-\sigma_{\alpha} \sigma_{\beta})}{\sqrt{K_\alpha K_\beta}}  \left\{ \frac{K_\beta}{4 N}   \left[ \delta_{\alpha}^{\bf 8_{^a}} ( 2 N - C_\beta)^2 + \delta_{\alpha}^{\bf 10} \, C_\beta \, (3 N - C_\beta) \right] +  (\alpha \leftrightarrow \beta)  \right\} \, .
\ee
It is possible to express the matrix elements of $T_t^2 - T_u^2$ solely in terms of the invariants $\sigma_\alpha$, $C_\alpha$ and $K_\alpha$ (given in \eq{invariant-sigma}--\eq{invariant-dimension}). This can be done by expressing the Kronecker delta $\delta_{\alpha}^{\bf 8_{^a}}$ and $\delta_{\alpha}^{\bf 10}$ as\footnote{The last factors $\propto K_\alpha$ in \eq{Kronecker} are introduced in order to make the expression \eq{Tt2minusTu2-mat-elements} explicitly symmetric under $\alpha \leftrightarrow \beta$.}
\be
\label{Kronecker}
\delta_{\alpha}^{\bf 8_{^a}} = \frac{1-\sigma_{\alpha}}{2} \, \frac{2N-C_\alpha}{N} \, \frac{K_\alpha}{K_A} \ ; \ \ \ \delta_{\alpha}^{\bf 10} =  \frac{1-\sigma_{\alpha}}{2} \, \frac{C_\alpha - N}{N} \, \frac{K_\alpha}{K_{\bf 10}} \, , 
\ee
where we denote $K_A \equiv K_{\bf 8} = N^2-1$ the dimension of the adjoint representation, and $K_{\bf 10}$ is given in \eq{invariant-dimension}. Using \eq{Kronecker} in \eq{Tt2minusTu2-mat-elements-0} we obtain 
\bea
\bra{\alpha} (T_t^2 - T_u^2) \ket{\beta} = - \frac{4N}{K_A} \, \sqrt{K_\alpha K_\beta} \, \frac{1-\sigma_{\alpha}\sigma_{\beta}}{2} \times \hskip 5cm && \nn \\ 
\times \left\{ \frac{1-\sigma_{\alpha}}{2}  \left[ (2-\frac{C_\alpha}{N}) (1 - \frac{C_\beta}{2N})^2 + \frac{1}{N^2-4} (\frac{C_\alpha}{N} -1) \, \frac{C_\beta}{2N} \, (3 - \frac{C_\beta}{N}) \right] +  (\alpha \leftrightarrow \beta)  \right\} \, . \hskip 1cm &&
\label{Tt2minusTu2-mat-elements}
\eea

Using the signums \eq{invariant-sigma} and Casimirs \eq{invariant-casimir}, the operator $T_t^2 - T_u^2$ in the basis \eq{ortho-basis} thus reads, in matrix form,
\be
\label{Tt2minusTu2-ortho-basis}
T_t^2 - T_u^2 =  - \frac{4 N}{K_A} \mbox{\fontsize{8}{2}\selectfont $
 \left( 
\begin{array}{cc|cccc}
0 & 0 & \sqrt{K_1 K_3}  &  \frac{\sqrt{K_1 K_4}}{4}   &  \frac{\sqrt{K_1 K_5}}{N^2}   & \frac{\sqrt{K_1 K_6 }}{N^2} \\[2ex]
0 & 0 &  0   &  \frac{\sqrt{K_2 K_4} }{N^2-4}  &  \frac{(N+1) \sqrt{K_2 K_5} }{N^2(N+2)} & \frac{(N-1) \sqrt{K_2 K_6 }}{N^2(N-2)} \\
[2ex] \hline 
& & & & & \\
\sqrt{K_1 K_3} & 0 & 0 & 0 & 0 & 0 \\
[2ex] \frac{\sqrt{K_1 K_4} }{4} & \frac{\sqrt{K_2 K_4}}{N^2-4}    & 0 & 0 & 0 & 0 \\
[2ex] \frac{\sqrt{K_1 K_5}}{N^2} & \frac{(N+1) \sqrt{K_2 K_5}}{N^2(N+2)}  &0 &0 & 0 & 0 \\
[2ex] \frac{\sqrt{K_1 K_6 }}{N^2} & \frac{(N-1)\sqrt{K_2 K_6}}{N^2(N-2)} &0 &0 & 0& 0 
\end{array}
\right) $} \, .
\ee

Finally, the anomalous dimension matrix ${\cal Q}$ defined in \eq{Q-def} reads (use \eq{sumT}) 
\be
\label{eq:Qgggg}
\bra{\alpha} {\cal Q} \ket{\beta} =  \frac{4 N - C_\alpha}{2N}  \, \delta_{\alpha \beta} + \frac{X_t - X_u}{X_t + X_u} \, \frac{\bra{\alpha} (T_t^2 - T_u^2) \ket{\beta}}{2N} \, ,
\ee
which using \eq{Tt2minusTu2-ortho-basis} can be verified to agree with the results of Refs.~\cite{Dokshitzer:2005ek,Nikolaev:2005zj}.

\section{Recovering four-point ${\bar q}q {\bar q}q$ and $qq {\bar q}{\bar q}$ correlators}
\label{app:4-point}

In this appendix we quote the explicit analytical form of $e^{\B}$ (in the basis \eq{ortho-basis}) following from \eq{B-aq}, in the $\bar{q}q\to \bar{q}q$ and $qq \to qq$ cases. 

In order to exponentiate the matrix $\B$, it is convenient to decompose $\B$ as 
\be
\B = \hat{\B} + U \, \unit \ ; \ \ \ \hat{\B} \equiv \B - \frac{\tr{\B}}{\tr{\unit}} \, \unit \ ; \ \ \ U \equiv \frac{\tr{\B}}{\tr{\unit}} \, ,
\label{B-decompo}
\ee
where the matrix $\hat{\B}$ is traceless, and to introduce the average dimension and Casimir
\be
\bar{K} \equiv \frac{\sum K_{\alpha}}{\tr{\unit}} \ ; \ \ \ \bar{C} = \frac{\sum C_{\alpha}}{\tr{\unit}} \, ,
\label{ave-inv}
\ee
where $\tr{\unit} = \delta_{\alpha}^{\alpha}$ is the number of $s$-channel irreps ($\tr{\unit} = 2$ for $\bar{q}q$ and $qq$ pairs). Using \eq{B-aq} we obtain  
\bea
\bra{\alpha} \hat{\B} \ket{\beta} = \delta_{\alpha \beta} \left[ \frac{K_\alpha - \bar{K}}{2 K_a} (X_t - X_u) -  \frac{\bar{C} - C_\alpha}{2} X_u \right]  + (1-\delta_{\alpha \beta})  \frac{\sqrt{K_\alpha K_\beta}}{2 K_a} (X_t - X_u) \, , \hskip 8mm && \label{Bhat} \\ 
U = \frac{\bar{K}}{2 K_a} (X_t - X_u) - \half \left[ C_a X_1 + C_F X_2 + N (X_t - X_u) - \bar{C} X_u \right] \, . \hskip 20mm && \label{Ufact}
\eea

For $\bar{q}q \to \bar{q}q$ and $qq \to qq$, we have $K_a = N$, $C_a = C_F$ and the (symmetric traceless) $2 \times 2$ matrix $\hat{\B}$ of the form 
\be
\hat{\B} =  \mbox{\fontsize{10}{2}\selectfont $
\left( 
\begin{array}{cc}
d & c  \\[2ex]
c & -d  
\end{array}
\right) $}   \equiv \hat{\B}(c,d) 
\ee
is easily shown to exponentiate as
\be
e^{\hat{\B}(c,d)} = \mbox{\fontsize{10}{2}\selectfont $
\left( 
\begin{array}{cc}
\cosh{\sqrt{c^2+d^2}} + d \, \frac{\sinh{\sqrt{c^2+d^2}}}{\sqrt{c^2+d^2}}  & c \, \frac{\sinh{\sqrt{c^2+d^2}} }{\sqrt{c^2+d^2}}  \\[2ex]
c\, \frac{\sinh{\sqrt{c^2+d^2}} }{\sqrt{c^2+d^2}} & \cosh{\sqrt{c^2+d^2}} - d \, \frac{\sinh{\sqrt{c^2+d^2}}}{\sqrt{c^2+d^2}}   
\end{array}
\right) $} \, .
\label{expBhat}
\ee

For $\bar{q}q \to \bar{q}q$, ordering the $s$-channel orthonormal basis $\ket{\alpha}$ as $\left\{ \ket{\bf 1}, \ket{\bf 8} \right\}$ we have $K_\alpha = \left\{ 1, N^2-1\right\}$ and $C_\alpha = \left\{ 0, N \right\}$, and from \eq{Bhat}, \eq{Ufact}, \eq{expBhat} we obtain the ${\bar q}q {\bar q}q$ correlator 
\bea
&& \hskip 3cm e^{\B_{{\bar q}q {\bar q}q}} = e^U \, e^{\hat{\B}(c,d)} \label{eB-qbarq} \\
U &=& \left( \frac{N}{4} - C_F \right) (\hat{\Gamma}_{12} + \hat{\Gamma}_{34} + \hat{\Gamma}_{14} + \hat{\Gamma}_{23} ) + \left( C_F - \frac{N}{2} \right) (\hat{\Gamma}_{13} + \hat{\Gamma}_{24}) \, , \\
&& \hskip 5mm c = \frac{\sqrt{N^2-1}}{2N} (X_t -X_u) \ ; \ \ \ d = \frac{X_t -X_u}{2N} -\frac{N}{4} X_t \, , \\
&&  \hskip 12mm \sqrt{c^2+d^2} = \frac{1}{4} \sqrt{4 X_u (X_u -X_t) +N^2 X_t^2} \label{lambda-qbarq} \, ,
\eea
which can be checked to coincide with the result of Ref.~\cite{Fukushima:2007dy}.

For $qq \to qq$, in the $s$-channel basis $\left\{ \ket{\bf \bar{3}}, \ket{\bf 6} \right\}$ we have $K_\alpha = \left\{ \frac{N(N-1)}{2},  \frac{N(N+1)}{2} \right\}$ and $C_\alpha = \left\{ \frac{(N+1)(N-2)}{N}, \frac{(N-1)(N+2)}{N} \right\}$, leading to the $qq {\bar q}{\bar q}$ correlator 
\bea
&& \hskip 3cm e^{\B_{qq {\bar q}{\bar q}}} = e^U \, e^{\hat{\B}(c,d)}  \label{eB-qq} \\
U &=& \left( \frac{N}{4} - C_F \right) (\hat{\Gamma}_{13} + \hat{\Gamma}_{24} + \hat{\Gamma}_{14} + \hat{\Gamma}_{23} ) + \left( C_F - \frac{N}{2} \right) (\hat{\Gamma}_{12} + \hat{\Gamma}_{34}) \, , \\
&&  \hskip 5mm c = \frac{\sqrt{N^2-1}}{4} (X_t -X_u) \ ; \ \ \ d = - \frac{X_t + X_u}{4} \, , \\
&&  \hskip 12mm  \sqrt{c^2+d^2} = \frac{1}{4} \sqrt{4 X_t  X_u +N^2 (X_t-X_u)^2} \label{lambda-qq} \, .
\eea

The expressions \eq{eB-qbarq}--\eq{lambda-qbarq} and \eq{eB-qq}--\eq{lambda-qq} encode the broadening properties of a $q {\bar q}$ pair and of a diquark, respectively. 

Note that the $qq {\bar q}{\bar q}$ correlator can be obtained from the ${\bar q}q {\bar q}q$ correlator without any additional effort (simply using the expression \eq{B-aq}  and replacing the $s$-channel ${\rm SU}(N)$ invariants). We also remark that these two correlators are {\it not} simply related by $1 \leftrightarrow 4$ exchange. Although the {\it operators} $\B$ (defined by \eq{B-def}) corresponding to $\bar{q}q \to \bar{q}q$ and $qq \to qq$ are indeed related by $1 \leftrightarrow 4$ (as can be easily checked), the expressions \eq{eB-qbarq}--\eq{lambda-qbarq} and \eq{eB-qq}--\eq{lambda-qq} give the {\it matrices} of $e^{\B}$ in {\it different bases}, namely the $s$-channel bases $\ket{\alpha}_{\bar{q}q} \equiv \ket{\alpha} = \left\{ \ket{\bf 1}, \ket{\bf 8} \right\}$ and $\ket{\alpha}_{qq} \equiv \ket{\alpha'} =\left\{ \ket{\bf \bar{3}}, \ket{\bf 6} \right\}$ of the $\bar{q}q \to \bar{q}q$ and $qq \to qq$ processes, respectively.\footnote{Note that the $s$-channel basis of $qq \to qq$ is the $u$-channel basis of $\bar{q}q \to \bar{q}q$.} Thus, in order to obtain the matrix $e^\B$ in the $qq \to qq$ case from that in the $\bar{q}q \to \bar{q}q$ case, one must combine the $1 \leftrightarrow 4$ exchange with the change of basis from $\ket{\alpha}$ to $\ket{\alpha'}$, 
\be
\bra{\alpha'} e^{\B_{qq}} \ket{\beta'} = \bra{\alpha'} e^{\B_{\bar{q}q}(1 \leftrightarrow 4)} \ket{\beta'} = \sum_{\alpha, \beta} \, \langle \alpha' \vert \alpha \rangle \bra{\alpha} e^{\B_{\bar{q}q}(1 \leftrightarrow 4)} \ket{\beta} \langle \beta \vert \beta' \rangle \, .
\ee
In matrix form, this can be written as 
\be
\label{qq-qbarq-relation}
e^{\B_{qq}}  = M \cdot e^{\B_{\bar{q}q}(1 \leftrightarrow 4)} \cdot M^{-1} \, ,
\ee
where the matrices $e^\B$ are implicitly expressed in the $s$-channel basis of the corresponding process, and  the matrix elements of the transition matrix $M$ read
\be
M_{\alpha' \alpha} = \langle \alpha' \vert \alpha \rangle = \frac{1}{\sqrt{K_{\alpha} K_{\alpha'}}}  \ \OBBraQQ(\alpha',14,-55) \hskip -2mm \Crossing (14,-55) \hskip -7mm \OBKetQbarQ(\alpha,14,-55) \, .
\ee
Using pictorial representations~\cite{Cvitanovic:2008zz,Dokshitzer:1995fv} of the projectors ${\cal P}_{\bf \bar{3}}$, ${\cal P}_{\bf 6}$ and ${\cal P}_{\bf 1}$,  ${\cal P}_{\bf 8}$ we easily obtain
\be
\label{mat-M}
M = \mbox{\fontsize{8}{2}\selectfont $
\left( 
\begin{array}{cc}
- \sqrt{\frac{N-1}{2N}} &  \sqrt{\frac{N+1}{2N}}   \\[2ex]
 \sqrt{\frac{N+1}{2N}} &   \sqrt{\frac{N-1}{2N}} 
\end{array}
\right) $} = M^{-1} \, .
\ee
Using \eq{mat-M} we can check that the matrices $e^\B$ given by \eq{eB-qbarq}--\eq{lambda-qbarq} and \eq{eB-qq}--\eq{lambda-qq} satisfy the relation \eq{qq-qbarq-relation}. We note that $U$ and $\sqrt{c^2+d^2}$, being related respectively to the trace of $\B$ and eigenvalues $\lambda = \pm \sqrt{c^2+d^2}$ of $\hat{\B}$, are basis-independent, and thus simply related by $1 \leftrightarrow 4$ exchange (implying $X_t \leftrightarrow X_t -X_u$ and $X_u \leftrightarrow -X_u$, see \eq{Xt}--\eq{Xu}) when going from one case to the other. 

Finally, as mentioned in~\cite{Fukushima:2007dy}, the specific ${\bar q}q {\bar q}q$ 4-point correlator considered in~\cite{Blaizot:2004wv} corresponds to the situation where the $q\bar{q}$ pair arises from $g \to q \bar{q}$ and is thus in an initial color octet state $\ket{i} = \ket{\bf 8}$, and where a sum over final quark and antiquark color indices is performed, which amounts to consider the final state 
\be
\label{app:color-sum}
\ket{f}= \sum_{\alpha} \OBKetQbarQ(\alpha,12,-55) = \sum_{\alpha} \sqrt{K_{\alpha}} \, \ket{\alpha} = \sqrt{K_{\bf 1}} \, \ket{\bf 1} + \sqrt{K_{\bf 8}} \, \ket{\bf 8} \, .
\ee
The ${\bar q}q {\bar q}q$ correlator calculated in~\cite{Blaizot:2004wv} thus corresponds to the following linear combination of matrix elements of $e^\B$ given in \eq{eB-qbarq}--\eq{lambda-qbarq}, 
\be
\label{corr-BGV}
\bra{i} e^\B \ket{f} = \bra{\bf 8} e^\B \ket{\bf 1} + \sqrt{N^2-1} \  \bra{\bf 8} e^\B \ket{\bf 8} \, .
\ee

\section{Broadening of a parton pair produced in a hard process: examples}
\label{app:broadening-examples}

Here we briefly derive the $g +{\rm A}\to gg + {\rm X}$ cross section in the compact pair limit in two cases: when the final $gg$ pair is produced in the color state $\beta = {\bf 27}$, and when it is produced in the same color state as the incoming  state, $\beta = \alpha = {\bf 8_a}$. 

In the first case, only the first term in the bracket of \eq{a2ag-cross} contributes. In the compact pair limit, $|\pvec_1| \gg |\qvec|$, the integral is dominated by $|\xvec_{12}|, |\xvec_{34}| \ll |\xvec_{23}|$, and the matrix element $ \bra{\alpha} e^{\B(\xvec_1,\xvec_2;\xvec_3,\xvec_4)} \ket{\beta}$ can be expanded as in section~\ref{sec:color-transitions} using \eq{compact-linear1}. Using also \eq{X-compact} we arrive at
\bea
\frac{\dd \sigma(g +{\rm A}\to gg[{\bf 27}] + {\rm X})}{ S_{\perp} \dd z_1 \dd^2 \pvec_1 \dd^2 \qvec} = \alpha_s \Phi_g^{g}(z_1)  \, {\cal F}_{\alpha \beta} \frac{\bar{Q}^2}{4} \! \left[ \int \! \mbox{\fontsize{13}{2}\selectfont $ \frac{\dd^2 \xvec_{12} \, \dd^2 \xvec_{34}}{(2\pi)^4} \, e^{i \pvec_1 \cdot (\xvec_{12}+\xvec_{34})} \, \frac{(\xvec_{12} \cdot \xvec_{34})^2}{\xvec_{12}^2 \, \xvec_{34}^2} $} \right] \, &&  \nn \\
\times \int_0^1 \! \!  \dd \delta \int \! \frac{\dd^2 \xvec_{23}}{(2 \pi)^2} \log\left(\frac{1}{\mu^2 \xvec_{23}^2}\right) e^{i \qvec \xvec_{23}} \, e^{- \bar{C}_{\delta} \hat{\Gamma}_{23}}  \, , \hskip 10mm &&
\label{app:broad-27}
\eea
where $\bar{C}_{\delta} \equiv \delta C_{\alpha} + (1-\delta)  C_{\beta}$, and for $\alpha = {\bf 8_a}$ and $\beta = {\bf 27}$ we have ${\cal F}_{\alpha \beta} = \sqrt{\frac{N+3}{4(N+1)}}$ (see   \eq{case-2} and \eq{Tt2minusTu2-ortho-basis}). Using now the identity\footnote{\label{foot:J-integral}The integral \eq{J-integral} can be conveniently performed by replacing a single factor $(\xvec_{12} \cdot \xvec_{34})$ by the expression $\int \! \frac{\dd \phi_a}{\pi} \, (\xvec_{12} \cdot \hat{\avec}) \, (\xvec_{34} \cdot \hat{\avec})$, where $\hat{\avec}$ is a unit vector of azimutal angle $\phi_a$. Then \eq{J-integral} becomes $\frac{1}{(2 \pi)^2} \int \! \frac{\dd \phi_a}{\pi} [ \int \frac{\dd^2 \xvec}{2 \pi \xvec^2} \, e^{i \pvec_1 \cdot \xvec} \, (\xvec \cdot \hat{\avec}) \, \xvec]^2$. Then, use $\int \frac{\dd^2 \xvec}{2 \pi \xvec^2} \, e^{i \pvec_1 \cdot \xvec} (\xvec \cdot \hat{\avec}) \, \xvec =\nablavec_{\pvec_1} \int \frac{\dd^2 \xvec}{2 \pi i \, \xvec^2} \, e^{i \pvec_1 \cdot \xvec} \xvec \cdot \hat{\avec} = \nablavec_{\pvec_1} \frac{\pvec_1 \cdot \hat{\avec}}{\pvec_1^2}  = \frac{\hat{\avec} - 2 (\hat{\pvec}_1 \cdot \hat{\avec})  \hat{\pvec}_1}{\pvec_1^2}$.}
\be
\int \! \frac{\dd^2 \xvec_{12} \, \dd^2 \xvec_{34}}{(2\pi)^4} \, e^{i \pvec_1 \cdot (\xvec_{12}+\xvec_{34})} \, \frac{(\xvec_{12} \cdot \xvec_{34})^2}{\xvec_{12}^2 \, \xvec_{34}^2} = \frac{1}{2 \pi^2 |\pvec_1|^4} \, ,
\label{J-integral}
\ee
the expression \eq{app:broad-27} becomes 
\bea
\frac{\dd \sigma(g +{\rm A}\to gg[{\bf 27}] + {\rm X})}{S_{\perp} \dd z_1 \dd^2 \pvec_1 \dd^2 \qvec} =  \frac{\alpha_s \Phi_g^{g}(z_1)  {\cal F}_{\alpha \beta} \bar{Q}^2}{8 \pi^2 |\pvec_1|^4} \! \int_0^1 \! \! \dd \delta \! \int \! \frac{\dd^2 \xvec_{23}}{(2 \pi)^2}  \log\left(\frac{1}{\mu^2 \xvec_{23}^2}\right) e^{i \qvec \xvec_{23}} \, e^{- \bar{C}_{\delta} \hat{\Gamma}_{23}}   \, . \nn \\ 
\label{app:g2gg-cross-27}
\eea
The dependence on transverse momentum imbalance $\qvec$ is the same as for an asymptotic compact gluon pair undergoing a color transition, see \eq{fqL-compact}. Thus, $q_\perp$-broadening of a gluon pair produced in the color state $\beta = {\bf 27}$ scales as $(C_A+C_{27})/2$ (see the end of section~\ref{sec:color-transitions}). 

We now consider the case $\beta = \alpha = {\bf 8_a}$. We start from \eq{a2ag-cross}, where now the four terms in the bracket contribute. Expanding the operator $e^{\B(\xvec_1,\xvec_2;\xvec_3,\xvec_4)}$ in the compact pair limit using \eq{eq:compact-pair-expansion}, we see that the second term of \eq{eq:compact-pair-expansion} vanishes (since $\alpha = \beta$), and the third term is suppressed. The operator $e^{\B(\xvec_1,\xvec_2;\xvec_3,\xvec_4)}$ can thus be replaced by $e^{D(\xvec_1,\xvec_2;\xvec_3,\xvec_4)}$, where $D$ is the diagonal part of $\B$. The latter is obtained from \eq{B-gg} and we obtain 
\be
\label{F-def}
e^{D(\xvec_1,\xvec_2;\xvec_3,\xvec_4)} = \exp{\left\{ -\frac{N}{2} \left[ \hat{\Gamma}_{12} + \hat{\Gamma}_{34}  +\frac{1}{2} (\hat{\Gamma}_{14} + \hat{\Gamma}_{23} + \hat{\Gamma}_{13} + \hat{\Gamma}_{24})  \right] \right\}}  \equiv F(\xvec_{23}; \xvec_{12}, \xvec_{34}) \, ,
\ee
defined as a function of the three variables  $\xvec_{23}, \xvec_{12}, \xvec_{34}$. The bracket in \eq{a2ag-cross} is the combination ($z \equiv z_1$) 
\be
\label{F-comb}
F(\xvec_{23}; \xvec_{12}, \xvec_{34}) - F(\xvec_{23}+z\xvec_{12} ; \zerovec, \xvec_{34}) - F(\xvec_{23}+z\xvec_{34} ;  \xvec_{12}, \zerovec)  +  F(\xvec_{23}+z\xvec_{12} +z\xvec_{34} ; \zerovec, \zerovec) \, .
\ee
In the compact pair limit, $|\xvec_{12}|, |\xvec_{34}| \ll |\xvec_{23}|$, the Taylor expansion of \eq{F-comb} up to second order in $\xvec_{12}$ and $\xvec_{34}$ gives 
\be
\label{F-comb-expanded}
\left[ (\xvec_{12} \cdot \nablavec_2 - z \, \xvec_{12} \cdot \nablavec_1) \, (\xvec_{34} \cdot \nablavec_3 - z \, \xvec_{34} \cdot \nablavec_1)  F \right]_{\xvec_{12} = \xvec_{34} = \zerovec} \, ,
\ee
where $\nablavec_i$ denotes the gradient operator w.r.t.~the $i^{\rm th}$ argument of $F(\xvec_{23}; \xvec_{12}, \xvec_{34})$, \ie,  $\nablavec_1 = \nablavec_{\xvec_{_{23}}}$, $\nablavec_2 = \nablavec_{\xvec_{_{12}}}$, $\nablavec_3 = \nablavec_{\xvec_{_{34}}}$. The calculation of \eq{F-comb-expanded} with the function $F$ defined by \eq{F-def} yields
\be
\label{F-comb-expanded-2}
- N e^{-N \hat{\Gamma}_{23}} (1-2z)^2 \left\{ \xvec_{12} \cdot \xvec_{23} \,  \xvec_{34} \cdot \xvec_{23} \left[ \hat{\Gamma}''(\xvec_{23}^2) - N \, \hat{\Gamma}'(\xvec_{23}^2)^2 \right] + \xvec_{12} \cdot \xvec_{34} \frac{\hat{\Gamma}'(\xvec_{23}^2)}{2} \right\} \, .
\ee
Now we insert \eq{F-comb-expanded-2} in \eq{a2ag-cross} and integrate over $\xvec_{12}$ and $\xvec_{34}$ using \eq{J-integral} and the identity
\be
\int \! \frac{\dd^2 \xvec_{12} \, \dd^2 \xvec_{34}}{(2\pi)^4} \, e^{i \pvec_1 \cdot (\xvec_{12}+\xvec_{34})} \, \frac{\xvec_{12} \cdot \xvec_{34}}{\xvec_{12}^2 \, \xvec_{34}^2} \, (\xvec_{12} \cdot \hat{\xvec}_{23}) \, (\xvec_{34} \cdot \hat{\xvec}_{23})= \frac{1}{4 \pi^2 |\pvec_1|^4} \, .
\label{K-integral}
\ee
We arrive at
\bea
\frac{\dd \sigma(g +{\rm A}\to gg[{\bf 8_a}] + {\rm X})}{S_{\perp} \dd z_1 \dd^2 \pvec_1 \dd^2 \qvec} =  \frac{\alpha_s \Phi_g^{g}(z_1)}{16 \pi^2 |\pvec_1|^4} (1-2z)^2  \! \int \! \frac{\dd^2 \xvec_{23}}{(2 \pi)^2} \, e^{i \qvec \cdot \xvec_{23}} \times && \nn \\  
\times 4 N \left\{ \xvec_{23}^2 \left[ \hat{\Gamma}''(\xvec_{23}^2) - N \, \hat{\Gamma}'(\xvec_{23}^2)^2 \right]  + \hat{\Gamma}'(\xvec_{23}^2) \right\} e^{- N \hat{\Gamma}_{23}} &&   \, . 
\label{app:g2gg-cross-8}
\eea
The second line of \eq{app:g2gg-cross-8} simplifies to $-\nablavec_1^2 \, e^{- N \hat{\Gamma}_{23}}$, and integrating by parts we finally obtain
\be
\frac{\dd \sigma(g +{\rm A}\to gg[{\bf 8_a}] + {\rm X})}{S_{\perp} \dd z_1 \dd^2 \pvec_1 \dd^2 \qvec} =  \frac{\alpha_s \Phi_g^{g}(z_1)}{16 \pi^2 |\pvec_1|^4} (1-2z)^2 \, \qvec^2 \int \! \frac{\dd^2 \xvec_{23}}{(2 \pi)^2} \, e^{i \qvec \cdot \xvec_{23}} \, e^{- N\hat{\Gamma}_{23}} \, ,
\label{app:g2gg-cross-8-final}
\ee
exhibiting a $q_\perp$-broadening similar to that of a gluon, see \eq{f-density}.


\begin{thebibliography}{10}

\bibitem{Collins:1989gx}
J.~C. Collins, D.~E. Soper and G.~F. Sterman, \emph{{Factorization of Hard
  Processes in QCD}},
  \href{https://doi.org/10.1142/9789814503266_0001}{\emph{Adv. Ser. Direct.
  High Energy Phys.} {\bfseries 5} (1989) 1--91},
  [\href{https://arxiv.org/abs/hep-ph/0409313}{{\ttfamily hep-ph/0409313}}].

\bibitem{Nikolaev:1990ja}
N.~N. Nikolaev and B.~G. Zakharov, \emph{{Color transparency and scaling
  properties of nuclear shadowing in deep inelastic scattering}},
  \href{https://doi.org/10.1007/BF01483577}{\emph{Z. Phys.} {\bfseries C49}
  (1991) 607--618}.

\bibitem{Mueller:1999wm}
A.~H. Mueller, \emph{{Parton saturation at small x and in large nuclei}},
  \href{https://doi.org/10.1016/S0550-3213(99)00394-6}{\emph{Nucl. Phys.}
  {\bfseries B558} (1999) 285--303},
  [\href{https://arxiv.org/abs/hep-ph/9904404}{{\ttfamily hep-ph/9904404}}].

\bibitem{Mueller:2001fv}
A.~H. Mueller, \emph{{Parton saturation: An Overview}},  in \emph{{QCD
  perspectives on hot and dense matter. Proceedings, NATO Advanced Study
  Institute, Summer School, Cargese, France, August 6-18, 2001}}, pp.~45--72,
  2001, \href{https://arxiv.org/abs/hep-ph/0111244}{{\ttfamily
  hep-ph/0111244}}.

\bibitem{Iancu:2003xm}
E.~Iancu and R.~Venugopalan, \emph{{The Color glass condensate and high-energy
  scattering in QCD}}.
\newblock In *Hwa, R.C. (ed.) et al.: Quark gluon plasma* 249-3363, 2003,
  \href{https://doi.org/10.1142/9789812795533-0005}{10.1142/9789812795533-0005}.

\bibitem{Gelis:2010nm}
F.~Gelis, E.~Iancu, J.~Jalilian-Marian and R.~Venugopalan, \emph{{The Color
  Glass Condensate}},
  \href{https://doi.org/10.1146/annurev.nucl.010909.083629}{\emph{Ann. Rev.
  Nucl. Part. Sci.} {\bfseries 60} (2010) 463--489},
  [\href{https://arxiv.org/abs/1002.0333}{{\ttfamily 1002.0333}}].

\bibitem{Albacete:2014fwa}
J.~L. Albacete and C.~Marquet, \emph{{Gluon saturation and initial conditions
  for relativistic heavy ion collisions}},
  \href{https://doi.org/10.1016/j.ppnp.2014.01.004}{\emph{Prog. Part. Nucl.
  Phys.} {\bfseries 76} (2014) 1--42},
  [\href{https://arxiv.org/abs/1401.4866}{{\ttfamily 1401.4866}}].

\bibitem{Armesto:2006ph}
N.~Armesto, \emph{{Nuclear shadowing}},
  \href{https://doi.org/10.1088/0954-3899/32/11/R01}{\emph{J. Phys.} {\bfseries
  G32} (2006) R367--R394},
  [\href{https://arxiv.org/abs/hep-ph/0604108}{{\ttfamily hep-ph/0604108}}].

\bibitem{Zakharov:1996fv}
B.~G. Zakharov, \emph{{Fully quantum treatment of the Landau-Pomeranchuk-Migdal
  effect in QED and QCD}}, \href{https://doi.org/10.1134/1.567126}{\emph{JETP
  Lett.} {\bfseries 63} (1996) 952--957},
  [\href{https://arxiv.org/abs/hep-ph/9607440}{{\ttfamily hep-ph/9607440}}].

\bibitem{Zakharov:1997uu}
B.~G. Zakharov, \emph{{Radiative energy loss of high-energy quarks in finite
  size nuclear matter and quark - gluon plasma}},
  \href{https://doi.org/10.1134/1.567389}{\emph{JETP Lett.} {\bfseries 65}
  (1997) 615--620}, [\href{https://arxiv.org/abs/hep-ph/9704255}{{\ttfamily
  hep-ph/9704255}}].

\bibitem{Baier:1996kr}
R.~Baier, Y.~L. Dokshitzer, A.~H. Mueller, S.~Peign\'e and D.~Schiff,
  \emph{{Radiative energy loss of high-energy quarks and gluons in a finite
  volume quark-gluon plasma}},
  \href{https://doi.org/10.1016/S0550-3213(96)00553-6}{\emph{Nucl. Phys.}
  {\bfseries B483} (1997) 291--320},
  [\href{https://arxiv.org/abs/hep-ph/9607355}{{\ttfamily hep-ph/9607355}}].

\bibitem{Baier:1996sk}
R.~Baier, Y.~L. Dokshitzer, A.~H. Mueller, S.~Peign\'e and D.~Schiff,
  \emph{{Radiative energy loss and $p_T$-broadening of high-energy partons in
  nuclei}}, \href{https://doi.org/10.1016/S0550-3213(96)00581-0}{\emph{Nucl.
  Phys.} {\bfseries B484} (1997) 265--282},
  [\href{https://arxiv.org/abs/hep-ph/9608322}{{\ttfamily hep-ph/9608322}}].

\bibitem{Gyulassy:1999zd}
M.~Gyulassy, P.~Levai and I.~Vitev, \emph{{Jet quenching in thin quark gluon
  plasmas. 1. Formalism}},
  \href{https://doi.org/10.1016/S0550-3213(99)00713-0}{\emph{Nucl. Phys.}
  {\bfseries B571} (2000) 197--233},
  [\href{https://arxiv.org/abs/hep-ph/9907461}{{\ttfamily hep-ph/9907461}}].

\bibitem{Gyulassy:2000er}
M.~Gyulassy, P.~Levai and I.~Vitev, \emph{{Reaction operator approach to
  non-Abelian energy loss}},
  \href{https://doi.org/10.1016/S0550-3213(00)00652-0}{\emph{Nucl. Phys.}
  {\bfseries B594} (2001) 371--419},
  [\href{https://arxiv.org/abs/nucl-th/0006010}{{\ttfamily nucl-th/0006010}}].

\bibitem{Kovner:2003zj}
A.~Kovner and U.~A. Wiedemann, \emph{{Gluon radiation and parton energy loss}},
   \href{https://arxiv.org/abs/hep-ph/0304151}{{\ttfamily hep-ph/0304151}}.

\bibitem{Arleo:2010rb}
F.~Arleo, S.~Peign\'e and T.~Sami, \emph{{Revisiting scaling properties of
  medium-induced gluon radiation}},
  \href{https://doi.org/10.1103/PhysRevD.83.114036}{\emph{Phys. Rev.}
  {\bfseries D83} (2011) 114036},
  [\href{https://arxiv.org/abs/1006.0818}{{\ttfamily 1006.0818}}].
  
  \bibitem{Armesto:2012qa}
N.~Armesto, H.~Ma, M.~Martinez, Y.~Mehtar-Tani and C.~A. Salgado,
  \emph{{Interference between initial and final state radiation in a QCD
  medium}}, \href{https://doi.org/10.1016/j.physletb.2012.09.039}{\emph{Phys.
  Lett.} {\bfseries B717} (2012) 280--286},
  [\href{https://arxiv.org/abs/1207.0984}{{\ttfamily 1207.0984}}].

\bibitem{Armesto:2013fca}
N.~Armesto, H.~Ma, M.~Martinez, Y.~Mehtar-Tani and C.~A. Salgado,
  \emph{{Coherence Phenomena between Initial and Final State Radiation in a
  Dense QCD Medium}},
  \href{https://doi.org/10.1007/JHEP12(2013)052}{\emph{JHEP} {\bfseries 12}
  (2013) 052}, [\href{https://arxiv.org/abs/1308.2186}{{\ttfamily 1308.2186}}].

\bibitem{Peigne:2014uha}
S.~Peign\'e, F.~Arleo and R.~Kolevatov, \emph{{Coherent medium-induced gluon
  radiation in hard forward $1 \to 1$ partonic processes}},
  \href{https://doi.org/10.1103/PhysRevD.93.014006}{\emph{Phys. Rev.}
  {\bfseries D93} (2016) 014006},
  [\href{https://arxiv.org/abs/1402.1671}{{\ttfamily 1402.1671}}].

\bibitem{Peigne:2014rka}
S.~Peign\'e and R.~Kolevatov, \emph{{Medium-induced soft gluon radiation in
  forward dijet production in relativistic proton-nucleus collisions}},
  \href{https://doi.org/10.1007/JHEP01(2015)141}{\emph{JHEP} {\bfseries 01}
  (2015) 141}, [\href{https://arxiv.org/abs/1405.4241}{{\ttfamily 1405.4241}}].

\bibitem{Qiu:1990xxa}
J.-w. Qiu and G.~F. Sterman, \emph{{Power corrections in hadronic scattering.
  1. Leading 1/Q**2 corrections to the Drell-Yan cross-section}},
  \href{https://doi.org/10.1016/0550-3213(91)90503-P}{\emph{Nucl. Phys.}
  {\bfseries B353} (1991) 105--136}.

\bibitem{Qiu:1990xy}
J.-w. Qiu and G.~F. Sterman, \emph{{Power corrections to hadronic scattering.
  2. Factorization}},
  \href{https://doi.org/10.1016/0550-3213(91)90504-Q}{\emph{Nucl. Phys.}
  {\bfseries B353} (1991) 137--164}.

\bibitem{Qiu:2001hj}
J.-w. Qiu and G.~F. Sterman, \emph{{QCD and rescattering in nuclear targets}},
  \href{https://doi.org/10.1142/S0218301303001235}{\emph{Int. J. Mod. Phys.}
  {\bfseries E12} (2003) 149},
  [\href{https://arxiv.org/abs/hep-ph/0111002}{{\ttfamily hep-ph/0111002}}].

\bibitem{Luo:1993ui}
M.~Luo, J.-w. Qiu and G.~F. Sterman, \emph{{Twist four nuclear parton
  distributions from photoproduction}},
  \href{https://doi.org/10.1103/PhysRevD.49.4493}{\emph{Phys. Rev.} {\bfseries
  D49} (1994) 4493--4502}.

\bibitem{Wang:2001ifa}
X.-N. Wang and X.-f. Guo, \emph{{Multiple parton scattering in nuclei: Parton
  energy loss}},
  \href{https://doi.org/10.1016/S0375-9474(01)01130-7}{\emph{Nucl. Phys.}
  {\bfseries A696} (2001) 788--832},
  [\href{https://arxiv.org/abs/hep-ph/0102230}{{\ttfamily hep-ph/0102230}}].

\bibitem{Majumder:2007hx}
A.~Majumder and B.~M\"uller, \emph{{Higher twist jet broadening and classical
  propagation}}, \href{https://doi.org/10.1103/PhysRevC.77.054903}{\emph{Phys.
  Rev.} {\bfseries C77} (2008) 054903},
  [\href{https://arxiv.org/abs/0705.1147}{{\ttfamily 0705.1147}}].

\bibitem{Majumder:2007ne}
A.~Majumder, R.~J. Fries and B.~M\"uller, \emph{{Photon bremsstrahlung and
  diffusive broadening of a hard jet}},
  \href{https://doi.org/10.1103/PhysRevC.77.065209}{\emph{Phys. Rev.}
  {\bfseries C77} (2008) 065209},
  [\href{https://arxiv.org/abs/0711.2475}{{\ttfamily 0711.2475}}].

\bibitem{Mueller:2012bn}
A.~H. Mueller and S.~Munier, \emph{{$p_{\perp}$-broadening and production
  processes versus dipole/quadrupole amplitudes at next-to-leading order}},
  \href{https://doi.org/10.1016/j.nuclphysa.2012.08.005}{\emph{Nucl. Phys.}
  {\bfseries A893} (2012) 43--86},
  [\href{https://arxiv.org/abs/1206.1333}{{\ttfamily 1206.1333}}].

\bibitem{Liou:2014rha}
T.~Liou and A.~H. Mueller, \emph{{Parton energy loss in high energy hard
  forward processes in proton-nucleus collisions}},
  \href{https://doi.org/10.1103/PhysRevD.89.074026}{\emph{Phys. Rev.}
  {\bfseries D89} (2014) 074026},
  [\href{https://arxiv.org/abs/1402.1647}{{\ttfamily 1402.1647}}].

\bibitem{Munier:2016oih}
S.~Munier, S.~Peign\'e and E.~Petreska, \emph{{Medium-induced gluon radiation in
  hard forward parton scattering in the saturation formalism}},
  \href{https://doi.org/10.1103/PhysRevD.95.014014}{\emph{Phys. Rev.}
  {\bfseries D95} (2017) 014014},
  [\href{https://arxiv.org/abs/1603.01028}{{\ttfamily 1603.01028}}].

\bibitem{Accardi:2009qv}
A.~Accardi, F.~Arleo, W.~K. Brooks, D.~D'Enterria and V.~Muccifora,
  \emph{{Parton Propagation and Fragmentation in QCD Matter}},
  \href{https://doi.org/10.1393/ncr/i2009-10048-0}{\emph{Riv. Nuovo Cim.}
  {\bfseries 32} (2010) 439--553},
  [\href{https://arxiv.org/abs/0907.3534}{{\ttfamily 0907.3534}}].

\bibitem{Blaizot:2013vha}
J.-P. Blaizot, F.~Dominguez, E.~Iancu and Y.~Mehtar-Tani, \emph{{Probabilistic
  picture for medium-induced jet evolution}},
  \href{https://doi.org/10.1007/JHEP06(2014)075}{\emph{JHEP} {\bfseries 06}
  (2014) 075}, [\href{https://arxiv.org/abs/1311.5823}{{\ttfamily 1311.5823}}].

\bibitem{Liou:2013qya}
T.~Liou, A.~H. Mueller and B.~Wu, \emph{{Radiative $p_\bot$-broadening of
  high-energy quarks and gluons in QCD matter}},
  \href{https://doi.org/10.1016/j.nuclphysa.2013.08.005}{\emph{Nucl. Phys.}
  {\bfseries A916} (2013) 102--125},
  [\href{https://arxiv.org/abs/1304.7677}{{\ttfamily 1304.7677}}].

\bibitem{Kang:2008us}
Z.-B. Kang and J.-W. Qiu, \emph{{Transverse momentum broadening of vector boson
  production in high energy nuclear collisions}},
  \href{https://doi.org/10.1103/PhysRevD.77.114027}{\emph{Phys. Rev.}
  {\bfseries D77} (2008) 114027},
  [\href{https://arxiv.org/abs/0802.2904}{{\ttfamily 0802.2904}}].

\bibitem{Johnson:2006wi}
M.~B. Johnson, B.~Z. Kopeliovich, M.~J. Leitch, P.~L. McGaughey, J.~M. Moss,
  I.~K. Potashnikova et~al., \emph{{Nuclear broadening of transverse momentum
  in Drell-Yan reactions}},
  \href{https://doi.org/10.1103/PhysRevC.75.035206}{\emph{Phys. Rev.}
  {\bfseries C75} (2007) 035206},
  [\href{https://arxiv.org/abs/hep-ph/0606126}{{\ttfamily hep-ph/0606126}}].

\bibitem{Albacete:2010sy}
J.~L. Albacete, N.~Armesto, J.~G. Milhano, P.~Quiroga-Arias and C.~A. Salgado,
  \emph{{AAMQS: A non-linear QCD analysis of new HERA data at small-x including
  heavy quarks}},
  \href{https://doi.org/10.1140/epjc/s10052-011-1705-3}{\emph{Eur. Phys. J.}
  {\bfseries C71} (2011) 1705},
  [\href{https://arxiv.org/abs/1012.4408}{{\ttfamily 1012.4408}}].

\bibitem{Albacete:2010bs}
J.~L. Albacete and C.~Marquet, \emph{{Single Inclusive Hadron Production at
  RHIC and the LHC from the Color Glass Condensate}},
  \href{https://doi.org/10.1016/j.physletb.2010.02.073}{\emph{Phys. Lett.}
  {\bfseries B687} (2010) 174--179},
  [\href{https://arxiv.org/abs/1001.1378}{{\ttfamily 1001.1378}}].

\bibitem{Albacete:2010pg}
J.~L. Albacete and C.~Marquet, \emph{{Azimuthal correlations of forward
  di-hadrons in d+Au collisions at RHIC in the Color Glass Condensate}},
  \href{https://doi.org/10.1103/PhysRevLett.105.162301}{\emph{Phys. Rev. Lett.}
  {\bfseries 105} (2010) 162301},
  [\href{https://arxiv.org/abs/1005.4065}{{\ttfamily 1005.4065}}].

\bibitem{Kang:2016ron}
Z.-B. Kang, J.-W. Qiu, X.-N. Wang and H.~Xing, \emph{{Next-to-leading order
  transverse momentum broadening for Drell-Yan production in p+A collisions}},
  \href{https://doi.org/10.1103/PhysRevD.94.074038}{\emph{Phys. Rev.}
  {\bfseries D94} (2016) 074038},
  [\href{https://arxiv.org/abs/1605.07175}{{\ttfamily 1605.07175}}].

\bibitem{Kopeliovich:2010aa}
B.~Z. Kopeliovich, I.~K. Potashnikova and I.~Schmidt, \emph{{Measuring the
  saturation scale in nuclei}},
  \href{https://doi.org/10.1103/PhysRevC.81.035204}{\emph{Phys. Rev.}
  {\bfseries C81} (2010) 035204},
  [\href{https://arxiv.org/abs/1001.4281}{{\ttfamily 1001.4281}}].

\bibitem{Arleo:2012rs}
F.~Arleo and S.~Peign\'e, \emph{{Heavy-quarkonium suppression in p-A collisions
  from parton energy loss in cold QCD matter}},
  \href{https://doi.org/10.1007/JHEP03(2013)122}{\emph{JHEP} {\bfseries 03}
  (2013) 122}, [\href{https://arxiv.org/abs/1212.0434}{{\ttfamily 1212.0434}}].

\bibitem{Arleo:2013zua}
F.~Arleo, R.~Kolevatov, S.~Peign\'e and M.~Rustamova, \emph{{Centrality and pT
  dependence of J/psi suppression in proton-nucleus collisions from parton
  energy loss}}, \href{https://doi.org/10.1007/JHEP05(2013)155}{\emph{JHEP}
  {\bfseries 05} (2013) 155},
  [\href{https://arxiv.org/abs/1304.0901}{{\ttfamily 1304.0901}}].

\bibitem{Basso:2016ulb}
E.~Basso, V.~P. Goncalves, M.~Krelina, J.~Nemchik and R.~Pasechnik,
  \emph{{Nuclear effects in Drell-Yan pair production in high-energy $pA$
  collisions}}, \href{https://doi.org/10.1103/PhysRevD.93.094027}{\emph{Phys.
  Rev.} {\bfseries D93} (2016) 094027},
  [\href{https://arxiv.org/abs/1603.01893}{{\ttfamily 1603.01893}}].

\bibitem{Gelis:2001da}
F.~Gelis and A.~Peshier, \emph{{Probing colored glass via $q {\bar q}$
  photoproduction}},
  \href{https://doi.org/10.1016/S0375-9474(01)01264-7}{\emph{Nucl. Phys.}
  {\bfseries A697} (2002) 879--901},
  [\href{https://arxiv.org/abs/hep-ph/0107142}{{\ttfamily hep-ph/0107142}}].

\bibitem{JalilianMarian:2004da}
J.~Jalilian-Marian and Y.~V. Kovchegov, \emph{{Inclusive two-gluon and valence
  quark-gluon production in DIS and pA}},
  \href{https://doi.org/10.1103/PhysRevD.71.079901,
  10.1103/PhysRevD.70.114017}{\emph{Phys. Rev.} {\bfseries D70} (2004) 114017},
  [\href{https://arxiv.org/abs/hep-ph/0405266}{{\ttfamily hep-ph/0405266}}].

\bibitem{Blaizot:2004wv}
J.~P. Blaizot, F.~Gelis and R.~Venugopalan, \emph{{High-energy pA collisions in
  the color glass condensate approach. 2. Quark production}},
  \href{https://doi.org/10.1016/j.nuclphysa.2004.07.006}{\emph{Nucl. Phys.}
  {\bfseries A743} (2004) 57--91},
  [\href{https://arxiv.org/abs/hep-ph/0402257}{{\ttfamily hep-ph/0402257}}].

\bibitem{Qiu:2004da}
J.-w. Qiu and I.~Vitev, \emph{{Coherent QCD multiple scattering in
  proton-nucleus collisions}},
  \href{https://doi.org/10.1016/j.physletb.2005.10.073}{\emph{Phys. Lett.}
  {\bfseries B632} (2006) 507--511},
  [\href{https://arxiv.org/abs/hep-ph/0405068}{{\ttfamily hep-ph/0405068}}].

\bibitem{Kharzeev:2004bw}
D.~Kharzeev, E.~Levin and L.~McLerran, \emph{{Jet azimuthal correlations and
  parton saturation in the color glass condensate}},
  \href{https://doi.org/10.1016/j.nuclphysa.2004.10.031}{\emph{Nucl. Phys.}
  {\bfseries A748} (2005) 627--640},
  [\href{https://arxiv.org/abs/hep-ph/0403271}{{\ttfamily hep-ph/0403271}}].

\bibitem{Baier:2005dv}
R.~Baier, A.~Kovner, M.~Nardi and U.~A. Wiedemann, \emph{{Particle correlations
  in saturated QCD matter}},
  \href{https://doi.org/10.1103/PhysRevD.72.094013}{\emph{Phys. Rev.}
  {\bfseries D72} (2005) 094013},
  [\href{https://arxiv.org/abs/hep-ph/0506126}{{\ttfamily hep-ph/0506126}}].

\bibitem{Marquet:2007vb}
C.~Marquet, \emph{{Forward inclusive dijet production and azimuthal
  correlations in pA collisions}},
  \href{https://doi.org/10.1016/j.nuclphysa.2007.09.001}{\emph{Nucl. Phys.}
  {\bfseries A796} (2007) 41--60},
  [\href{https://arxiv.org/abs/0708.0231}{{\ttfamily 0708.0231}}].

\bibitem{Tuchin:2009nf}
K.~Tuchin, \emph{{Rapidity and centrality dependence of azimuthal correlations
  in Deuteron-Gold collisions at RHIC}},
  \href{https://doi.org/10.1016/j.nuclphysa.2010.06.001}{\emph{Nucl. Phys.}
  {\bfseries A846} (2010) 83--94},
  [\href{https://arxiv.org/abs/0912.5479}{{\ttfamily 0912.5479}}].

\bibitem{Stasto:2011ru}
A.~Stasto, B.-W. Xiao and F.~Yuan, \emph{{Back-to-Back Correlations of
  Di-hadrons in dAu Collisions at RHIC}},
  \href{https://doi.org/10.1016/j.physletb.2012.08.044}{\emph{Phys. Lett.}
  {\bfseries B716} (2012) 430--434},
  [\href{https://arxiv.org/abs/1109.1817}{{\ttfamily 1109.1817}}].

\bibitem{Kang:2011bp}
Z.-B. Kang, I.~Vitev and H.~Xing, \emph{{Dihadron momentum imbalance and
  correlations in d+Au collisions}},
  \href{https://doi.org/10.1103/PhysRevD.85.054024}{\emph{Phys. Rev.}
  {\bfseries D85} (2012) 054024},
  [\href{https://arxiv.org/abs/1112.6021}{{\ttfamily 1112.6021}}].

\bibitem{Lappi:2012nh}
T.~Lappi and H.~M\"antysaari, \emph{{Forward dihadron correlations in
  deuteron-gold collisions with the Gaussian approximation of JIMWLK}},
  \href{https://doi.org/10.1016/j.nuclphysa.2013.03.017}{\emph{Nucl. Phys.}
  {\bfseries A908} (2013) 51--72},
  [\href{https://arxiv.org/abs/1209.2853}{{\ttfamily 1209.2853}}].

\bibitem{Nikolaev:2005qs}
N.~N. Nikolaev, W.~Sch\"afer and B.~G. Zakharov, \emph{{Nonuniversality aspects
  of nonlinear $k_\perp$-factorization for hard dijets}},
  \href{https://doi.org/10.1103/PhysRevLett.95.221803}{\emph{Phys. Rev. Lett.}
  {\bfseries 95} (2005) 221803},
  [\href{https://arxiv.org/abs/hep-ph/0502018}{{\ttfamily hep-ph/0502018}}].

\bibitem{Nikolaev:2005dd}
N.~N. Nikolaev, W.~Sch\"afer, B.~G. Zakharov and V.~R. Zoller, \emph{{Nonlinear
  $k_\perp$-factorization for quark-gluon dijet production off nuclei}},
  \href{https://doi.org/10.1103/PhysRevD.72.034033}{\emph{Phys. Rev.}
  {\bfseries D72} (2005) 034033},
  [\href{https://arxiv.org/abs/hep-ph/0504057}{{\ttfamily hep-ph/0504057}}].

\bibitem{Nikolaev:2005zj}
N.~N. Nikolaev, W.~Sch\"afer and B.~G. Zakharov, \emph{{Nonlinear
  $k_\perp$-factorization for gluon-gluon dijets produced off nuclear
  targets}}, \href{https://doi.org/10.1103/PhysRevD.72.114018}{\emph{Phys.
  Rev.} {\bfseries D72} (2005) 114018},
  [\href{https://arxiv.org/abs/hep-ph/0508310}{{\ttfamily hep-ph/0508310}}].

\bibitem{McLerran:1993ni}
L.~D. McLerran and R.~Venugopalan, \emph{{Computing quark and gluon
  distribution functions for very large nuclei}},
  \href{https://doi.org/10.1103/PhysRevD.49.2233}{\emph{Phys. Rev.} {\bfseries
  D49} (1994) 2233--2241},
  [\href{https://arxiv.org/abs/hep-ph/9309289}{{\ttfamily hep-ph/9309289}}].

\bibitem{McLerran:1998nk}
L.~D. McLerran and R.~Venugopalan, \emph{{Fock space distributions, structure
  functions, higher twists and small x}},
  \href{https://doi.org/10.1103/PhysRevD.59.094002}{\emph{Phys. Rev.}
  {\bfseries D59} (1999) 094002},
  [\href{https://arxiv.org/abs/hep-ph/9809427}{{\ttfamily hep-ph/9809427}}].

\bibitem{Peigne:2008wu}
S.~Peign\'e and A.~V. Smilga, \emph{{Energy losses in a hot plasma revisited}},
  \href{https://doi.org/10.3367/UFNe.0179.200907a.0697}{\emph{Phys. Usp.}
  {\bfseries 52} (2009) 659--685},
  [\href{https://arxiv.org/abs/0810.5702}{{\ttfamily 0810.5702}}].

\bibitem{Cvitanovic:2008zz}
P.~Cvitanovi\'c, \emph{{Group theory: Birdtracks, Lie's and exceptional groups}}.
\newblock Princeton, USA, Univ. Pr. (2008) 273 p, 2008.

\bibitem{Dokshitzer:1995fv}
{\relax Yu}.~L. Dokshitzer, \emph{{Perturbative QCD (and beyond)}},
  \href{https://doi.org/10.1007/BFb0105858}{\emph{Lect. Notes Phys.} {\bfseries
  496} (1997) 87--135}.

\bibitem{Keppeler:2017kwt}
S.~Keppeler, \emph{{Birdtracks for SU(N)}},  in \emph{{QCD Master Class 2017
  Saint-Jacut-de-la-Mer, France, June 18-24, 2017}}, 2017,
  \href{https://arxiv.org/abs/1707.07280}{{\ttfamily 1707.07280}},
  \href{https://inspirehep.net/record/1611314/files/arXiv:1707.07280.pdf}{https://inspirehep.net/record/1611314/files/arXiv:1707.07280.pdf}.

\bibitem{Keppeler:2012ih}
S.~Keppeler and M.~Sj\"odahl, \emph{{Orthogonal multiplet bases in SU(Nc) color
  space}}, \href{https://doi.org/10.1007/JHEP09(2012)124}{\emph{JHEP}
  {\bfseries 09} (2012) 124},
  [\href{https://arxiv.org/abs/1207.0609}{{\ttfamily 1207.0609}}].

\bibitem{Baier:1998kq}
R.~Baier, Y.~L. Dokshitzer, A.~H. Mueller and D.~Schiff, \emph{{Medium induced
  radiative energy loss: Equivalence between the BDMPS and Zakharov
  formalisms}},
  \href{https://doi.org/10.1016/S0550-3213(98)00546-X}{\emph{Nucl. Phys.}
  {\bfseries B531} (1998) 403--425},
  [\href{https://arxiv.org/abs/hep-ph/9804212}{{\ttfamily hep-ph/9804212}}].

\bibitem{Dokshitzer:2005ek}
{\relax Yu}.~L. Dokshitzer and G.~Marchesini, \emph{{Hadron collisions and the
  fifth form-factor}},
  \href{https://doi.org/10.1016/j.physletb.2005.10.009}{\emph{Phys. Lett.}
  {\bfseries B631} (2005) 118--125},
  [\href{https://arxiv.org/abs/hep-ph/0508130}{{\ttfamily hep-ph/0508130}}].

\bibitem{Botts:1989kf}
J.~Botts and G.~F. Sterman, \emph{{Hard Elastic Scattering in QCD: Leading
  Behavior}}, \href{https://doi.org/10.1016/0550-3213(89)90372-6}{\emph{Nucl.
  Phys.} {\bfseries B325} (1989) 62--100}.

\bibitem{Sotiropoulos:1993rd}
M.~G. Sotiropoulos and G.~F. Sterman, \emph{{Color exchange in near forward
  hard elastic scattering}},
  \href{https://doi.org/10.1016/0550-3213(94)90357-3}{\emph{Nucl. Phys.}
  {\bfseries B419} (1994) 59--76},
  [\href{https://arxiv.org/abs/hep-ph/9310279}{{\ttfamily hep-ph/9310279}}].

\bibitem{Contopanagos:1996nh}
H.~Contopanagos, E.~Laenen and G.~F. Sterman, \emph{{Sudakov factorization and
  resummation}},
  \href{https://doi.org/10.1016/S0550-3213(96)00567-6}{\emph{Nucl. Phys.}
  {\bfseries B484} (1997) 303--330},
  [\href{https://arxiv.org/abs/hep-ph/9604313}{{\ttfamily hep-ph/9604313}}].

\bibitem{Kidonakis:1998nf}
N.~Kidonakis, G.~Oderda and G.~F. Sterman, \emph{{Evolution of color exchange
  in QCD hard scattering}},
  \href{https://doi.org/10.1016/S0550-3213(98)00441-6}{\emph{Nucl. Phys.}
  {\bfseries B531} (1998) 365--402},
  [\href{https://arxiv.org/abs/hep-ph/9803241}{{\ttfamily hep-ph/9803241}}].

\bibitem{Oderda:1999kr}
G.~Oderda, \emph{{Dijet rapidity gaps in photoproduction from perturbative
  QCD}}, \href{https://doi.org/10.1103/PhysRevD.61.014004}{\emph{Phys. Rev.}
  {\bfseries D61} (2000) 014004},
  [\href{https://arxiv.org/abs/hep-ph/9903240}{{\ttfamily hep-ph/9903240}}].

\bibitem{Bonciani:2003nt}
R.~Bonciani, S.~Catani, M.~L. Mangano and P.~Nason, \emph{{Sudakov resummation
  of multiparton QCD cross-sections}},
  \href{https://doi.org/10.1016/j.physletb.2003.09.068}{\emph{Phys. Lett.}
  {\bfseries B575} (2003) 268--278},
  [\href{https://arxiv.org/abs/hep-ph/0307035}{{\ttfamily hep-ph/0307035}}].

\bibitem{Appleby:2003hp}
R.~B. Appleby, \emph{{Rapidity gap physics at contemporary colliders}}, Ph.D.
  thesis, Manchester U., 2003.
\newblock \href{https://arxiv.org/abs/hep-ph/0311210}{{\ttfamily
  hep-ph/0311210}}.

\bibitem{Banfi:2004yd}
A.~Banfi, G.~P. Salam and G.~Zanderighi, \emph{{Principles of general
  final-state resummation and automated implementation}},
  \href{https://doi.org/10.1088/1126-6708/2005/03/073}{\emph{JHEP} {\bfseries
  03} (2005) 073}, [\href{https://arxiv.org/abs/hep-ph/0407286}{{\ttfamily
  hep-ph/0407286}}].

\bibitem{Kyrieleis:2005dt}
A.~Kyrieleis and M.~H. Seymour, \emph{{The Colour evolution of the process $q q
  \to q q g$}},
  \href{https://doi.org/10.1088/1126-6708/2006/01/085}{\emph{JHEP} {\bfseries
  01} (2006) 085}, [\href{https://arxiv.org/abs/hep-ph/0510089}{{\ttfamily
  hep-ph/0510089}}].

\bibitem{Sjodahl:2008fz}
M.~Sj\"odahl, \emph{{Color evolution of 2 $\to$ 3 processes}},
  \href{https://doi.org/10.1088/1126-6708/2008/12/083}{\emph{JHEP} {\bfseries
  12} (2008) 083}, [\href{https://arxiv.org/abs/0807.0555}{{\ttfamily
  0807.0555}}].

\bibitem{Forshaw:2008cq}
J.~R. Forshaw, A.~Kyrieleis and M.~H. Seymour, \emph{{Super-leading logarithms
  in non-global observables in QCD: Colour basis independent calculation}},
  \href{https://doi.org/10.1088/1126-6708/2008/09/128}{\emph{JHEP} {\bfseries
  09} (2008) 128}, [\href{https://arxiv.org/abs/0808.1269}{{\ttfamily
  0808.1269}}].

\bibitem{Seymour:2005ze}
M.~H. Seymour, \emph{{Symmetry of anomalous dimension matrices for colour
  evolution of hard scattering processes}},
  \href{https://doi.org/10.1088/1126-6708/2005/10/029}{\emph{JHEP} {\bfseries
  10} (2005) 029}, [\href{https://arxiv.org/abs/hep-ph/0508305}{{\ttfamily
  hep-ph/0508305}}].

\bibitem{Seymour:2008xr}
M.~H. Seymour and M.~Sj\"odahl, \emph{{Symmetry of anomalous dimension matrices
  explained}}, \href{https://doi.org/10.1088/1126-6708/2008/12/066}{\emph{JHEP}
  {\bfseries 12} (2008) 066},
  [\href{https://arxiv.org/abs/0810.5756}{{\ttfamily 0810.5756}}].

\bibitem{Fukushima:2007dy}
K.~Fukushima and Y.~Hidaka, \emph{{Light projectile scattering off the color
  glass condensate}},
  \href{https://doi.org/10.1088/1126-6708/2007/06/040}{\emph{JHEP} {\bfseries
  06} (2007) 040}, [\href{https://arxiv.org/abs/0704.2806}{{\ttfamily
  0704.2806}}].

\bibitem{Kovner:2001vi}
A.~Kovner and U.~A. Wiedemann, \emph{{Eikonal evolution and gluon radiation}},
  \href{https://doi.org/10.1103/PhysRevD.64.114002}{\emph{Phys. Rev.}
  {\bfseries D64} (2001) 114002},
  [\href{https://arxiv.org/abs/hep-ph/0106240}{{\ttfamily hep-ph/0106240}}].

\bibitem{Marquet:2010cf}
C.~Marquet and H.~Weigert, \emph{{New observables to test the Color Glass
  Condensate beyond the large-$N_c$ limit}},
  \href{https://doi.org/10.1016/j.nuclphysa.2010.05.056}{\emph{Nucl. Phys.}
  {\bfseries A843} (2010) 68--97},
  [\href{https://arxiv.org/abs/1003.0813}{{\ttfamily 1003.0813}}].

\bibitem{Dominguez:2011wm}
F.~Dominguez, C.~Marquet, B.-W. Xiao and F.~Yuan, \emph{{Universality of
  Unintegrated Gluon Distributions at small x}},
  \href{https://doi.org/10.1103/PhysRevD.83.105005}{\emph{Phys. Rev.}
  {\bfseries D83} (2011) 105005},
  [\href{https://arxiv.org/abs/1101.0715}{{\ttfamily 1101.0715}}].

\bibitem{Iancu:2013dta}
E.~Iancu and J.~Laidet, \emph{{Gluon splitting in a shockwave}},
  \href{https://doi.org/10.1016/j.nuclphysa.2013.07.012}{\emph{Nucl. Phys.}
  {\bfseries A916} (2013) 48--78},
  [\href{https://arxiv.org/abs/1305.5926}{{\ttfamily 1305.5926}}].

\bibitem{Lepage:1980fj}
G.~P. Lepage and S.~J. Brodsky, \emph{{Exclusive Processes in Perturbative
  Quantum Chromodynamics}},
  \href{https://doi.org/10.1103/PhysRevD.22.2157}{\emph{Phys. Rev.} {\bfseries
  D22} (1980) 2157}.

\bibitem{Dokshitzer:1991wu}
Y.~L. Dokshitzer, V.~A. Khoze, A.~H. Mueller and S.~I. Troian, \emph{{Basics of
  perturbative QCD}}.
\newblock Gif-sur-Yvette, France: Ed. Frontieres (1991) 274 p. (Basics of),
  1991.

\bibitem{Binosi:2008ig}
D.~Binosi, J.~Collins, C.~Kaufhold and L.~Theussl, \emph{{JaxoDraw: A Graphical
  user interface for drawing Feynman diagrams. Version 2.0 release notes}},
  \href{https://doi.org/10.1016/j.cpc.2009.02.020}{\emph{Comput. Phys. Commun.}
  {\bfseries 180} (2009) 1709--1715},
  [\href{https://arxiv.org/abs/0811.4113}{{\ttfamily 0811.4113}}].

\end{thebibliography}
%

\providecommand{\href}[2]{#2}\begingroup\raggedright\endgroup

\end{document}